\documentclass{article}
\usepackage{authblk}
\usepackage[english]{babel}
\usepackage[a4paper,top=1.5cm,bottom=1.5cm,left=1.5cm,right=1.5cm]{geometry}

\usepackage{hyperref}
\usepackage{xcolor}
\usepackage{graphicx}
\usepackage{caption}
\usepackage{subcaption}
\usepackage[T1]{fontenc}
\usepackage{setspace}
\usepackage{csquotes}
\usepackage{amsmath}
\usepackage{amsfonts}
\usepackage{amssymb}
\usepackage{amsthm}
\usepackage{braket}
\usepackage[numbers,sort&compress]{natbib}
\usepackage{layout}
\usepackage{comment}
\usepackage{url}
\usepackage{tabularx}
\newcommand{\Bovergam}{20}
\newcommand{\Bovergamg}{3.8}
\newcommand{\WilsonLo}{75.3\%}
\newcommand{\WilsonHi}{94.1\%}
\newcommand{\NumAPC}{0.2}
\newcommand{\NumAPIDD}{0.833}
\newcommand{\NumAPAAGH}{0.0067}
\newcommand{\NumAPAHH}{0.077}
\newcommand{\NumAPEA}{0.40}
\newcommand{\NumAPF}{0.5}
\newcommand{\NumAPAB}{0.01}
\newcommand{\NumAPI}{0.8}
\newcommand{\NumBPA}{1.0}
\newcommand{\NumBPC}{1.2}
\newcommand{\NumBPE}{1.4}
\newcommand{\NumBPG}{1.6}
\newcommand{\NumAPB}{0.1}
\newcommand{\NumAPBF}{0.15}
\newcommand{\NumAPAEC}{0.042}
\newcommand{\NumAPAF}{0.05}
\newcommand{\NumAPBA}{0.10}
\newcommand{\NumAPCA}{0.20}
\newcommand{\NumAPCF}{0.25}
\newcommand{\NumHAPID}{70.83}
\newcommand{\NumIBPCF}{81.25}
\newcommand{\NumFIPDD}{58.33}
\newcommand{\NumIHPF}{87.5}

\newif\ifREVIEWCOPY
\REVIEWCOPYfalse
\ifREVIEWCOPY
  \usepackage[left,mathlines]{lineno}
  \AtBeginDocument{\doublespacing\renewenvironment{multicols}[1]{}{}\linenumbers}
\fi

\newtheorem{theorem}{Theorem}

\newtheorem{corollary}{Corollary}

\usepackage{multicol}
\newcommand{\dd}{\,\mathrm{d}}

\definecolor{LightGray}{gray}{0.9}
\hypersetup{
  colorlinks=true,
  linkcolor={blue},
  filecolor={maroon},
  citecolor={blue},
  urlcolor={blue}}

\title{A Single Atom in Front of a Mirror is a Universal Reservoir Computer}
\author[1,*,$\dagger$]{Peter J. Ehlers}
\author[1,*,$\dagger$]{Phi Hung Nguyen}
\author[1,2]{Kanu Sinha}
\author[1]{Noelle Daigle}
\author[1]{Travis W. Sawyer}
\author[3]{Hendra I. Nurdin}
\author[1,*]{Daniel Soh}
\affil[1]{\small Wyant College of Optical Sciences, University of Arizona, Tucson, AZ, USA}
\affil[2]{Department of Physics, University of Arizona, Tucson, AZ, USA}
\affil[3]{School of Electrical Engineering and Telecommunications, University of
New South Wales, Sydney, Australia}
\affil[*]{Corresponding authors: ehlersp@arizona.edu, hpnguye7@arizona.edu, danielsoh@optics.arizona.edu}
\affil[$\dagger$]{These authors contributed equally.}
\date{}

\begin{document}

\maketitle

\begin{abstract}
Universal approximation in reservoir computing is typically associated with a class of reservoirs. We show that universality can be associated with a single reservoir, considering a minimal setup of a single atom in front of a mirror. In its linear-transducer limit, our reservoir is a universal approximator of fading-memory maps under an operating class of checkable conditions, with a rate constant measured at the operating point. A given reservoir can reach arbitrary accuracy by changing measurement settings. The proof gives an explicit recipe: for a target accuracy, it specifies the required physical resources and resonator modes. Enlarging the number of accessible modes increases the matchable kernel span without reducing capability. Beyond the linear limit, the atom’s saturation replaces high-order polynomial readouts, and the device operates on real-world tasks alongside classical baselines. Our results highlight an example of universality with a minimal quantum setup.

\end{abstract}

\begin{multicols}{2}

\section{Introduction}

How much hardware does a universal learning machine require? In reservoir computing---a fixed dynamical system replacing trained recurrent connections, only a linear readout trained \cite{jaeger2004harnessing, maass2002realtime}---the answer has always been: more than one machine.

Universality has been proven for echo-state networks \cite{GRIGORYEVA2018495, GONON202110}, state-affine systems \cite{JMLR:v19:18-020}, stochastic echo-state networks \cite{EhlersNurdinSoh2025}, and Ising, spin-ensemble, and Gaussian quantum reservoirs \cite{chen2019learning, chen2020temporal, nokkala2021gaussian, fujii2017harnessing}---always associated with a \emph{class}, by ranging over weights, couplings, or realizations, with no single member carrying the property. The two known single-member constructions \cite{Cuchiero, gauthier2021next} obtain it by making the reservoir digital, so it is no longer physical hardware at all. Physically realized reservoirs \cite{tanaka2019recent} scale expressivity by adding components; the counter-trend, the single-node delay-line reservoir \cite{appeltant2011, larger2017high}, has been hardware reservoir computing's workhorse for over a decade with no universality guarantee ever established for it (Supplementary Table~S.5). Supplementary Table~S.5 states the basis of that claim from a literature search across the delay-reservoir theory line \cite{grigoryeva2014tdr, koster2021master, ortin2015unified, stelzer2021deep}: capacity analyses, class-level equivalences and folded-in-time emulations, never a universality theorem attached to one fixed physical member.

\begin{figure*}[h]
    \centering
    \includegraphics[width=\textwidth]{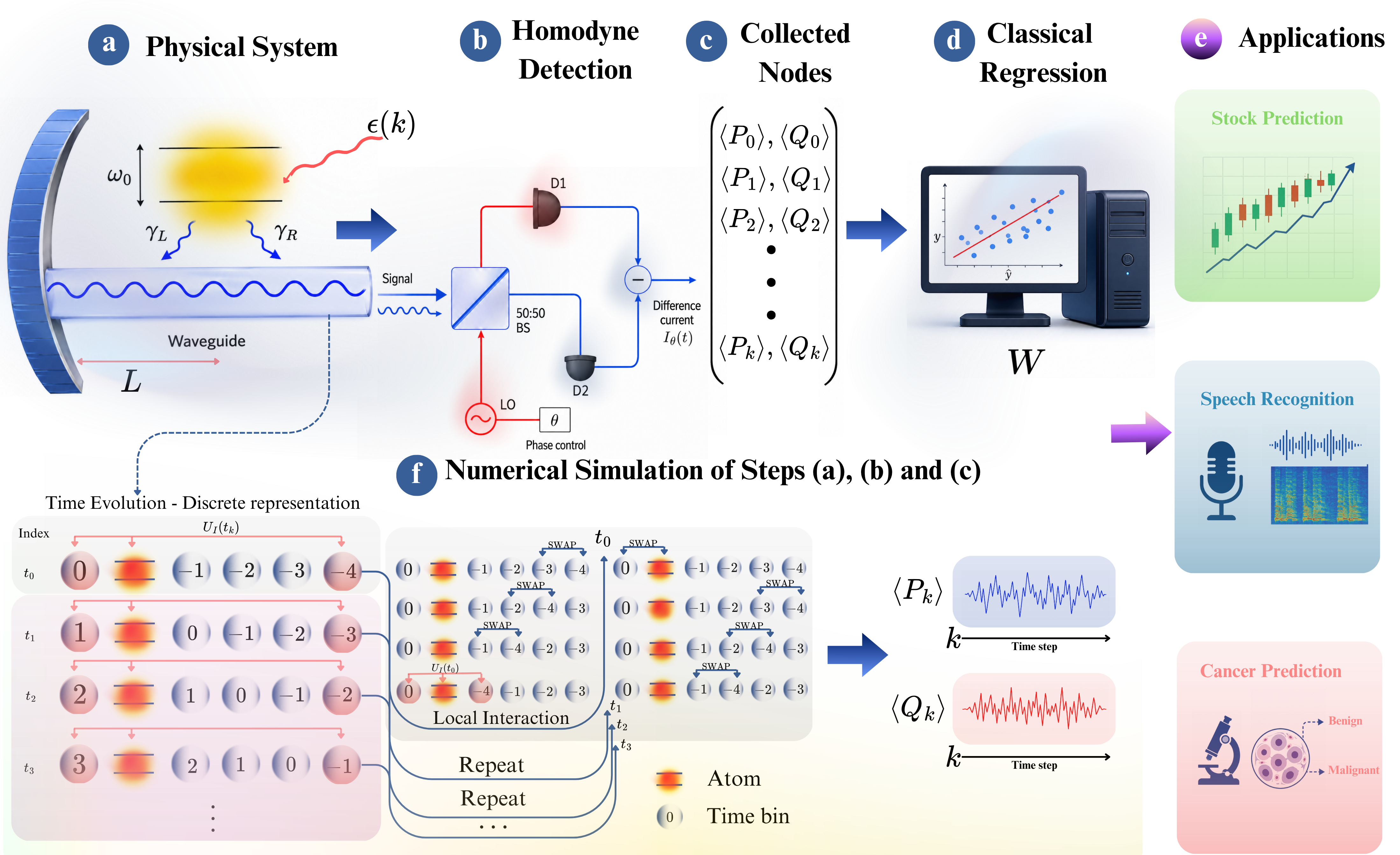}
    \caption{Overview of the minimal quantum reservoir computer performing real-world tasks.
    \textbf{a}, A single atom with transition energy $\hbar\omega_0$ sits a distance $L$ in front of a mirror, driven by the discrete input $\epsilon(k)$ and decaying into the waveguide to the left and right at rates $\gamma_L,\gamma_R$. The round-trip delay $\tau=2L/v$ produces coherent feedback.  
    \textbf{b--c}, Homodyne detection collects quadrature nodes in each time bin.
    \textbf{d}, A classical linear readout $W$ predicts the target $y_k$.  
    \textbf{e}, The three real-world tasks: financial prediction, speech recognition, and tumor classification.
    \textbf{f}, Time-bin matrix-product-state scheme (Methods). With delay $\ell=4$, at $k=0$ a sequence of $\mathcal{SWAP}$ gates brings time bin $-4$ adjacent to the atom, the local evolution $\hat{U}_I(t_0)$ is applied, and the bins are returned to order; a final $\mathcal{SWAP}$ advances $k=0\to1$. The process repeats to the final time. 
    }
    \label{fig:model}
\end{figure*}

In this work we show the minimal architecture can carry the guarantee: a single atom before a mirror is a provably universal reservoir computer. ``Atom'' means any two-level emitter coupled to the guided field: trapped atom or ion, quantum dot, colour centre, or superconducting artificial atom. The device (Fig.~\hyperref[fig:model]{\ref*{fig:model}a}) has one geometric parameter entering the feedback structure, the atom--mirror distance $L$, alongside the fixed band envelope and mode spacing that define the device, and one scaling resource, the number of field modes it accesses, set by geometry and bandwidth. We prove universality of this single machine in a precisely stated limit with explicit rates; capability monotone, and the matchable kernel span strictly growing, with mode number; and genericity of the conditions, holding at the exact parameters of every simulation here. It is Supplementary Table~S.5's only entry combining physical hardware, a single-member guarantee and strict scaling. The theorem is proven for the architecture with its node operated as a linear element and the nonlinearity carried by the readout, and is available precisely because the atom--mirror loop then decomposes into independent linear modes---the structure the classical device's nonlinear node destroys.

The result is possible because quantum optics supplies, in one passive object, every ingredient reservoir computing otherwise assembles from parts (Box~1): the standing-wave modes are the network, the atom's saturable response the activation function, the round-trip delay $\tau=2L/v$ the recurrence \cite{Laser, Collective, PhysRevA.90.012113}---a geometry realized with trapped atoms \cite{eschner2001light} and superconducting artificial atoms \cite{hoi2015probing}. Nothing inside is trained, wired, or manufactured as a network.  

\begin{figure*}[t]
\centering
\fbox{\parbox{0.96\linewidth}{\small
\textbf{Box 1 $|$ The device in plain terms.}
A reservoir computer needs three things: a large state space to hold information, a nonlinearity to mix it, and recurrence to remember it. This device gets all three from geometry. \emph{Memory:} light emitted by the atom travels to the mirror and comes back after a delay $\tau=2L/v$; what the atom does now therefore depends on what it did one round trip ago, and, through repeated reflections, on progressively fainter echoes of its earlier history---a memory that fades at a rate set by how fast light leaks out rather than by any clock or register. \emph{State space:} between atom and mirror the field forms standing waves, one for each frequency the waveguide supports; each behaves as an independent oscillator, and together they play the role of the hidden neurons of a recurrent network---added by widening the usable bandwidth rather than by fabricating components. \emph{Nonlinearity:} a two-level atom saturates (it cannot absorb a second photon while holding one) which is the activation function, built into the atom by quantum mechanics. Training touches none of this: one measures the outgoing light and fits a linear map from measurement records to targets, the only learned object in the machine.}}
\end{figure*}

We claim no quantum computational advantage: quantum mechanics enables \emph{minimality}, collapsing the recurrent, high-dimensional, nonlinear core of a learning machine into the passive dynamics of one atom, with only a linear readout trained classically. Correspondingly the empirical standard for the demonstrations is \emph{parity} with mature classical methods, which is exactly what ``minimal hardware suffices'' predicts.

``Non-Markovian'' refers to the \emph{reduced} atomic dynamics: the atom interacts with its own past through returning photons, while the joint system is Markovian to the axioms' standard (Methods; Supplement Sec.~S.1).

\section{Results}

\subsection{Mode-space representation of the non-Markovian reservoir}
\label{sec:exact}

The description on which every claim rests: the atom--mirror reservoir, with all its delayed feedback, is captured by a single Lindblad generator on the atom plus its field modes, the feedback carried exactly---relative to axioms A1--A6 of Supplement Sec.~S.1 (rotating wave; flat coupling $g(\omega)\approx\sqrt{\gamma/2\pi}$ over bandwidth $B$; unidirectional outcoupling; classical drive; idealized measurement).  

The mirror boundary selects standing-wave modes; passing to this basis is a \emph{unitary transformation of the free field}---no auxiliary or lossy degrees of freedom, no pseudomodes---and the coupling acquires the form factor $\tilde{g}(\omega)\propto\sin(\omega\tau/2)$ sampling the standing wave at the atom. All dependence on $L$, hence all delay and feedback, resides in this coherent structure, and the retarded self-interaction is recovered identically from its frequency integral (Methods). The reservoir is described by
\begin{align}
    \label{eq:LBDN}
    \dot{\rho}(t) &= \left[\mathbf{\Gamma}_g\cdot a\rho(t), a^\dagger\right] + \left[a, \rho(t) a^\dagger\cdot\mathbf{\Gamma}_g^\dagger\right] \\ \nonumber  
    &\quad -i\eta u(t)[(a^\dagger\cdot\alpha)+(\alpha^*\cdot a),\rho(t)],
\end{align}
in which $a$ is the vector of mode annihilation operators, $u(t)$ the scalar input drive with coupling amplitude $\eta$, and $\mathbf{\Gamma}_g=i\mathbf{\Omega}+\tfrac{\gamma}{2}(v_\mathrm{geo}\otimes v_\mathrm{geo}^*)+\tfrac{\gamma_g}{2}(\alpha\otimes\alpha^*)$, with $\mathbf{\Omega}$ the diagonal matrix of standing-wave frequencies, $\alpha$ ($\alpha_k\propto\sin(\omega_k\tau/2)$) the normalized emission profile, $v_\mathrm{geo}$ the geometry-fixed, flat profile of the readout leakage channel ($v_{\mathrm{geo},k}=1/\sqrt K$), and $\gamma$, $\gamma_g$ the readout and drive-channel rates. The anti-Hermitian part carries the coherent feedback; the Hermitian rank-one terms carry the Markovian measurement- and drive-induced evolution; the homodyne local oscillator selects only which functional of the \emph{record} is computed (full reading in Supplement Sec.~S.1). A discrete $\mathbf{\Omega}$ is a comb of spacing $\Delta_0=\pi v/\ell$ with quantization length $\ell$ an operating parameter; no continuum limit enters the proof. 

\subsection{Universality in the linear-transducer limit}
\label{sec:universality}

The output is obtained from powers of the measured quadrature,
\begin{align}
    \label{eq:output}
    \hat{y}_k &= \sum_{n=0}^N\int_0^{T_\mathrm{off}}W_n(t)\,\mathrm{Tr}[q_\mathrm{geo}^n e^{\mathcal{L}_0 t}\rho_\zeta(t_k)]\dd t,
\end{align}
with $q_\mathrm{geo}=\mathrm{Re}[v_\mathrm{geo}^*\cdot a]$ the measured quadrature of the geometric leakage channel, $W_n(t)$ classical readout weights, and $\mathcal{L}_0$ the generator at $u=0$; the drive alternates input windows $T_\mathrm{on}$ and measurement windows $T_\mathrm{off}$, with readout period $T=T_\mathrm{on}+T_\mathrm{off}$. The powers come from the same homodyne record but cost $(2n-1)!!\,(v_0+q_{\max}^2)^n/\sigma^2$ shots at precision $\sigma$: the factorial price of Gaussian-limit operation, and exactly what the nonlinearity transfer of Sec.~\ref{sec:transfer} eliminates (Methods; Supplement Sec.~S.7).  

The proof rests on the \emph{adiabatic elimination of the atom}---a reduction of the atom rather than of the environment---which locks the saturable dipole into following the field and replaces it by a linear transducer, with controlled small parameter the saturation $s\sim(\varepsilon_\mathrm{INPUT}/\gamma_g)^2$ (Methods). We call this the linear-transducer (Gaussian) limit.  

\paragraph{Three kinds of resource.} The theorem quantifies over one fixed machine: \emph{fabricated hardware} never changes with the target; \emph{geometric} and \emph{measurement settings} are selected after the accuracy is announced, fabricating nothing. Measurement settings alone reach every accuracy down to an explicit envelope of the current geometry; the geometric dial, provably never wasted, moves the envelope itself (full taxonomy: Supplement Sec.~S.2.5). A network whose width is a runtime setting still stores per-width coupling data; every mode-number description of this device follows by one rule from one set of physical data (Supplement Sec.~S.1.1).

\begin{theorem}[Constructive universality with explicit rates]
    \label{thm:univ}
    Consider the class $\mathcal{C}$ of reservoir computers with dynamics Eq.~\eqref{eq:LBDN} and outputs Eq.~\eqref{eq:output} in the linear-transducer limit, realized by a single atom--mirror device operated at the explicitly constructed point of Supplement Sec.~S.2.5:
    \begin{itemize}
        \item \emph{Overlap by construction:} $\tau$ avoids the explicit node set, so $\alpha_k\neq0$, and the geometry-fixed coupler is flat, $v_{\mathrm{geo},k}=1/\sqrt K$; the overlap conditions $v_\mathrm{geo}^*\cdot v_{R,k}\neq0$, $v_{L,k}^*\cdot\alpha\neq0$ (eigendata of $\mathbf{\Gamma}_g$) then hold, with explicit margins $(1-\varepsilon_v)/(2\sqrt K)$ on the coupler side and $1/(2q\sqrt K)$ on the drive side certified by the comb-spacing condition of the next bullet, where $\varepsilon_v$ is the coupler-flatness error of Supplement Sec.~S.1 and enters every lower bound below as a subtraction;  
        \item \emph{Weak-dressing comb:} an equidistant comb with spacing $\Delta_0\geq8q\sqrt{K}\,\bar e$, where $q$ is the prime of the delay lock of the third bullet, $\bar e\geq\Vert\mathbf E\Vert$ is the dressing constant of Supplement Sec.~S.2.5 and $\mathbf E$ the Hermitian dressing of $\mathbf{\Gamma}_g$, large enough that second-order dressing shifts fall below the node tolerance, and locked to the readout period (a timing condition on $T$ alone, condition D2 of Supplement Sec.~S.2.5) so the nodes $e^{-\lambda_kT}$ sit at the $M$-th roots of unity of common radius $\bar r=e^{-\lambda_\star T}$, $\lambda_\star=(\gamma+\gamma_g)/2K$;  
        \item \emph{Graded defects and generic drive spread:} one exponentially graded defect hierarchy (Supplement Sec.~S.2.5, D5) separates all conjugation-resolved eigenvalue-sum points of order $n\leq N$ with distinct signed mode content, and the drive channel's decay-rate spread, floored by a delay lock placing the atom at a rational point of the standing wave, separates the remainder with an explicit margin, so every sum point is distinct or exactly conjugate and the higher-order selection is well posed;
        \item \emph{Weak drive, flat coupler:} conditions D4 and D6 of Supplement Sec.~S.2.5 cap the drive-channel and coupler-flatness spreads within the node-placement budget; D6 alone is checked rather than set, the coupler flatness being fabricated data.
    \end{itemize}
    Conditions D1--D4 are geometric or measurement settings of the one fixed device, while condition D5 specifies the resonator spectrum, likewise a geometric setting of the same fixed device, as Supplementary Table~S.2 records, and each is proven satisfiable and sufficient with explicit constants. Then for every continuous fading-memory target $y_k=y(u_k,u_{k-1},\dots)$ with a convergent Volterra expansion on $\mathcal{K}_{u_{\max}}=[-u_{\max},u_{\max}]^{\mathbb{Z}_-}$ and every $\epsilon>0$, there exist finite $N$, memory depth $M$, mode number $K=M$, readout period $T=O(M\ln M+\ln\epsilon^{-1})/(\gamma+\gamma_g)$, and weights $\{W_n(t)\}$ such that the single device achieves $\vert y_k-\hat{y}_k\vert\leq\epsilon$ on $\mathcal{K}_{u_{\max}}$, with the explicit bound  
    \begin{align*}
        \vert y_k-\hat y_k\vert \;&\leq\; \delta_{MN} \;+\; A(M,N)\,\big(2M\,\bar r^{\,M}+e\,M\,\bar\varepsilon\big),\\
        \bar r^{\,M}&=e^{-(\gamma+\gamma_g)T/2}.
    \end{align*}
    Here $\delta_{MN}$ is the target's Boyd--Chua truncation error, $A(M,N)=\sum_{n=1}^{N}n\,2^{2n}u_{\max}^{n}M^{n-1}\Vert h^{(M,N)}_n\Vert_\infty$ is set by the target's truncated kernels and input bound alone, $\bar\varepsilon$ is the node-placement error driven below any tolerance by the comb spacing, and $\bar r^{\,M}$ recovers the full trace budget over the memory span. No weight norm, revival time, or loop-stability constant enters; no infinite-mode limit is invoked; parameters are chosen once and in order, $(M,N)$ then $T$ then the comb, with no circular dependence (Supplement Sec.~S.2.5). The rate is claimed for Volterra-convergent targets; merely continuous fading-memory targets are reached through the density route of Supplement Sec.~S.3.2, without a rate.  
\end{theorem}

The certificate's price belongs beside the theorem. The designed operating point demands a comb spacing growing exponentially with memory depth, a relative timing precision falling exponentially with it, a drive weak enough to inflate shot budgets by orders of magnitude, and a rate ordering $\gamma_g\ll\lambda_\star$ (condition D4) outside the adiabatic derivation of the working generator from the physical atom, so that at this point the theorem is a statement about the Gaussian generator of Eq.~\eqref{eq:LBDN} (Methods; Supplement Sec.~S.2.5). Its reach, bounded in Methods, does not extend beyond $M=2$--$3$ on any platform. Every simulation in this paper instead operates at generic points whose distinctness conditions are verified rather than designed, a relation the Supplement makes precise: the certificate proves existence with explicit constants, and generic operation is how the device is used.

The full proof is in Supplement Sec.~S.2, its logic in Methods. Two features shape it. A uniform spectral gap is provably impossible---the Hermitian part of $\mathbf{\Gamma}_g$ has rank two, so the slowest mode closes as $1/K$---and the proof turns this into the rate's engine, the fixed budget $(\gamma+\gamma_g)/2$ over $K=M$ modes being recovered in full over the memory span. And the bound never touches the matched weights: once kernels are matched at lags $\leq M$, deeper response is fixed by the spectrum alone (extrapolation identity, Sec.~S.2.5). Delay stability is the condition under which the \emph{device} has fading memory (Secs.~S.4, S.10), never an input to the bound.

Each condition is physical: $v_{L,k}^*\cdot\alpha\neq0$ says no standing wave has a node at the atom, and $v_\mathrm{geo}^*\cdot v_{R,k}\neq0$ is immediate for the flat coupler. Delay stability $D(c_0)$---roots of $s+\tfrac{\gamma}{2}(1+re^{i\phi}e^{-s\tau})=0$ with $\mathrm{Re}[s]\leq-c_0$, the quantum echo-state condition, failing only at the dark state---holds with wide margins ($c_0=0.62\gamma$, $0.36\gamma$ at $\gamma\tau=0.5$, $1$) and is not consumed by the bound.  

A structural consequence is monotonic scaling.
\begin{corollary}[Strict scaling]
    \label{cor:scaling}  
    For $M\leq M'$ and $N\leq N'$, every accuracy achievable by the $(M,N)$-member of $\mathcal{C}$ on every target is achievable by the $(M',N')$-member (Supplement Sec.~S.3.1, capability containment): enlarging the accessible mode number of the single device---the nested family of Supplement Sec.~S.1.1, no spectral relation across mode numbers assumed---never decreases capability. The exactly matchable kernel span strictly grows: at $(M',N')$ every block of the enlarged index set is realizable, while blocks supported outside $[1,M]^n$ are not indexable at $(M,N)$ (Supplement Sec.~S.3.1, Proposition~2). The statement is about spans; a capability-level separation is not claimed.
\end{corollary}
Universality \emph{itself} is also reachable by the standard Stone--Weierstrass route~\cite{GRIGORYEVA2018495, chen2019learning}; we prove Theorem~\ref{thm:univ} constructively because only the kernel route certifies scaling is never wasted, with explicit rates no density argument produces (Supplementary Discussion Sec.~S.12; numerics, Sec.~S.11).

Gaussian-reservoir \emph{class} universality~\cite{nokkala2021gaussian}, the nearest prior art, ranges over trainable couplings; here the swept resource is one passive geometric parameter of one fixed device, and the guarantee arrives with explicit rates and a strict-scaling corollary (Supplement Sec.~S.8).

What does the work is \emph{linear multimode} structure with generically non-resonant frequencies, supplied here by quantization but not exclusive to it. The dichotomy is \emph{nonlinear single node with virtual nodes}---the classical delay line, and why it resisted a theorem---versus \emph{linear field with independent modes}, this device; one atom packages encoding, mode structure and saturable nonlinearity in one passive component. The resulting objection and our answer are in Supplementary Discussion Sec.~S.12.

Finally, the theorem--device gap is bounded: the saturable device's kernels deviate from the Gaussian-limit kernels by $O(s)$, a computable number at each simulated operating point (Supplement Sec.~S.5).

\subsection{Genericity of the universality conditions}
\label{sec:genericity}

Theorem~\ref{thm:univ} is only useful if realizable devices meet its conditions, and they hold for almost every mirror distance: each resonance function $R_{\{n\}}(\tau)=\sum_kn_k\lambda_k(\tau)$ is analytic and provably not identically zero, with no arithmetic condition on the spectrum (the pointwise shortcut is simply false; Supplement Sec.~S.6), so the bad delays form a countable union of discrete sets; overlaps fail only at standing-wave nodes, delay stability only at the dark state---likewise measure zero.  

We verified this on the exact device spectra: over a dense $\tau$ sweep the eigenvalues are distinct with positive real parts, overlaps stay bounded away from zero except at predicted node distances, finite-order non-resonance surrogates stay positive away from isolated $\tau$, and the delay-stability margin is $c_0\geq0.36\gamma$ throughout the simulated regime. The Gaussian-limit devices of Fig.~\ref{fig:convergence} and of the diagonalization study are verified to satisfy the overlap and non-resonance conditions on which Theorem~\ref{thm:univ} rests, while the saturable devices of the remaining figures are related to the theorem through the kernel-continuity bound of Supplement Sec.~S.5 rather than through membership, with measured margins reported in Supplement Sec.~S.6; the designed conditions D1--D5 are a separate constructive certificate, and the simulated devices are generically fabricated and do not meet them.

\subsection{One device, one axis: convergence with mode number}
\label{sec:convergence}

\begin{figure*}[t]
    \centering
    \includegraphics[width=\textwidth]{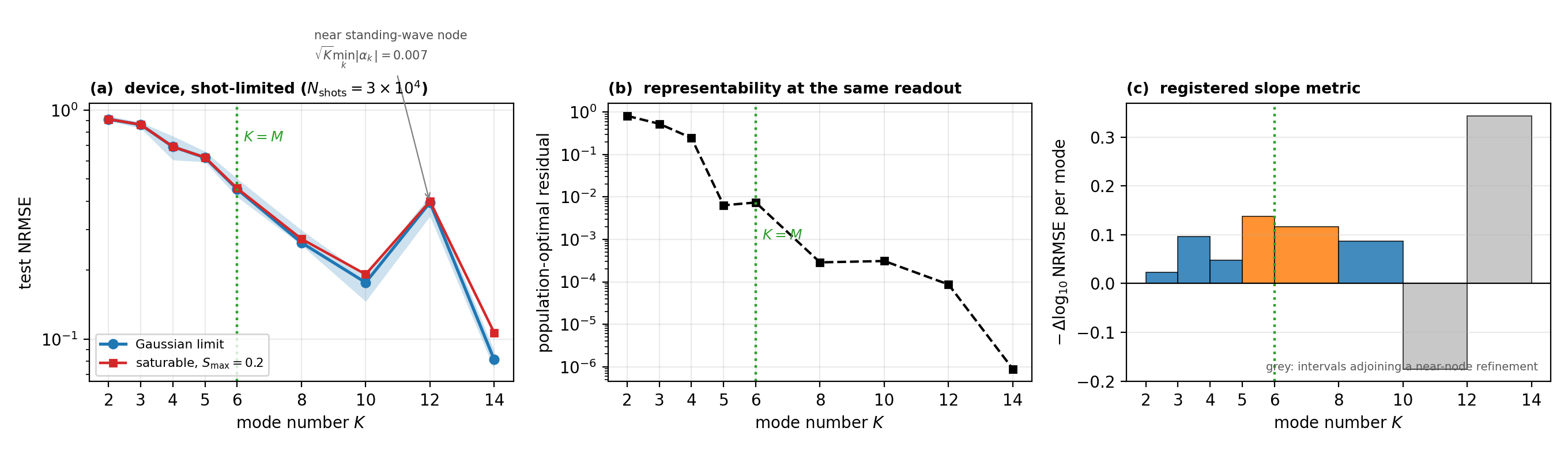}
    \caption{Convergence of a single device with mode number. \textbf{a}, Test-set approximation error (median over $12$ input realizations; band, interquartile range) on a target with known Volterra kernels ($M=6$, $N=2$, random bounded kernels) versus the mode number $K$ of one device with a nested dispersive comb at fixed retained band and fixed mirror distance, so that $\Delta_0=B/K$ and the readout node count is held fixed across the sweep. Blue: linear-transducer (Gaussian) regime, exact closed-form dynamics; red: the \emph{saturable} device on the identical protocol (displaced-frame quasi-static Bloch route of Methods, quasi-static Bloch computation; $S_{\max}=\NumAPC$, $\langle\sigma_z\rangle$ driven from $-1$ to $-\NumAPIDD$). Both carry shot noise at $N_\mathrm{shots}=3\times10^{4}$, the one-percent relative precision of Methods. The non-monotonicity at $K=12$ is a device effect rather than scatter: at that comb refinement one mode sits near a standing-wave node, $\sqrt K\min_k|\alpha_k|=\NumAPAAGH$ against $\NumAPAHH$ at the next lowest refinement and $\NumAPEA$ at $K=10$, the isolated overlap failure the genericity analysis predicts. In \textbf{c} the two intervals adjoining that refinement are greyed; excluding it on the overlap diagnostic alone leaves the maximum on $K=5\to6$. \textbf{b}, The population-optimal residual over the same feature set, i.e.\ what the device can represent rather than what it achieves: it falls by roughly six orders of magnitude across the sweep, dropping sharply at $K=5$ and running from two to five orders below the shot-limited error of \textbf{a} thereafter, the exact-realizability statement of the theorem. Beyond that point the measured error in \textbf{a} is limited by the shot budget rather than by mode number. \textbf{c}, The registered slope metric, $-\Delta\log_{10}$NRMSE per mode across the sampled grid; the maximum falls in the region $K\approx M$ (orange). The metric, the grid, the seed count and the claim under test were fixed before the sweep was run; the regulariser is selected on a validation split and never on test.}  
    \label{fig:convergence}
\end{figure*}

The theorem and its corollary make a testable prediction: for a target of known Volterra structure, the single device's error should fall as accessible mode number grows, dropping sharply once $K$ reaches $M$ and tracking the proof's envelope. Figure~\ref{fig:convergence} tests this in the linear-transducer regime, where the test is exact: a fixed target ($M=6$, $N=2$, random bounded kernels), one device at fixed mirror distance, mode number swept, weights trained by linear regression. Under a slope metric registered before the sweep---the change in $\log_{10}$NRMSE per mode across the sampled grid---the measured error falls by a factor $11$. The registered claim, that the steepest per-mode descent falls in the region $K\approx M$, does \emph{not} hold on the full grid: the maximum sits on the last interval, and it does so because of a single comb refinement.

At $K=12$ one mode lies near a standing-wave node (the per-mode overlap values are given in the caption of Fig.~\ref{fig:convergence}); the error rises there and the recovery beyond it produces the largest slope. Excluding refinements flagged by that overlap diagnostic alone---a criterion independent of the measured error, though one we did not register in advance---the maximum falls on the interval $K=5\to6$, exactly at $K=M$. We report both. The episode is itself the genericity statement of Sec.~\ref{sec:genericity} observed directly: overlaps fail only at isolated node distances, and one such distance is sampled here. What the device can represent and what it achieves separate there: the population-optimal residual at the same readout falls by roughly six orders of magnitude across the sweep (Fig.~\ref{fig:convergence}b), so beyond the knee the device represents the target far more accurately than it achieves it and the residual error is set by the shot budget rather than by mode number.

Overlaid is the same protocol on the \emph{saturable} device (Bloch response, saturation depth $S_{\max}=0.2$): it converges along the same axis, tracking the Gaussian curve closely and separating from it only beyond the knee, with an eight-fold error reduction. Its floor here is higher, as it must be: the target's kernels are drawn from the family the Gaussian limit represents exactly, so the residual gap measures kernel distortion rather than any failure of mode scaling; on tasks outside the Gaussian family the ordering reverses (Fig.~\ref{fig:withwithout}), which is the nonlinearity-transfer result of Sec.~\ref{sec:transfer}.  

Two insights surfaced (Methods): a trained readout needs only nonzero splittings, converging well below the sufficient bound, and dispersion lifts the fatal exact degeneracies.

\subsection{Nonlinearity transfer: feedback replaces polynomial readout}
\label{sec:transfer}

The proof lives in the Gaussian limit, where all nonlinearity must be supplied by the readout's powers $q_\mathrm{geo}^n$. Restoring the atom's saturable response moves the nonlinearity from the measurement into the hardware: delayed self-interference composed with a saturable dipole generates nonlinear dependence on input history that a purely \emph{linear} readout harvests. Theorem and mechanism are two endpoints of one dial---where the nonlinearity resides.

\begin{figure*}[t]
    \centering
    \begin{minipage}{0.94\linewidth}
        \centering
        \includegraphics[width=\linewidth]{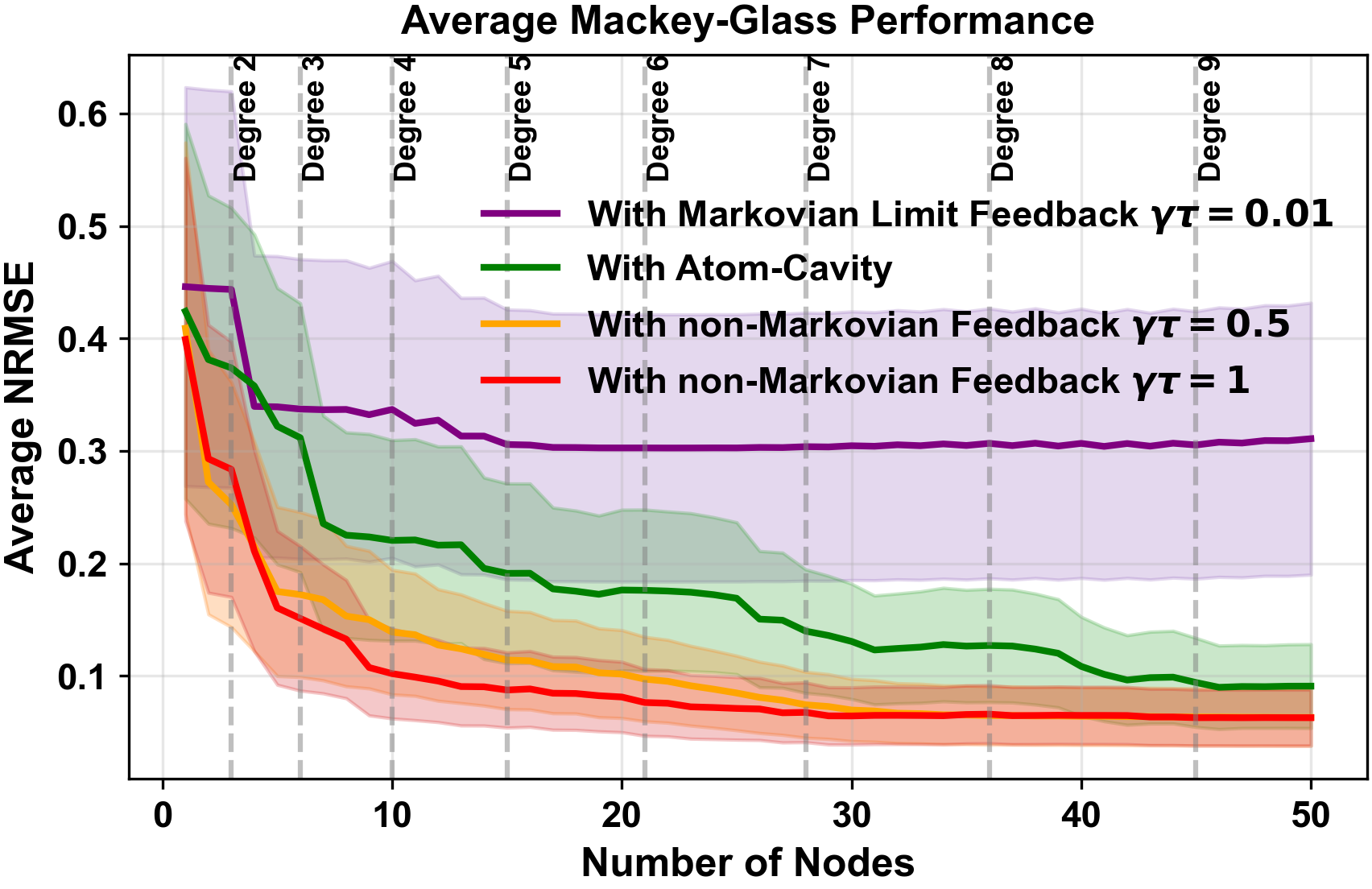}
    \end{minipage}
    \hfill
    \begin{minipage}{0.94\linewidth}
        \centering
        \includegraphics[width=\linewidth]{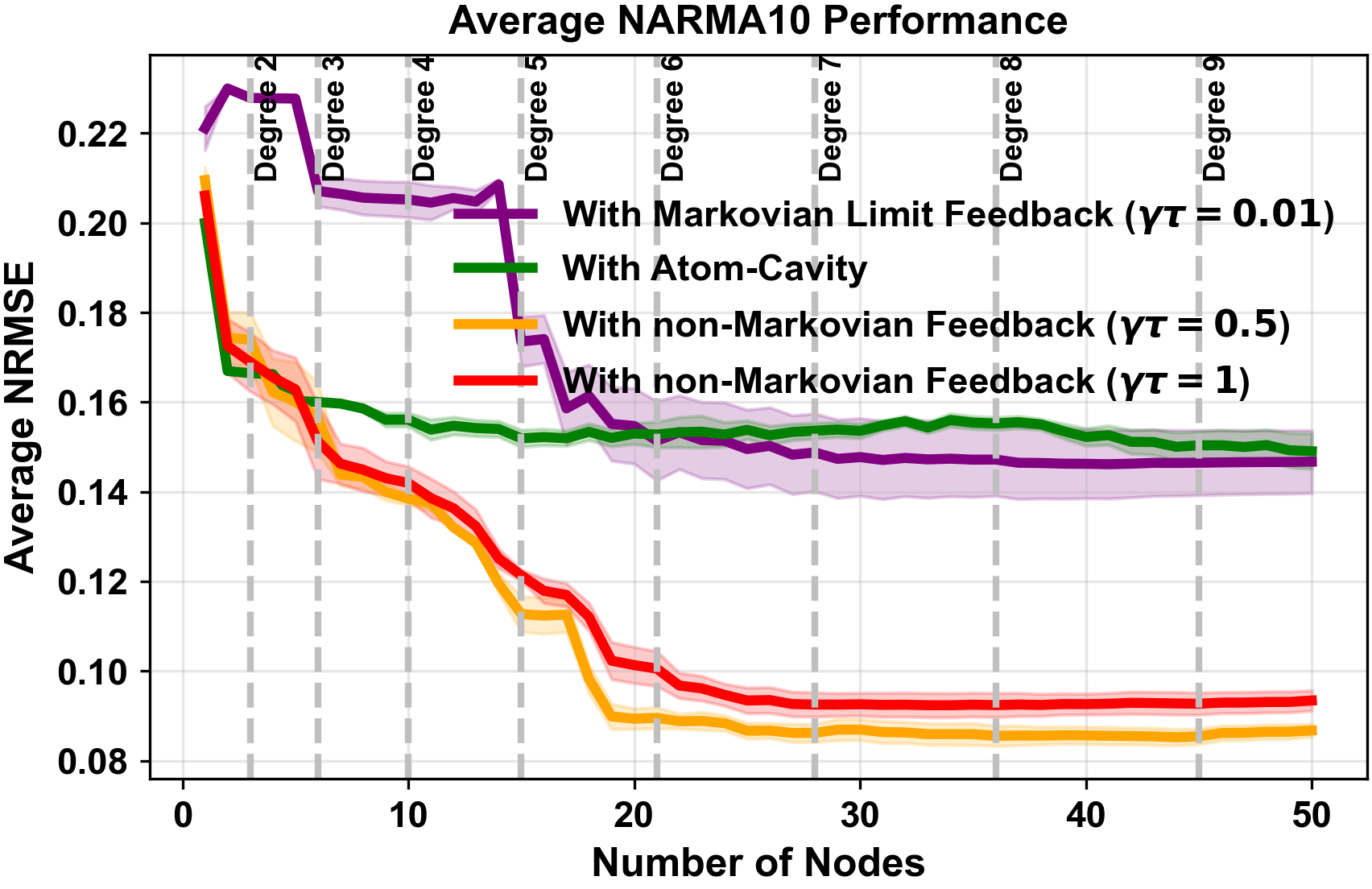}
    \end{minipage}
    \caption{Nonlinearity transfer. A non-Markovian reservoir with a purely linear readout (feedback $\gamma\tau=\NumAPF,1$) is compared against a Markovian atom--cavity reservoir trained to increasing polynomial order (vertical guides mark the polynomial degree at which the Markovian device gains each corresponding node) and against the Markovian-limit feedback $\gamma\tau=\NumAPAB$, on Mackey--Glass (top) and NARMA10 (bottom). Parameters: $\Delta t=1$, $\theta=\phi=\pi/3$, $g=(\gamma/2(1+\cos\phi),\gamma/4(1+\cos\phi))$ with $\phi=\pi-\omega_L\tau$ the round-trip phase entering the comparator's mirror-modified decay (Methods, ``Feedback-free comparison model''), averaged over $\varepsilon_\mathrm{INPUT}=(\NumAPI,\NumBPA,\NumBPC,\NumBPE,\NumBPG)\gamma$. The linear-readout feedback reservoir matches or surpasses ninth-order polynomial readout. Solid lines: mean NRMSE over the five input strengths $\varepsilon_\mathrm{INPUT}$; shaded bands: $\pm1$ s.e.m.\ over those strengths. This spread is over a swept device parameter rather than over repeated runs; no seed-resolved variability is reported for this figure, and the comparison is stated as an ordering rather than as a tested difference.}  
    \label{fig:withwithout}
\end{figure*}

Figure~\ref{fig:withwithout} makes this quantitative: a non-Markovian reservoir with a \emph{linear} readout against a Markovian atom--cavity reservoir trained to higher polynomial order, on Mackey--Glass and NARMA10, as node count grows. (One feedback node exposes two quadrature features against the comparator's one; the resource-matched comparison, an axis relabeling anchored at the plateau, is in Methods.) Despite purely linear measurements, the feedback reservoir matches or surpasses the polynomial-readout comparator, reaching by moderate node counts what the Markovian device attains only at high order or not at all; feedback strength is non-monotonic, a moderate delay optimal, since excessive memory retains stale information. Non-Markovian feedback thus substantially reduces, and can eliminate, the demanding requirement of high-order polynomial readouts. No tuned, noise-matched comparison against a classical echo-state network is claimed here; the reasons, and the noiseless classical reference reported with the financial study, are set out in Supplement Sec.~S.9.

This figure also answers the sharpest objection to any minimal-hardware claim---that the regression rather than the physics is the computer. If so, the dynamics would be immaterial; instead the same linear readout on the same feature count performs very differently with feedback ($\gamma\tau=0.01$ versus $0.5,1$).

\subsection{Operation on real-world tasks}
\label{sec:realworld}

We evaluated the reservoir on spoken-digit recognition and tumor classification (Fig.~\hyperref[fig:NMR_figure2]{\ref*{fig:NMR_figure2}b,c}); financial forecasting (panel a) is analyzed in Supplement Sec.~S.9, its one-step-ahead structure being closely tracked by a persistence predictor. Under the minimality thesis the evidentiary standard is \emph{parity}: matching mature classical methods is exactly what ``minimal hardware suffices'' predicts, and we benchmark against standard methods and trivial baselines, reporting ties and shortfalls plainly. The standard is falsifiable: parity failure at achievable node counts, or performance insensitive to feedback strength, would count against the thesis. The falsifier is tested and rejected twice, by Fig.~\ref{fig:withwithout} and by the feedback-loss sweep of Supplement Sec.~S.10, where severing the loop collapses performance to the feedback-free plateau.

\begin{figure*}
    \centering
    \includegraphics[width=\textwidth]{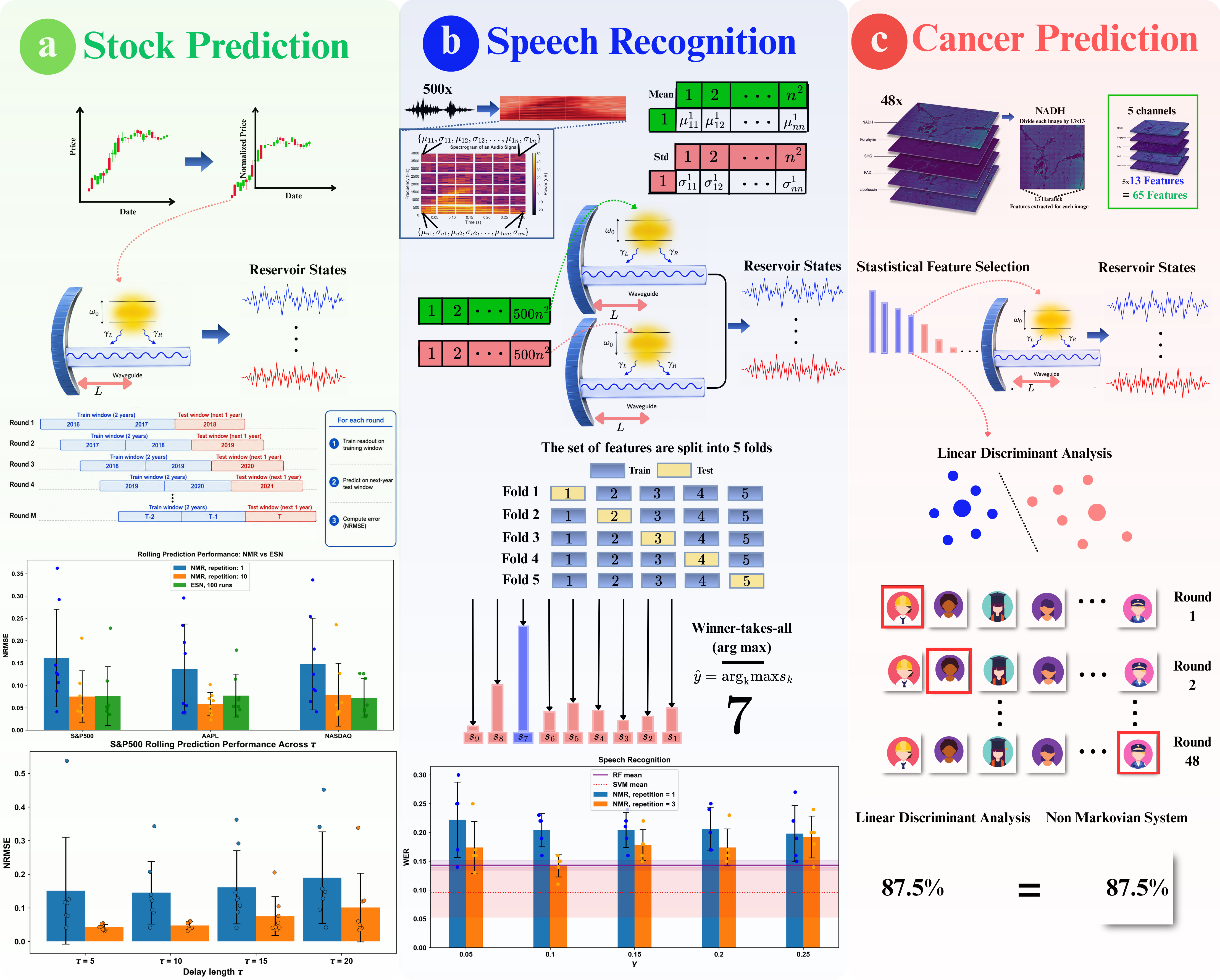}
     \caption{The single-atom reservoir across three real-world tasks. \textbf{a}, Financial prediction with a rolling two-year-train / one-year-test protocol, reported by NRMSE; parameters $\Delta t=1$, $\gamma=\NumAPB$, $\epsilon=\NumAPBF$, $\theta=\phi=\pi/3$, $\tau=15$, with a delay sweep $\tau=5,10,15,20$. The quoted $\tau=15$ is the setting shared with panels (b,c) rather than this task's optimum---shorter delays score better ($\NumAPAEC\pm0.008$ at $\tau=5$; Supplement Sec.~S.9)---and is retained so the three panels report one device at one operating point. A persistence baseline $\hat y_t=y_{t-1}$ scores $\approx0.037$ on these identical windows, below both the reservoir and the echo-state comparator, which is why this task is a consistency check between methods rather than a pillar (Supplement Sec.~S.9). \textbf{b}, Speech recognition on the Free Spoken Digit Dataset, reported by word error rate (WER); pipeline in Methods. $\gamma=(\NumAPAF,\NumAPBA,\NumAPBF,\NumAPCA,\NumAPCF)$, other parameters as in (a) with $\tau=15$. \textbf{c}, Pancreatic neuroendocrine-tumor classification from multiphoton-microscopy Haralick features ($13$ per channel, $65$ per sample; correlation-pruned to $22$, then the four LDA-optimal four-feature subsets retained and encoded), leave-one-out evaluation; $\gamma=\NumAPB$, $\epsilon=\NumAPB$, $\theta=\phi=\pi/3$. Across the four LDA-selected subsets $\tau=10$ is the robust setting ($\NumHAPID$--$\NumIBPCF\%$) relative to $\tau=5$ ($\NumFIPDD$--$\NumIHPF\%$). Because the subsets are chosen to optimize the classical baseline, the comparison disadvantages the reservoir.}  
    \label{fig:NMR_figure2}
\end{figure*}

For speech recognition (pipeline in Methods) the reservoir reached a word error rate of $0.142\pm0.019$, matching a random forest ($0.143\pm0.009$) and trailing a support-vector machine ($0.096\pm0.043$). We report this plainly: parity with one standard method, short of the strongest. No formal significance claim is made, and none is attainable---at five paired folds the exact two-sided Wilcoxon test's smallest achievable $p$ is $0.0625$, so the fold count itself precludes significance (Supplement Sec.~S.9).
Varying $\gamma$ over $0.05$--$0.25$ gave word error rates of $0.174$, $0.142$, $0.178$, $0.174$, $0.192$ ($\pm0.02$--$0.05$), with no clear monotonic trend in $\gamma$, which sets the response time.

For tumor classification (multiphoton-microscopy images, $48$ FFPE samples near-evenly split between tumor and normal, five channels) features were prepared with the classically optimized pipeline of Ref.~\cite{10.1117/1.BIOS.2.4.045001} (Haralick textures, correlation pruning, exhaustive LDA subset search; Methods)---a protocol that \emph{inherently disadvantages} the reservoir, which receives features selected to favor the classical baseline. Performance is delay-dependent: at the robust operating point $\tau=10$ the reservoir classified $70.83$--$81.25\%$ across subsets, below the LDA best, whereas $\tau=5$ ranged $58.33$--$87.5\%$, its low end no better than the majority-class baseline. On one configuration ($\tau=5$, one subset) it matched the best classical accuracy, $87.5\%$ ($42$ of $48$)---a descriptive, best-configuration tie rather than a tested equivalence (Wilson $95\%$ interval $[\WilsonLo,\WilsonHi]$, $n=48$). The paired comparison on that configuration is nonetheless exact: both classifiers scoring $42$ of $48$ forces the discordant pairs into balance, so McNemar's exact test gives $p=1.0$ identically, confirming no detectable difference at this sample size. Because the subset search runs on the full sample, the absolute accuracies of both classifiers carry selection bias; the shared selection step makes the reservoir--LDA comparison, rather than either absolute number, the valid object (Supplement Sec.~S.9).
We note the limits plainly: $n=48$ under one filter setting, with no stability sweep claimed. Together the tasks show a single non-Markovian atom operating alongside standard baselines: reaching the classical best on one imaging configuration (with the robust setting below it), and at parity with a random forest on acoustic data (trailing the strongest kernel method), at minimal hardware cost.

\section{Discussion}

The results assemble into one statement about the hardware floor of machine intelligence. A single atom before a mirror---one active component, one geometric knob---is a provably universal approximator in its linear-transducer limit, capability never decreasing along one physical axis and the matchable span strictly growing; the saturable device converges along the same axis and outperforms the Gaussian regime on non-Gaussian benchmarks. The modes are the network, the atom the nonlinearity, the mirror delay the recurrence. The delay-line architecture, without a universality theorem in fifteen years of use, acquires one here for a linear-node instance with the nonlinearity restored at the readout; Supplementary Discussion Sec.~S.12 identifies the obstruction that has kept the nonlinear-node original from one.

Because the scaling resource is central: modes are added by geometry, so the machine stays one piece of hardware as it scales. The theorem's limit refines the mode \emph{spacing}; a second budget governs \emph{bandwidth} via the flat-coupling error $\varepsilon_\mathrm{flat}(B)$. A superconducting atom ($\gamma/2\pi\sim10$\,MHz, flat to a few percent over $\sim1$\,GHz) supports $\sim10^2$ comb modes at percent-level $\varepsilon_\mathrm{flat}$, which fixes the mode spacing at $\Delta_0/2\pi\sim10$\,MHz and hence a quantization length $\ell=\pi v/\Delta_0\approx6$\,m at $v\approx0.4c$---an order of magnitude beyond our simulations, though not unboundedly so (budget arithmetic, optimum, and precision costs: Methods). This arithmetic describes generic operating points, whose distinctness margins are measured rather than prescribed; the certificate's spacing requirements are excluded from it and are unreachable at these platform parameters beyond trivial depth (Methods; Supplement Sec.~S.2.5). ``Arbitrarily many modes'' is thus an idealization with a stated budget, as ``arbitrarily many neurons'' is classically. Table~\ref{tab:resources} summarizes the claim; the structural point is arithmetic-free: the recurrent core has \emph{zero} trained parameters and one fabricated atom.

\begin{table*}[t]
\centering
\small
\setlength{\tabcolsep}{5pt}
\renewcommand{\arraystretch}{1.2}
\begin{tabularx}{\textwidth}{@{}l >{\raggedright\arraybackslash}X >{\raggedright\arraybackslash}X >{\raggedright\arraybackslash}X@{}}
\hline
 & trained params (core) & trained params (readout) & fabricated units\\
\hline
This device & $0$ & $\sim10^2$--$10^3$ (linear) & 1 atom, 1 mirror, 1 detector, 1 fixed band envelope\\
ESN ($N{=}50$--$100$) & $0$ (random, stored $\sim10^3$--$10^4$) & $\sim10^2$--$10^3$ & $N$ digital neurons\\
\hline
\end{tabularx}
\caption{Resource accounting for the minimality claim. The accuracy comparisons are reported in Results and are not repeated here. Against a Markovian atom--cavity comparator with polynomial readout, the device with a purely linear readout matches or surpasses ninth-order readout on Mackey--Glass and NARMA10 (Fig.~\ref{fig:withwithout}); no tuned, noise-matched comparison against a classical echo-state network is claimed (Supplement Sec.~S.9). We do not tabulate energy per inference: energy accounting for analog physical reservoirs (cryogenics amortization, detection electronics) is unsettled in the literature, and the structural comparison this table supports (what is trained, what is fabricated) does not depend on it.}
\label{tab:resources}
\end{table*}

Two questions remain open: extending Theorem~\ref{thm:univ} to the saturable atom (perturbatively in $s$, or via analyticity in $g$), motivated by the numerical finding that this regime outperforms the Gaussian one; and whether the constructive resources---readout time $T=O(M\ln M+\ln\epsilon^{-1})/(\gamma+\gamma_g)$ and the exponentially graded design separations whose realization costs shots---can be brought to their information-theoretic floor.

The architecture maps onto existing platforms. Demonstrated \emph{jointly}: the atom-before-mirror geometry with feedback-modified emission \cite{hoi2015probing, eschner2001light}, in the simulated regime $\gamma\tau\sim0.5$--$1$; separately: encoded coherent driving and homodyne detection at the required bandwidths. Tolerances are comfortable (Supplement Sec.~S.10): under 8\% error at $\gamma_\phi/\gamma=10^{-2}$, jitter to $\sim$0.5\,rad, $20$\,dB feedback loss recovering most of the benefit. Platform arithmetic, including the $\approx1\,\mu$s per-symbol clock, is in Methods.  

One structural limit belongs here. The contraction the approximation bound needs and the sub-revival horizon the uniform fading-memory bound needs cannot both hold at any mode number, the trace budget and the revival time scaling identically in it; the device is operated beyond its own revival, and the bound survives because no step of it uses fading memory (Supplement Sec.~S.4). Stated conservatively, the contribution is threefold: an existence-and-structure theorem for one fixed machine; a generic operating regime in which every simulated device meets the theorem's conditions with verified margins; and a proposed experiment assembled from demonstrated components. The certified constants belong to the theorem rather than to the laboratory, and the difference is priced above.

To close with the opening question in its defensible form: no quantum computational advantage is claimed and training remains classical, but the recurrent, high-dimensional, nonlinear core of a learning machine can in principle be replaced by the passive dynamics of one atom and one mirror, capability purchased by bandwidth rather than component count. The purchase is metered: reaching accuracy through higher readout order in the Gaussian limit costs shots factorially in that order (Supplement Sec.~S.7), the price the saturable device's nonlinearity transfer removes. The hardware floor for universal temporal learning is very low, and quantum physics puts it there.

\section{Methods}
\subsection{Hamiltonian with non-Markovian feedback}
The laser-driven atom coupled to the waveguide was previously modeled in \cite{Photonics}; we adopt the same numerical framework with a time-dependent input drive. The joint state evolves under $i\hbar\,d_t\ket{\Psi(t)}=\hat{H}_\mathrm{TOT}\ket{\Psi(t)}$ with $\hat{H}_\mathrm{TOT}=\hat{H}_A+\hat{H}_F+\hat{H}_\mathrm{INT}$. The input is encoded in the drive on the atom,  
\begin{align}
    \hat{H}_A=\omega_0\hat\sigma_+\hat\sigma_- - \tfrac{\varepsilon_\mathrm{INPUT}}{2}\epsilon(t)(e^{i\omega_L t}\hat\sigma_-+\mathrm{h.c.}),
\end{align}
with normalized input $\epsilon(t)$ and strength $\varepsilon_\mathrm{INPUT}$. The field Hamiltonian is $\hat{H}_F=\int_B d\omega\,\omega\,\hat{a}^\dagger(\omega)\hat{a}(\omega)$ with $[\hat{a}(\omega),\hat{a}^\dagger(\omega')]=\delta(\omega-\omega')$, and the interaction is  
\begin{align}
    \hat{H}_\mathrm{INT}=&\,i\int_B d\omega\,[(g_R(\omega)e^{-i\omega\tau/2}\\ &-g_L(\omega)e^{i\omega\tau/2})\hat{a}^\dagger(\omega)\hat\sigma_- -\mathrm{h.c.}],  
\end{align}
with $\tau=2L/v$ the round-trip delay. The idealizations require the scale hierarchy $\omega_0,\omega_L \gg B \gg \gamma,\gamma_g$, with $1/\tau$ inside $B$ so the feedback is resolved by the retained band; at the simulated parameters ($\gamma=0.1$, $\tau=10$, $B\sim2$, carrier several orders above $B$) each inequality holds with at least an order of magnitude of margin. Regarding the input--output geometry: the mirror terminates the waveguide on one side and supports no outgoing channel; the detected field exits through the open end only, so the returning feedback (standing-wave structure) and the outgoing signal (unidirectional continuum past the atom) are distinct channels, which is what licenses the exact Lindblad form of the measured channel in Sec.~\ref{sec:exact}; residual loss, where present, is a separate unidirectional channel. Under the flat spectral density approximation $g(\omega)\approx\sqrt{\gamma/2\pi}$, moving to the interaction picture and the rotating frame yields
\begin{align}
    \hat{H}_{\mathrm{INT},I}(t)=i[(\sqrt{\gamma_L}\hat{a}^\dagger(t)+\sqrt{\gamma_R}\hat{a}^\dagger(t-\tau)e^{i\phi})\hat\sigma_- -\mathrm{h.c.}],
\end{align}
with $\Delta=\omega_L-\omega_0$ and phase $\phi=\pi-\omega_L\tau$ between the time-bin operators ($\phi$ is the round-trip phase in the frame rotating at the drive; the undriven loop's $\pi-\omega_0\tau$ of Supplement Sec.~S.4 is the same quantity, the two coinciding on resonance). The discretized propagator is
\begin{align}
\label{eq:evolution}
\hat{U}_I(t_k,\Delta t)=&\exp\big(-i\hat{H}_{A,I}(t_k)\Delta t \\ \nonumber
&+[(\sqrt{\gamma_L}\Delta\hat{A}^\dagger(t_k)+\sqrt{\gamma_R}\Delta\hat{A}^\dagger(t_{k-\ell})e^{i\phi})\hat\sigma_- \\ \nonumber
&\qquad\quad -\,\mathrm{h.c.}]\big),
\end{align}
with $\Delta\hat{A}(t_k)=\int_{t_k}^{t_{k+1}}dt\,\hat{a}(t)$ the discretized time-bin operator.

\subsection{Readout construction and shot budget}
In Eq.~\eqref{eq:output}, $t_k$ is the start of the $k$-th measurement window and $\rho_\zeta(t_k)$ the coherently driven reservoir state there, the subscript denoting the displaced trajectory below. The output is built from powers of the measured quadrature. The higher powers
$q_\mathrm{geo}^n$ are obtained from the same homodyne record used for the linear quadrature, with one
care required by continuous measurement: the raw photocurrent contains white shot noise, whose pointwise
powers are not well defined, so the current is first integrated against the readout window for each sample
time, yielding a well-defined, noise-corrupted outcome per window per repetition; powers of these
\emph{binned} outcomes are then averaged (Supplement Sec.~S.7). Readout begins after an initial washout
interval---standard in reservoir computing---so that the transient of the stable dynamics has decayed and
the binned outcomes carry no correlations from the initial state (Supplement Sec.~S.7). No additional apparatus is required: the
shot-noise contribution to these moment estimates is Gaussian and is absorbed exactly into the readout
weights, and the number of shots required for precision $\sigma$ on the order-$n$ moment is bounded by
$(2n-1)!!\,(v_0+q_{\max}^2)^n/\sigma^2$ (Supplement Sec.~S.7).

\subsection{Adiabatic elimination and the linear-transducer limit}
Passing from Eq.~\eqref{eq:LBDN} to the working generator of the proof is not an approximation of the
environment (which the model's axioms already render Markovian) but the adiabatic elimination of the atom.
The induced rate is a physical quantity, fixed by the coupling and the flat band rather than by any
numerical window: the band of width $B$ has correlation time $\Delta t\equiv1/B$, the interval over which
the atom absorbs and re-emits before the field can respond, and the golden-rule rate of the
atom-mediated drive channel over that band is $\gamma_g=2g^2\Delta t=2g^2/B$ (correspondingly
$\eta=\gamma_g/2g=g/B$). In the simulations the time-bin width is of order this correlation time at the retained band, so the
pipeline's operational $2g^2\Delta t$ realizes the physical rate there, and no proof constant depends on
the discretization. The limit's hierarchy is two-sided, and we name each scale. On the fast side the band is broad, $B\gg\gamma,\gamma_g$: the field's correlation time $1/B$ is the shortest scale, which is what licenses the golden-rule coarse-graining above, and at the simulated operating points ($\gamma=0.1$, $B\sim2$) the ratios are $B/\gamma=\Bovergam$ and $B/\gamma_g=\Bovergamg$ at the drive-channel rate of Supplement Sec.~S.2.5, so the fast-side hierarchy is the weakest of the stated inequalities at that operating point. On the slow side the dipole must follow the field's envelope: its induced relaxation $\gamma_g$ must exceed the rates at which the driven field amplitude it follows evolves (the per-mode decay $\lambda_\star$ and the drive-envelope rate $1/T_\mathrm{on}$), the same separation, in the same role, as the quasi-static condition of the Bloch route (Sec.~\ref{sec:bloch}). When both sides hold, the saturable dipole adiabatically follows the field and is replaced by a linear transducer: a c-number drive
$\eta u(t)\alpha$ together with the induced decay channel $\tfrac{\gamma_g}{2}(\alpha\otimes\alpha^*)$.
Two scopes should be kept apart. The Gaussian-limit simulations instantiate the mode-space generator directly, so for them the hierarchy is a statement about which hardware realizes that generator rather than an approximation internal to the numerics. The designed certificate of Supplement Sec.~S.2.5 is another matter: its condition D4 drives $\gamma_g$ far below $\lambda_\star$ ($\gamma_g/\lambda_\star\approx10^{-2}$ at the deposited operating points), placing the certificate outside the slaving ordering; this is stated with the certificate's other physical costs (Supplement Sec.~S.2.5, physical resources), and generic operating points are not so constrained.
The controlled small parameter is the saturation $s\sim(\varepsilon_\mathrm{INPUT}/\gamma_g)^2$, and the
property the proof actually uses is the resulting \emph{Gaussianity}:
$\ket{\zeta(t)}$ with $\zeta(t_k)=\sum_{n\ge1}u_{k-n}\zeta_n$, the moments $\langle  
q_\mathrm{geo}^n\rangle$ are polynomials in $(\zeta,\zeta^*)$, and the Heisenberg evolution maps
$a\mapsto e^{-\mathbf{\Gamma}_g t}\cdot a$. The retarded self-interaction responsible for the delayed
feedback is recovered identically from the frequency integral of the form factor
$\tilde g(\omega)\propto\sin(\omega\tau/2)$, whose stationary-phase contributions occur at the present
time and one round trip in the past.

\subsection{Logic of the universality proof}
In the Gaussian limit the output is a finite sum of Volterra terms whose kernels factor into products of single-mode exponentials $e^{-\lambda_k(m-1)(T_\mathrm{on}+T_\mathrm{off})}$. The freedom in the time-dependent weights $W_n(t)$ is exactly the freedom to fix these kernels: integrating a weight against $e^{-\Lambda t}$ over $T_\mathrm{off}$ realizes any function with a well-defined Taylor expansion at the eigenvalue sums, provided those sums are distinct---which the non-resonance condition guarantees; since only finitely many Taylor coefficients are ever prescribed, polynomial weight functions $W_n(t)$ on $[0,T_\mathrm{off}]$ suffice, and no exotic function space is invoked. (Throughout, $\mathbf{\Gamma}_g$ is taken at each finite $K$ in its biorthogonal spectral resolution $\mathbf{\Gamma}_g=\sum_k\lambda_k\,v_{R,k}v_{L,k}^*$---guaranteed by eigenvalue distinctness under non-resonance---and no infinite-mode decomposition is invoked: the many-mode limit enters only through the uniformly decaying kernel $h(t)$ below, so convergence of the eigenvector expansion as $K\to\infty$ is never needed; the finite-$K$ family itself is defined from the single physical device, by a single rule and with no spectral nesting assumed, in Supplement Sec.~S.1.1.) The dependence on memory index reduces to a Vandermonde matrix in $x_k=e^{-\lambda_k(T_\mathrm{on}+T_\mathrm{off})}$, invertible because the eigenvalues are distinct, so all kernels up to order $N$ and memory depth $M$ are matched exactly whenever the device supports at least $M$ modes. Controlling the error is where the structure of the proof is decisive, and two structural facts shape it. No mode-space spectral gap uniform in $K$ exists: the Hermitian part of $\mathbf{\Gamma}_g$ has rank two, so the summed decay rates are fixed at $(\gamma+\gamma_g)/2$ and the slowest mode closes as $1/K$ (Supplement, trace-identity lemma, Sec.~S.2). And any bound routed through the norm of the matched weights inherits exponential growth from the node geometry (Supplement, Sec.~S.2.5, Fig.~S.2).  

The proof instead rests on a structural identity: once the kernels are matched at lags $\leq M$, the realized kernels at all deeper lags are the spectral extrapolation of the matched block, determined by the eigenvalues alone and independent of the weights (Supplement, Extrapolation identity). The designed operating point---comb spacing locked to the readout period so the spectral nodes sit at the $M$-th roots of unity of a common radius $\bar r=e^{-\lambda_\star T}$---then makes the extrapolation tail explicitly small, $2M\bar r^{\,M}+e\,M\,\bar\varepsilon$, with $\bar r^{\,M}=e^{-(\gamma+\gamma_g)T/2}$ recovering the full trace budget over the memory span and $\bar\varepsilon$ the node-placement error controlled by the comb spacing (Supplement Sec.~S.2.5). The parameters are chosen once and in order---$(M,N)$ from the target, then the readout period, then the comb---and no infinite-mode limit is invoked: the constructed device has exactly $K=M$ modes.

\subsection{Readout resolution and the role of dispersion}
The constructive proof selects eigenvalue-sum points individually, which is sufficient when
$T_\mathrm{off}\gtrsim\pi/\delta_{\min}$ with $\delta_{\min}$ the minimal sum splitting; a trained  
least-squares readout is less demanding---it requires only that the splittings be nonzero, so that the
feature span is not collapsed---and indeed Fig.~\ref{fig:convergence} converges at $T_\mathrm{off}=60$
although the worst-case order-two splitting of the comb corresponds to
$\pi/\delta_{\min}\approx1.7\times10^{2}$ already at $K=6$. What is fatal is \emph{exact} degeneracy:
for an equidistant comb the order-two sums are degenerate at leading order (split only by the
non-Hermitian dressing, $O((\gamma+\gamma_g)/K)$), the achievable feature span collapses, and performance
plateaus far above the envelope. Waveguide dispersion lifts these degeneracies at leading order: a chirped
comb $\omega_k=\omega_0+\Delta_0 k+\Delta_2 k|k|$ splits the sums at $O(\Delta_2)$. In the constructive  
theorem this role is played instead by a single explicitly graded hierarchy of comb-frequency
defects (Supplement Sec.~S.2.5, D5), which makes the required eigenvalue-sum points with distinct
signed mode content separate by construction, the generic drive-channel spread separating
the remainder with measured margins; natural dispersion remains the practical mechanism on
undesigned devices, and the trained-readout operation of Fig.~\ref{fig:convergence} exploits it.

\subsection{Detection efficiency versus feedback loss}
Two efficiencies must be kept apart. The feedback transmissivity $\eta$ of Supplement Sec.~S.10 attenuates
the \emph{returning} field and therefore changes the dynamics---it enters $r=\gamma_R/\gamma$ in the loop
characteristic equation and hence the stability margin $c_0$, which is why severing it collapses the
device to the feedback-free plateau. Detector inefficiency $\eta_\mathrm{det}$ acts on the \emph{outgoing}
channel only, after the dynamics: it admixes vacuum into the measured field, leaving the generator, the
input--output kernel and $c_0$ untouched, and costs only shots---the order-$n$ moment budget of Supplement
Sec.~S.7 inflates by at most $\eta_\mathrm{det}^{-n}$, benign at the $n=1$ readout used for every
real-world benchmark here (Supplement Sec.~S.7). Imperfect detection therefore degrades this device's
statistics rather than its physics; the loss that degrades its physics is loss in the feedback path, which is the
quantity Supplement Sec.~S.10 sweeps.

\subsection{Time-bin matrix-product-state simulation}
We simulate the device with a matrix-product-state representation \cite{Or_s_2014} using a time-bin (rather than spatial) decomposition, Fig.~\hyperref[fig:model]{\ref*{fig:model}f}. The Hilbert space is $\mathcal{H}=\mathcal{H}_S\bigotimes_{k\in\mathbb{Z}}\mathcal{H}_k$, with $\mathcal{H}_S$ the two-level atom and each $\mathcal{H}_k$ a Fock space per time bin, $\ket{i_k}=(\Delta\hat{A}^\dagger)^{i_k}\ket{0_k}/\sqrt{\Delta t^{i_k}i_k!}$. With $\tau=\ell\,\Delta t$, evolution to step $k$ entangles time-bin pairs $(t_p,t_{p-\ell})$ for $p<k$; later bins factor out as ground states, and the remaining state is decomposed into site tensors $\Gamma$ and singular-value matrices $\Lambda$. The long-range propagator is implemented with $\mathcal{SWAP}$ gates that make each interaction local \cite{Photonics}. At each step the quadratures $\hat{Q}_k=\hat{b}_k e^{-i\theta}+\hat{b}_k^\dagger e^{i\theta}$ and $\hat{P}_k=i(\hat{b}_k^\dagger e^{i\theta}-\hat{b}_k e^{-i\theta})$ with $\hat{b}=\Delta\hat{A}/\sqrt{\Delta t}$ are extracted (physically, by homodyne detection) and stacked into a feature matrix $X$; the linear readout $\hat{y}=WX$ is trained by the Moore--Penrose pseudoinverse.  
The MPS convergence settings used for these simulations (maximum bond dimension $\chi$, singular-value truncation cutoff, and time step $\Delta t$, together with the convergence checks under bond-dimension and time-step refinement) are recorded with the deposited pipeline configuration and are not reproduced here, and this manuscript makes no claim of convergence in any of them. The comparisons within each figure are between arms of the same pipeline at the same settings, so the orderings reported are the claims this evidence supports; the absolute error values are conditional on the truncation settings, and where they stand beside classical baselines that carry no truncation error the resulting parity statements should be read as conditional on those settings rather than as converged results.  

\subsection{Quasi-static Bloch computation of the saturable overlay (Fig.~\ref{fig:convergence})}
\label{sec:bloch}
The red curve of Fig.~\ref{fig:convergence} (the \emph{saturable} device on the convergence target) is computed without the MPS pipeline, by a displaced-frame quasi-static Bloch route that retains the atom's nonlinearity while remaining closed-form in the field. We record it here so the result is reproducible from the manuscript.

\emph{Displaced frame.} We work in the frame displaced by the coherent field amplitude $\beta(t)=-\tfrac{i}{g}\alpha\,u(t)$ that removes the atomic drive in favor of a field drive (as in the elimination of Supplement Sec.~S.1), but (unlike the linear-transducer limit) we do not slave the atom to the field. Instead the atom dipole is carried as a dynamical two-level Bloch vector $(\langle\hat\sigma_x\rangle,\langle\hat\sigma_y\rangle,\langle\hat\sigma_z\rangle)$ driven by the local field, and the field modes evolve under the Gaussian generator $\mathcal{L}_0$ of Eq.~\eqref{eq:LBDN} sourced by the atom dipole rather than by a c-number.  

\emph{Quasi-static approximation.} When the field response time is long compared with the atom's internal relaxation over a readout window (the same separation of scales that licenses the elimination, but retained to next order) the atom dipole follows its instantaneous steady state on the field's slow timescale. We solve the steady-state Bloch equations for $\langle\hat\sigma_-\rangle$ as a function of the instantaneous drive, including saturation through $\langle\hat\sigma_z\rangle=-1/(1+S)$ with $S$ the saturation set by the drive and detuning, and feed the resulting nonlinear dipole into the mode dynamics. The controlled small parameter is the ratio of field-response to atomic-relaxation time; at the operating drive the saturation depth is $S_{\max}=0.2$ and $\langle\hat\sigma_z\rangle$ is driven from $-1$ to $-0.833$ (an excited-state population of $0.083$; the atom remains predominantly in the ground state, and no inversion is implied), so the atom is driven appreciably out of the linear-transducer regime---which is the point of the overlay.

\emph{Validation.} The quasi-static dipole was checked against the full time-dependent solution of the coupled atom--field dynamics on a set of $(K,\text{drive})$ points of the sweep whose size and composition are recorded with the deposited script, so the deviation quoted is a maximum over that set rather than a bound over the sweep; the maximum relative deviation of the readout feature over the validation set is $4\times10^{-3}$, small compared with the eight-fold error reduction the curve reports, so the approximation does not affect the convergence conclusion. The script implementing this route and reproducing Fig.~\ref{fig:convergence} is included in the repository script-to-figure map.

\subsection{Feedback-free comparison model}
For the polynomial-readout comparison of Fig.~\ref{fig:withwithout} we use a Jaynes--Cummings reservoir in which the atom is coupled to a single cavity mode $\hat{c}$,
\begin{align}
    H_0&=\omega_c\hat{c}^\dagger\hat{c}+\Delta\hat\sigma_+\hat\sigma_-+g(\hat{c}^\dagger\hat\sigma_-+\hat{c}\hat\sigma_+),\\  
    H_1(t)&=i\tfrac{\varepsilon_\mathrm{INPUT}}{2}\epsilon(t)(\hat\sigma_--\hat\sigma_+). 
\end{align}
Consistent with the mirror device in its Markovian limit ($\gamma\tau\ll1$), the returning field is re-injected essentially instantaneously and the cavity mode decays at the mirror-modified rate
\begin{align}
    \kappa=\gamma(1+\cos\phi)=\gamma(1-\cos\omega_L\tau),  
\end{align}
so the master equation is $\dot\rho=-i[H_0+H_1(t),\rho]+\mathcal{D}[\sqrt{\kappa}\,\hat{c}]\rho$ with the dissipator acting on the \emph{measured cavity mode} $\hat{c}$, and the readout observables $\hat{Q}=\hat{c}+\hat{c}^\dagger$, $\hat{P}=i(\hat{c}-\hat{c}^\dagger)$, $\hat{P}^2$, $\hat{P}\hat{Q}$, $\hat{Q}^2,\dots$ are those of the same mode. A Lamb shift $\delta\omega=\tfrac{\gamma}{2}\sin\phi$ accompanies the decay, with $\phi=\pi-\omega_L\tau$ the round-trip phase defined above. This makes the dissipative channel and the measured observable consistent, which is required for a fair comparison against the feedback device. The comparator set tests the feedback ablation (this Markovian device) and a classical network reference (the echo-state network of Supplement Sec.~S.9); a classical single-node delay-line reservoir at matched virtual-node counts is a natural comparison not performed here, and the claim this work makes about that architecture concerns the availability of a single-member theorem rather than relative performance.

\subsection{Scope of the model and the sense of exactness}
A central technical point organizes the paper. ``Non-Markovian'' here refers to the \emph{reduced} dynamics of the atom: because photons return after a delay, the atom interacts with its own past. The \emph{joint} atom--field system evolves under a time-independent Hamiltonian, and once the field is written in the standing-wave modes selected by the mirror, delay and feedback appear entirely as coherent structure among those modes. The only irreversible processes (the field leaving through the open end, and the homodyne measurement) act on unidirectional continua that never return and are therefore Markovian to the same standard as the axioms themselves. Under the standard input--output (SLH) idealizations---stated as axioms A1--A6 in Supplement Sec.~S.1, and including the classical treatment of the input drive and the idealized continuous measurement---the device is consequently described by a single Lindblad generator that truncates none of the delayed feedback, and our theorem is a statement about that generator. We are careful about the sense of ``exact'': writing a Lindbladian presupposes those axioms rather than deriving them from an unspecifiable environment. What is not approximated, relative to the usual treatment of delayed feedback, is the memory itself---no Born--Markov truncation of the retarded self-interaction, no pseudomodes, no auxiliary lossy modes. The one further approximation is of the atom rather than of the environment, a linear-transducer (Gaussian) limit, and we show numerically that operating \emph{outside} it, where the atom's own nonlinearity is active, reduces the hardware's measurement requirements further on the tasks tested. Extending the proof to the saturable atom is an open problem we state explicitly.

\subsection{Resource matching for the Markovian comparator (Fig.~\ref{fig:withwithout})}
To be explicit about the resource counted, since the two devices expose different observables: for the feedback reservoir one ``node'' is one measured quadrature pair per time bin; for the comparator it is one polynomial readout feature. The map is: feedback reservoir, $n$ nodes $=n$ time bins $=2n$ features ($Q$ and $P$ per bin); comparator, $n$ nodes $=n$ features. Resource-matching is then an axis relabeling---equal feature count places $n$ feedback nodes against $2n$ comparator nodes, shifting the feedback curves by a factor of two along the node axis (rightward on the logarithmic axis of Fig.~\ref{fig:withwithout}), which any reader can perform from this map. Since the separation at large node count is an asymptotic vertical gap (both families having plateaued), a horizontal rescaling by two leaves the ordering, and the conclusion, unchanged. We anchor the claim at the plateau for exactly this reason: before both families saturate, a horizontal factor of two need not preserve the ordering.

\subsection{Repetitions}
A \emph{repetition} is one independent experimental shot of the input--readout cycle. Because the
readout estimates moments of a measured quadrature, each reported output sample is an average over
$R$ repetitions, and $R$ is a resource on the same footing as the node count: the shot budget of
Supplement Sec.~S.7 fixes $R$ from the target relative precision. Where a benchmark figure reports
more than one value of $R$, the value used for the quoted result is stated in the figure legend.

\subsection{Error metric}
Throughout, NRMSE denotes root-mean-square error normalized by the target range (maximum minus minimum) over the test set, the convention of Supplement Secs.~S.9--S.10 as well, so values are comparable across figures. Range normalization yields systematically smaller values than the standard-deviation normalization common in the benchmark literature, so comparison with published NARMA10 figures requires conversion. No published NARMA10 or Mackey--Glass value is invoked anywhere in this work: every comparator reported here is computed within this study under the identical normalization.

\subsection{Speech pipeline}
For speech recognition we used $500$ utterances of the Free Spoken Digit Dataset, converting each to a mel-spectrogram reduced to per-bin mean and standard-deviation streams injected into two reservoirs, with a $144$-dimensional concatenated state, ten one-vs-rest readouts, and a winner-takes-all decision under five-fold cross-validation. Word error rates, the $\gamma$ sweep, and the significance arithmetic are reported in Results, with the paired-analysis protocol of record in Supplement Sec.~S.9.

\subsection{Tumor feature pipeline}
For tumor classification, features were prepared following the feature pipeline of Ref.~\cite{10.1117/1.BIOS.2.4.045001}: $13$ Haralick texture features per channel, so $65$ features per sample; the unit of analysis throughout is the FFPE sample, and whether several samples derive from one individual is a property of the archived dataset that this secondary analysis does not resolve. The $65$ features were first reduced to $22$ by removing highly correlated features at a pairwise-correlation threshold of $0.45$; all $\binom{22}{4}=7{,}315$ four-feature subsets were then evaluated with linear discriminant analysis, and the four best-performing subsets were retained. Only these classically optimized subsets were subsequently encoded and processed by the reservoir.

\subsection{Physical budgets of the scaling resource}
Because the scaling resource is central, we state its budgets in full. The theorem's limit refines the mode \emph{spacing}: increasing the quantization length $\ell$ grows the mode number $K$ and the revival time $T_P=2\pi/\Delta_0$ together, which is what places the memory horizon inside the sub-revival window where the kernel bound holds; that budget is set by $\ell$ and by coherence over the round trip. A second, independent budget governs \emph{bandwidth}: the flat-coupling axiom degrades as the coupling density varies across the occupied band $B=K\Delta_0$, with relative error $\varepsilon_\mathrm{flat}(B)\sim\sup_{|\omega-\omega_0|\leq B/2}|g(\omega)/g(\omega_0)-1|$. Refining the spacing at fixed $B$ does not pay this cost; widening $B$ at fixed spacing does, and it caps how fine a comb a given bandwidth supports. We are precise about where this budget bites: the simulations reported here \emph{impose} the flat-coupling axiom exactly ($g(\omega)=\sqrt{\gamma/2\pi}$ over the retained band), so their operating $\varepsilon_\mathrm{flat}$ is zero by construction, and the numbers below are a statement about a physical device rather than a correction to our numerics. The device's intrinsic fading-memory rate $c_0$, set by loop stability rather than the spectrum, degrades with neither budget; the \emph{theorem's} convergence rate is a third quantity---$e^{-(\gamma+\gamma_g)T/2}$ over the memory span---purchased by readout time, and it too is independent of both budgets. A superconducting artificial atom with $\gamma/2\pi\sim10$\,MHz on a line flat to a few percent over $\sim1$\,GHz supports $\sim10^2$ comb modes before $\varepsilon_\mathrm{flat}$ reaches the percent level---an order of magnitude beyond the $K\leq14$ of our convergence experiment and the $\sim30$ nodes of our benchmarks, though not unboundedly so. 

Two companion statements complete the budget. First, the trade-off has an optimum: the constructive theorem's node-placement error $\bar\varepsilon$ falls as the comb spacing grows, but a wider comb occupies more band and pays $\varepsilon_\mathrm{flat}(B)$; the optimal operating point is where the falling $\bar\varepsilon$ meets the growing $\varepsilon_\mathrm{flat}(B)$, and it is the honest physical boundary of the idealization: the one step of the construction where the proof meets the hardware. Second, for the constructive certificate specifically: the designed operating point's defect-protection condition demands a comb spacing that grows exponentially with the memory depth, and its timing condition demands a relative precision of order $\bar\varepsilon/(\Delta_0T)$ (Supplement Sec.~S.2.5, scaling ledger); a generically fabricated device pays neither price, separating the same eigenvalue sums by its own verified conditions, which is how every simulation in this paper is operated. Growing precision requirements on settings are the standard cost of universal-approximation constructions, exactly as classical weight precision grows with accuracy, and we state them rather than leave them implicit.

The certificate's reach can be stated as one number. Even at its cheapest admissible selections ($N=1$, minimal readout period, maximal node budget $\bar\varepsilon_\mathrm{target}=1/4$), the binding entry of D1---the radial-protection entry, which exceeds the defect-protection entry by a factor growing like $(16/5)^M$---together with the band constraint $M\Delta_0\leq B$ requires $B/\gamma\geq8192\,M^{2}q^{2}\ln(2M)\,4^{\,q-2}$ with $q$ the least prime $\geq2K_\mathrm{phys}+1$: about $7.3\times10^{7}$ at $M=2$. The superconducting example above ($B/\gamma\approx10^{2}$) therefore supports the certified construction at no depth $M\geq2$; realizing even $M=2$ requires kHz-class linewidths under GHz-flat coupling, a combination not among the demonstrated platforms. The certificate is a mathematical existence statement with explicit constants; the physically operable regime is the generic one, which satisfies no designed spacing entry and whose conditions every simulation verifies with measured margins (Supplement Sec.~S.2.5, physical resources).

\subsection{Experimental feasibility and throughput}
To fix the scale: the simulated feedback regime $\gamma\tau\sim0.5$--$1$ at the superconducting example above ($\gamma/2\pi\sim10$\,MHz) corresponds to a round-trip delay $\tau\approx8$--$16$\,ns, i.e.\ an atom--mirror path $L=v\tau/2\approx0.5$--$1$\,m at a typical coplanar-waveguide phase velocity $v\approx0.4c$---a meandered delay line of ordinary size rather than an exotic requirement. Demonstrated \emph{jointly}, in single experiments: the atom-before-mirror geometry with distance-dependent, feedback-modified emission, in superconducting circuits \cite{hoi2015probing} and with a trapped atom before a mirror \cite{eschner2001light}, in the same feedback regime we simulate ($\gamma\tau\sim0.5$--$1$) \cite{Laser, Collective, PhysRevA.90.012113}. Demonstrated \emph{separately} on the same platform classes, but not yet combined with mirror feedback in one apparatus: coherent driving with encoded input sequences, and homodyne detection at the bandwidths our readout requires.

To be concrete about that requirement: the homodyne chain must pass the occupied comb band $B=K\Delta_0$ about the carrier---up to $\sim1$\,GHz at the $10^2$-mode ceiling above, correspondingly less at the simulated mode counts---together with windowed integration at the $\sim\mu$s symbol clock, both standard for microwave homodyne detection. Integrating these into one operated reservoir computer is the experimental step this proposal calls for; no individual capability is new. The scheme does not require the coherence budget of a qubit register, and its tolerance to imperfections is quantified directly in Supplement Sec.~S.10: dephasing degrades performance gracefully (under 8\% relative error at $\gamma_\phi/\gamma=10^{-2}$, within reach of the platforms above), phase jitter is tolerated to at least $\sim$0.5\,rad (a round-trip path stability of $\sim\lambda/13$, equivalently $\sim\lambda/25$ of the one-way distance $L$), and the feedback path is loss-tolerant---restoring 1\% of the returning power ($20$\,dB loss) recovers most of the benefit relative to a severed loop. That last sweep uses single deterministic-input realizations and carries no seed spread, so its loss tolerance is quoted qualitatively here and in the Supplement, with no interval attached.

Two of these are more than robustness checks: the performance floor over static phases falls exactly at the dark-state value $\phi=\pi$ predicted by the delay-stability analysis, a task-level confirmation of the stability landscape; and severing the loop executes the paper's own falsifier.

Throughput completes the practical picture. The per-symbol clock is $T=T_\mathrm{on}+T_\mathrm{off}$, dominated by the readout window; at the operating point of Fig.~\ref{fig:convergence}, $T_\mathrm{off}=60/\gamma$, which at the superconducting example's $\gamma/2\pi\sim10$\,MHz is $\approx1\,\mu$s per symbol, i.e.\ a repetition rate of order $10^6$ per second. At matched statistical precision this becomes an output rate of order $10^2$ samples per second for the linear readout at one-percent relative precision, since each output sample consumes $N_\mathrm{shots}$ repetitions; the single-shot photonic delay-line reservoir of Ref.~\cite{larger2017high} is quoted at the former rate and the comparison should be made at the latter. Two scalings qualify the number in opposite directions. Toward harder targets, the constructive theorem spends readout time $T=O(M\ln M+\ln\epsilon^{-1})/(\gamma+\gamma_g)$, set by the trace budget rather than the per-mode rate, so the clock grows as $\Theta(M\ln M)$ in the memory depth, as the scaling ledger records; and the readout-resolution analysis of Sec.~\ref{sec:convergence} shows that with dispersion the required $T_\mathrm{off}$ does not grow with $K$, so adding modes costs bandwidth rather than time. Toward statistics, the quoted rate is per experimental repetition: moment estimation at precision $\sigma$ multiplies wall-clock time by $N_\mathrm{shots}$ (Supplement Sec.~S.7), which is the same ensemble cost every analog reservoir pays and is kept at its $n=1$ floor here by the nonlinearity transfer.  

\section*{Data Availability}
The datasets are publicly available. The tumor dataset is at \url{https://doi.org/10.25422/azu.data.29983060}. The spoken-digit dataset is the Free Spoken Digit Dataset (FSDD), \url{https://github.com/Jakobovski/free-spoken-digit-dataset}, distributed under the Creative Commons Attribution--ShareAlike 4.0 International licence; FSDD is an open, growing corpus versioned by Zenodo DOI and git tag, and the version tag used here is recorded with the analysis code so the 500-utterance subset is reproducible. Financial data are daily closing prices for the tickers GSPC (S\&P~500), AAPL (Apple), and IXIC (NASDAQ) over 1~January 2014 to 1~January 2024, obtained from Yahoo Finance. Because financial data providers revise historical series, the fetch script, the ticker list, the adjustment settings, the download date, and SHA-256 checksums of the exact CSV extracts used here are deposited with the code, so a reader who obtains the deposited extracts can confirm they are the ones used here by comparing those checksums.

\section*{Code Availability}
All simulation and analysis code, including the scripts generating every figure (with fixed seeds and a script-to-figure map in the repository README), is available at \url{https://github.com/nhula01/nonMarkovianReservoirComputer}; this manuscript does not carry a release identifier, and a reader wishing to confirm which revision produced these results must match the deposited configuration against the repository history.

\end{multicols}
\clearpage

\section*{Declarations}
\paragraph{Ethics.} \sloppy The pancreatic-tumor study uses the previously published, publicly archived multiphoton-microscopy dataset cited in Methods; the original specimens were collected under the institutional approvals reported by that study: University of Michigan Endocrine Oncology Repository IRB \#HUM00115310, Icahn School of Medicine at Mount Sinai Biorepository and Pathology Core IRB STUDY-12-00145, and University of Arizona Tissue Acquisition and Cellular/Molecular Analysis Shared Resource IRB \#0600000609; and the present work performs secondary computational analysis of the de-identified public archive only, involving no new human-subject data collection.
\paragraph{Author contributions.} D.S. conceived and supervised the project and formulated the universality claim and the design conditions. P.J.E. and P.H.N. developed the theoretical framework and the exact mode-space representation, contributed equally, and share first authorship. P.H.N. implemented and ran the matrix-product-state simulation pipeline and the benchmark experiments. P.J.E. developed the proof of universality and the associated analysis. K.S. contributed to the waveguide-QED modeling. N.D. contributed to the numerical experiments. T.W.S. provided the multiphoton-microscopy tumor dataset and its clinical and provenance context. H.I.N. provided technical analysis of the reservoir-computing theory and the moment-readout treatment. All authors discussed the results and contributed to the manuscript.
\paragraph{Competing interests.} The authors declare no competing interests.
\paragraph{Acknowledgment.} This work was supported by the U.S. Department of Energy, Office of Science, Award No.\ DE-SC0025910 (D.S.), and by the National Science Foundation, Award No.\ 2529700 (D.S.).
\paragraph{Use of AI tools.} We used a large language model (Claude, Anthropic) to verify author-developed mathematical derivations and to improve the readability and grammar of the text. All the results presented were reviewed and approved by the authors, and the authors take full responsibility for all content.

\bibliographystyle{unsrtnat}
\bibliography{references}

\end{document}


\title{Supplementary Information for ``A Single Atom in Front of a Mirror is a Universal Reservoir Computer''}

\author{\fnm{Peter J.} \sur{Ehlers}}
\author{\fnm{Phi Hung} \sur{Nguyen}}
\author{\fnm{Kanu} \sur{Sinha}}
\author{\fnm{Noelle} \sur{Daigle}}
\author{\fnm{Travis W.} \sur{Sawyer}}
\author{\fnm{Hendra I.} \sur{Nurdin}}
\author{\fnm{Daniel} \sur{Soh}}

\maketitle

This Supplement provides the full derivations underlying the main text. Section~\ref{s:sec:exactderiv} derives the mode-space Lindbladian and the linear-transducer (Gaussian) limit, states the input--output axioms A1--A6 relative to which the delayed feedback is carried without truncation, identifies the adiabatic elimination of the atom as the one further approximation, and defines the nested finite-$K$ family from the single physical device by a single rule (Sec.~\ref{s:sec:family}), stating explicitly that no spectral nesting across $K$ is assumed and locating the two tiers (readout-side in-band selection, envelope-controlled band truncation) at which the operating mode number is set.

Section~\ref{s:sec:proof} gives the complete proof of universality. The error bound is shaped by a structural constraint on the device: the trace of the generator's Hermitian part is pinned, so the mode-space spectrum has no uniform gap and its slowest rate closes as $1/K$ (Lemma~\ref{s:lem:trace}). The proof accordingly rests on a weight-independent extrapolation identity (Proposition~\ref{s:prop:extrapolation}): once the kernels are matched at the working depth, the device's deeper response is fixed by its spectrum alone, and an explicitly constructed operating point (Sec.~\ref{s:sec:design}) makes that spectral tail exponentially small, with every constant explicit and no infinite-mode limit invoked. Delay stability is treated separately in Sec.~\ref{s:sec:fadingmemory}, as the fading-memory property of the device itself (Lemma~\ref{s:lem:kerneldecay}, Corollary~\ref{s:cor:fmp}); Sec.~\ref{s:sec:cuchiero} verifies the hypotheses of the density theorem we invoke.

Section~\ref{s:sec:saturation} bounds the deviation between the Gaussian-limit kernels and those of the fully saturable device by the saturation parameter, and defines the invariant bounded-excitation sector on which that bound is controlled. Section~\ref{s:sec:genericity} proves the genericity of the universality conditions (by an argument based on functional rather than pointwise independence) and records the numerical verification, including of the delay-stability margin. Section~\ref{s:sec:shotnoise} proves that shot-noise-corrupted moment estimates suffice and bounds the required shot number. Section~\ref{s:sec:incumbents} locates the result against prior constructions in tabular form. Section~\ref{s:sec:benchmarks} presents the benchmark numerics; Section~\ref{s:sec:robustness} quantifies robustness to atomic dephasing, round-trip phase jitter, and feedback loss, including a direct dynamical confirmation of the delay-stability landscape and a direct execution of the feedback-insensitivity falsifier; Section~\ref{s:sec:univexamples} presents the property-verification numerics. Section~\ref{s:sec:suppdisc} is a Supplementary Discussion carrying, in full, the Stone--Weierstrass route and the classical-linear-optics objection summarized in the main text.

\section{The Mode-Space Lindbladian and the Linear-Transducer Limit}
\label{s:sec:exactderiv}

The full system is a single atom in a multimode field described, in the standing-wave basis selected by the mirror, by the Lindbladian below, where $\varrho$ denotes the joint atom--field density operator:  
\begin{align}
    \label{s:eq:LBDNfull}
    \dot{\varrho} &= \left[\mathbf{\Gamma}\cdot a\varrho, a^\dagger\right] + \left[a, \varrho a^\dagger\cdot\mathbf{\Gamma}^\dagger\right] \\ \nonumber
    &\quad -i[g\sigma^+(\alpha^*\cdot a)+g\sigma^-(\alpha\cdot a^\dagger)+\epsilon\sigma^+\sigma^-,\varrho] + u(t)[\sigma^+-\sigma^-,\varrho],
\end{align}
where $a$ is the vector of mode annihilation operators, $\sigma^\pm$ the atomic operators, $g$ the atom--field coupling, $\alpha$ the emitted-mode profile with $\alpha_k\propto\sin(\omega_k\tau/2)$, and $\epsilon$ the excited-state energy. The matrix $\mathbf{\Gamma}=\tfrac{\gamma}{2}(v_\mathrm{geo}\otimes v_\mathrm{geo}^*)+i\mathbf{\Omega}$ contains the mode frequencies $\mathbf{\Omega}$ (Hermitian, diagonal) and the non-unitary evolution from continuous weak measurement of the channel $v_\mathrm{geo}^*\cdot a$.  

\paragraph{Epistemic status of Eq.~\eqref{s:eq:LBDNfull}.} Because the strength of the claim depends entirely on what is being held fixed, we state the model's axioms explicitly and then say precisely what is and is not exact relative to them. The axioms are the standard input--output idealizations of waveguide QED:
\begin{itemize}
\item[A1.] \emph{Rotating-wave approximation} for the atom--field coupling.
\item[A2.] \emph{Flat coupling density} over the relevant bandwidth $B$ about the atomic transition, $g(\omega)\approx\sqrt{\gamma/2\pi}$, with the fixed Gaussian edge rolloff of Lemma~\ref{s:lem:kerneldecay}; the residual variation across the band is the budget $\varepsilon_\mathrm{flat}(B)$ of the main text.  
\item[A3.] \emph{Unidirectional outcoupling}: the \emph{detected} and lost channels are outgoing continua past the atom that never return to it. The quantization length $\ell$ is a physical termination of the guided line rather than a bookkeeping device: it is what makes the retained spectrum a discrete comb of spacing $\Delta_0=\pi v/\ell$, and the conditions D1, D3 and D5 are conditions on those physical resonances. The far termination is part of the coherent structure the axiom does not exclude; what the axiom excludes is return through the measured port.  
\item[A4.] \emph{Structureless vacuum} for those outgoing channels: they are initially in vacuum, are uncorrelated with the atom--field system at the initial time, and carry no internal structure on the timescales resolved by $B$.
\item[A5.] \emph{Classical input drive}: the input field is treated as a c-number drive $u(t)$ on the atom, i.e.\ a coherent input in the standard input--output sense, with its quantum fluctuations subsumed into the vacuum of the drive channel.
\item[A6.] \emph{Idealized continuous measurement}: the homodyne detection of the outcoupled channel is described in the standard continuous-measurement framework---the detector's internal dynamics is eliminated, the local oscillator is classical, and the measured channel propagates unidirectionally to the detector.
\item[A7.] \emph{Two-level validity}: the emitter's nearest non-resonant transition is detuned from the retained band by much more than the band width, $B\ll|\Delta_\mathrm{anh}|$, with $\Delta_\mathrm{anh}$ the anharmonicity of an artificial atom or the nearest-level detuning of a natural emitter. The mode ceiling of a given platform is then the smaller of the flat-coupling ceiling and $c\,|\Delta_\mathrm{anh}|/\Delta_0$ with $c<1$; for a weakly anharmonic artificial atom the second branch binds, while for a natural emitter, ion or colour centre it is inactive.
\end{itemize}
These are exactly the assumptions of the SLH / input--output formalism of open quantum networks \cite{GardinerCollett1985, CombesSLH2017}, and we adopt them as such: the drive and measurement treatments of A5--A6 are idealizations of the same standing as A1--A4 rather than consequences of the model.
Given A1--A6, two facts fix the status of Eq.~\eqref{s:eq:LBDNfull}. First, the standing-wave decomposition is a unitary change of basis of the free field, so the delay and feedback are carried entirely by the coherent structure $(\mathbf{\Omega},\alpha)$ with no auxiliary or lossy modes. Second, the measured and outcoupled channels act on unidirectional continua that never return, so their delta-correlation---and hence the Lindblad form of the Hermitian rank-one terms---follows from A3, A4, and A6 jointly, and it is a statement inside the SLH framework rather than one that survives outside it.

The claim we make is therefore a relative one, and we are careful not to overstate it. Writing a Lindblad generator at all presupposes A1--A6; it is not derived from an unspecified environment, and no such derivation is available in principle, since the environment of a laboratory device is never fully specifiable. In particular, the input drive as written is not an exact description of the physical input field (A5 replaces the quantized input by its coherent amplitude), and the weak-measurement scheme removes the dynamics of the measurement apparatus and assumes unidirectional propagation to the detector (A6); both are of SLH type and are owned as axioms here. What is \emph{not} approximated in Eq.~\eqref{s:eq:LBDNfull}, relative to the usual treatment of a delayed-feedback atom \emph{within the same SLH framework}, is the memory itself: the mirror-mediated feedback is carried exactly by the coherent mode structure, with no Born--Markov truncation of the delayed self-interaction, no pseudomode fitting, and no auxiliary lossy degrees of freedom. This is the sense---and the only sense---in which we call Eq.~\eqref{s:eq:LBDNfull} exact, and the theorem below is a statement about that generator. How well the generator describes a given physical device is an empirical question about A1--A6, and the budgets that govern it ($\varepsilon_\mathrm{flat}$, coherence over the round trip) are stated in the Discussion of the main text.

The input is the continuous-time version of the discrete drive,
\begin{align}
    \label{s:eq:input}
    u(t)=\sum_{k=-\infty}^\infty u_k\,\mathbf{1}\!\big((k{+}1)T-t\big)\,\mathbf{1}\!\big(t-T_\mathrm{off}-kT\big),\qquad T=T_\mathrm{on}+T_\mathrm{off},  
\end{align}
so that $u(t)=u_k$ during the on-interval of step $k$ and $u(t)=0$ during the off-interval, ensuring measurements are not taken while input is injected.

\paragraph{The one approximation internal to the model: adiabatic elimination of the atom.}
Beyond the axioms A1--A6 themselves, the passage to the working generator involves exactly one further approximation, and it is not a Markov approximation of the environment: it is the elimination of the atom. Displacing Eq.~\eqref{s:eq:LBDNfull} by $\beta_{\mathrm d}(t)=-\tfrac{i}{g}\alpha u(t)$ cancels the atomic drive in favor of a field drive. When $g$ is large enough that the atom absorbs and re-emits within a window $\Delta t$ shorter than the field response time---quantified by the saturation $s\sim(\varepsilon_\mathrm{INPUT}/\gamma_g)^2\ll1$ with $\gamma_g=2g^2\Delta t$---the atom adiabatically follows the field and is reduced to a linear transducer. The window is set by the band rather than by the numerics: the flat coupling density of width $B$ has correlation time of order $1/B$, and we adopt the convention $\Delta t\equiv1/B$, so that $\gamma_g=2g^2/B$ is the golden-rule rate of the drive channel into the retained band up to the $O(1)$ factor this convention fixes; any other $O(1)$ choice of the window rescales $\gamma_g$ and $\eta$ by the same factor, and no conclusion depends on it. In the simulations the time-bin width is of order this correlation time at the retained band, so the pipeline's operational $2g^2\Delta t$ realizes the physical rate there; no proof constant depends on the discretization. Reversing the displacement gives  
\begin{align}
    \label{s:eq:LBDNMarkov}
    \dot{\varrho}(t) &= \mathcal{L}(t)\varrho(t) = \left[\mathbf{\Gamma}_g\cdot a\varrho(t), a^\dagger\right] + \left[a, \varrho(t) a^\dagger\cdot\mathbf{\Gamma}_g^\dagger\right] -i\eta u(t)[(a^\dagger\cdot\alpha)+(\alpha^*\cdot a),\varrho(t)], \\
    \mathbf{\Gamma}_g &= \mathbf{\Gamma}+\tfrac{\gamma_g}{2}(\alpha\otimes\alpha^*), \qquad \gamma_g=2g^2\Delta t=\tfrac{2g^2}{B}, \qquad \eta=\tfrac{\gamma_g}{2g}=g\Delta t=\tfrac{g}{B}.
\end{align}
The property used throughout the proof is that Eq.~\eqref{s:eq:LBDNMarkov} is \emph{Gaussian}: it is quadratic in the mode operators with a c-number drive, so its driven steady state is a coherent state (Sec.~\ref{s:sec:proof}). The corrections to this limit are $O(s)$ and reintroduce the atom's non-Gaussian (Mollow) statistics; extending the proof to finite $s$ is the open analytic problem stated in the main text. We refer to Eq.~\eqref{s:eq:LBDNMarkov} as the \emph{linear-transducer (Gaussian) limit} and reserve ``non-Markovian'' for the reduced atomic dynamics of Eq.~\eqref{s:eq:LBDNfull}.

\subsection{One device, one rule: definition of the nested finite-$K$ family}
\label{s:sec:family}

Throughout this Supplement, statements are made about a family of finite-dimensional generators $\{\mathbf{\Gamma}_g^{(K)}\}_K$, one for each mode number $K$, while the physical claim of the main text concerns a single device. This subsection makes the relation exact: the family is not postulated member by member, but \emph{derived}, by one rule, from a single set of physical data, and (this is the point of Remark~\ref{s:rem:nonesting} below) no relation between the spectral data of different members is ever assumed.

\paragraph{The physical data.} The single device is specified by: (i) a coupling density $g(\omega)$ carrying the fixed band envelope of Theorem~1, written explicitly as $g(\omega)^2=\tfrac{\gamma}{2\pi}\,\Theta_{B_\mathrm{env}}(\omega)$ with $\Theta_{B_\mathrm{env}}(\omega)=1$ for $|\omega-\omega_0|\leq B_\mathrm{env}/2$ and $\Theta_{B_\mathrm{env}}(\omega)=\exp[-(|\omega-\omega_0|-B_\mathrm{env}/2)^2/2\sigma_B^2]$ beyond it, normalized so that $\int\Theta=B_\mathrm{env}+\sqrt{2\pi}\,\sigma_B$: flat over the occupied band, with the Gaussian edge of fixed width $\sigma_B$ used in Lemma~\ref{s:lem:kerneldecay}; (ii) a quantization length $\ell$, fixing the standing-wave comb $\omega_k$ with spacing $\Delta_0=\pi v/\ell$; and (iii) the atom--mirror delay $\tau$, fixing the unnormalized emission profile through the mode functions at the atom, $\tilde\alpha_k\propto g(\omega_k)\sqrt{\Delta_0}\,\sin(\omega_k\tau/2)$---the per-mode coupling $\sim\sqrt{\Delta_0}$ being the continuum-consistent scaling already used after Lemma~\ref{s:lem:trace}. The outcoupling profile of the readout leakage channel is likewise fixed by the density and the outcoupling geometry, with the same $\sqrt{\Delta_0}$ scaling; over the retained band it is frequency-flat, $v_{\mathrm{geo},k}=1/\sqrt K$. Nothing about it is designed, tuned, or shaped; it belongs to the physical data of the device rather than to the design, and in particular $v_{\mathrm{geo},k}\neq0$ for every $k$, so the readout-side overlap condition holds identically. All spectral freedom of the homodyne detection resides in the local oscillator and the window functions $W_n(t)$, which act on the measurement record and never enter the generator. 

\paragraph{The rule.} The member $\mathbf{\Gamma}_g^{(K)}$ of the family is obtained from this single object by retaining the $K$ comb modes of the occupied band and repackaging at fixed total rates: writing $P_K$ for the coordinate projection onto the retained modes,  
\begin{align}
    \label{s:eq:familyrule}
    \begin{split}
    \alpha^{(K)}=\frac{P_K\tilde\alpha}{\Vert P_K\tilde\alpha\Vert},\qquad
    &v_\mathrm{geo}^{(K)}=\frac{P_K\tilde v}{\Vert P_K\tilde v\Vert},\\[2pt]
    \mathbf{\Gamma}_g^{(K)} = i\mathbf{\Omega}^{(K)}
    &+\tfrac{\gamma}{2}\big(v_\mathrm{geo}^{(K)}\otimes v_\mathrm{geo}^{(K)*}\big)
    +\tfrac{\gamma_g}{2}\big(\alpha^{(K)}\otimes\alpha^{(K)*}\big),
    \end{split}
\end{align}
with $\mathbf{\Omega}^{(K)}$ the retained diagonal block of the comb. Two consequences should be stated plainly. First, $\mathbf{\Gamma}_g^{(K)}$ is not the literal compression $P_K\mathbf{\Gamma}_g P_K$ of an infinite-mode operator: the fixed-rate convention (which the trace identity of Lemma~\ref{s:lem:trace} requires) rescales the two rank-one dressings by the retained weights $\Vert P_K\tilde\alpha\Vert^{-2}$ and $\Vert P_K\tilde v\Vert^{-2}$. The difference is controlled by the envelope: as the retained comb fills the fixed envelope's support, the retained weights converge to one at the Gaussian-tail rate (bounded symbolically in Lemma~\ref{s:lem:envtail} below, and consistent with the simulated comb, on which the deficit falls from $O(1)$ to $4\times10^{-4}$ to below $10^{-8}$ as the retained band passes the envelope support), so the fixed-rate members converge to the compression of the single physical generator, and no unbounded-operator limit is invoked at any point. Second, each finite $K$ is therefore a different \emph{retained-mode description} of the one device (a different differential equation, exactly as one should expect) generated by one stated rule from one physical object; the operating mode number is set by the band the drive and outcoupling occupy, an operating parameter of the device rather than a fabrication choice.

\begin{remark}[No spectral nesting]
\label{s:rem:nonesting}
No step of the proof assumes that the eigenvalues or the left/right eigenvectors of $\mathbf{\Gamma}_g^{(K)}$ and $\mathbf{\Gamma}_g^{(K')}$ are related for $K\neq K'$, nested or otherwise, and indeed they are not: the normalization in Eq.~\eqref{s:eq:familyrule} and the rank-two dressing shift all spectral data with $K$ (all spectral data are recomputed independently at each $K$). The finite-$K$ spectral resolution (Assumption~\ref{s:ass:spectral}) is invoked member by member; the Vandermonde construction of Secs.~S.2.3--S.2.4 is self-contained at each $K$; the strict-scaling corollary of the main text rests on kernel containment (Remark~\ref{s:rem:KgtM}), which compares realized input--output kernels and never the spectra of two members; and the many-mode limit enters solely through the scalar kernel convergence $h_K\to h_\infty$ of Lemma~\ref{s:lem:kerneldecay}, where the family is identified precisely by the fixed coupling-density envelope of the physical data above.  
\end{remark}

\paragraph{The plain reading.} Stated without formalism: the mode number is a setting of the machine rather than a piece of it. Nothing is fabricated when $K$ grows; the count is fixed by how many comb lines the occupied band contains, which the quantization length $\ell$ controls, so changing $K$ is turning a dial on hardware that already exists, in the same sense that a classical universality theorem selects the width of one network family after the accuracy target is announced. The family $\{\mathbf{\Gamma}_g^{(K)}\}$ is then nothing more than the list of descriptions of that one machine at its different settings, each generated by the single rule above from the single set of physical data. Remark~\ref{s:rem:nonesting} records that these descriptions owe each other nothing spectrally, and that the proof is built so this costs nothing: the only statement that ever compares two settings is the strict-scaling corollary, and it compares what the settings can do (their realized kernels, where containment holds) rather than what they are (their spectra, which reshuffle). The practical consequence is the monotonicity the main text advertises: turning the dial up can never reduce capability, and the constructive selection shows some setting of the dial reaches any accuracy, so a single device becomes more precise by being operated at more modes, with the price paid in geometry and precision budgets rather than in components.

The normalisation of the emission profile is needed before the envelope tail can be estimated, and it needs no design hypothesis: it holds for the equidistant comb as such. It is also what the radial floor of Sec.~\ref{s:sec:design} compares against.

\begin{lemma}[Profile normalisation without the delay lock]
\label{s:lem:profilegen}  
Let $\omega_k=\omega_0+\Delta_0k$, $k=0,\dots,K-1$, and $\theta\equiv\Delta_0\tau$. Then
\begin{align}
    \label{s:eq:dirichlet}
    D=\sum_{k=0}^{K-1}\sin^2\!\Big(\frac{\omega_k\tau}{2}\Big)
    =\frac K2-\frac12\cos\!\Big(\omega_0\tau+\frac{(K-1)\theta}{2}\Big)\frac{\sin(K\theta/2)}{\sin(\theta/2)} ,
\end{align}
and consequently $D\geq K/2-1/(2|\sin(\Delta_0\tau/2)|)$. In particular
\begin{align}
    \label{s:eq:Dquarter}
    \Big|\sin\frac{\Delta_0\tau}{2}\Big|\;\geq\;\frac2K
    \qquad\Longrightarrow\qquad
    D\;\geq\;\frac K4 .
\end{align}
\end{lemma}
\begin{proof}
Write $\sin^2 x=\tfrac12(1-\cos2x)$, so $D=K/2-\tfrac12\sum_k\cos(\omega_k\tau)$, and sum the geometric series $\sum_{k=0}^{K-1}e^{i(\omega_0\tau+k\theta)}=e^{i\omega_0\tau}(e^{iK\theta}-1)/(e^{i\theta}-1)$, whose real part is the displayed Dirichlet kernel. The bound follows from $|\sin(K\theta/2)|\leq1$ and the stated implication from $1/(2|\sin(\theta/2)|)\leq K/4$.
\end{proof}

\begin{lemma}[Envelope tail of the retained weights]
\label{s:lem:envtail}  
Let the coupling density carry the envelope of the physical data above, $g(\omega)^2=\tfrac{\gamma}{2\pi}\Theta_{B_\mathrm{env}}(\omega)$ with Gaussian edge width $\sigma_B$, let the retained band be $B_K=K\Delta_0\geq B_\mathrm{env}$, and let the comb be equidistant with $|\sin(\Delta_0\tau/2)|\geq2/K$ (Lemma~\ref{s:lem:profilegen}). Then  
\begin{align}
    \label{s:eq:envtail}
    1-\Vert P_K\tilde\alpha\Vert^2\;\leq\;\frac{4\sqrt{2\pi}\,\sigma_B}{K\Delta_0}\,
    \mathrm{erfc}\!\Big(\frac{K\Delta_0-B_\mathrm{env}}{2\sqrt2\,\sigma_B}\Big),
\end{align}
and the same bound with the factor $4$ replaced by $1$ holds for $1-\Vert P_K\tilde v\Vert^2$. Under the delay lock D3 in place of Lemma~\ref{s:lem:profilegen}, the factor $4$ is replaced by $q^2$.
\end{lemma}
\begin{proof}
For the numerator, $\tilde\alpha_k^2\leq g(\omega_k)^2\Delta_0$ by $\sin^2\leq1$, and the Gaussian edge is monotone in $|\omega-\omega_0|$ beyond $B_\mathrm{env}/2$, so each out-of-band term is at most the integral of $g^2$ over its own $\Delta_0$-cell; summing the two sides of the band,
\begin{align}
    \sum_{k\notin K}\tilde\alpha_k^2\;\leq\;\frac{\gamma}{2\pi}\cdot2\int_{B_K/2}^{\infty}e^{-(u-B_\mathrm{env}/2)^2/2\sigma_B^2}\dd u
    \;=\;\frac{\gamma}{2\pi}\,\sigma_B\sqrt{2\pi}\;\mathrm{erfc}\!\Big(\frac{K\Delta_0-B_\mathrm{env}}{2\sqrt2\,\sigma_B}\Big).
\end{align}
For the denominator, $\sum_{k\in K}\tilde\alpha_k^2\geq\tfrac{\gamma}{2\pi}\Delta_0D$ with $D\geq K/4$ by Eq.~\eqref{s:eq:Dquarter}. Dividing gives Eq.~\eqref{s:eq:envtail}. For $\tilde v$ the mode functions are flat over the retained band, so $D$ is replaced by $K$ exactly and the factor $4$ by $1$; under D3, $D\geq K/q^2$ by Lemma~\ref{s:lem:profile}(ii) and the factor is $q^2$.
\end{proof}

The difference between the fixed-rate member and the compression of the single physical generator is controlled in operator norm by Eq.~\eqref{s:eq:envtail} through the two rank-one dressings, so that difference vanishes at the Gaussian-tail rate as the retained band passes the envelope support, and no unbounded-operator limit is invoked at any point.

\paragraph{Physical selection of the operating modes.} The modes entering a given operating point are selected at two logically distinct tiers, and only the second is an approximation. \emph{Within the occupied band}, the retained modes are never physically decoupled and need not be: when the device supports $K$ in-band modes and the construction uses $M\leq K$ of them, the selection is performed entirely on the readout side, by the time-integration basis functions of Sec.~S.2.3, which are chosen to vanish on the eigenvalue sums of every unused mode (Remark~\ref{s:rem:KgtM}). No hardware acts on individual modes; the weight functions $W_n(t)$ do. \emph{At the band edge}, the restriction to the occupied band is physical and approximate: the drive spectrum is confined to the band, and the coupling of out-of-band modes is suppressed by the fixed envelope (they cannot be switched off, only made envelope-small) and the error of exactly this truncation is what the envelope-convolution step of Lemma~\ref{s:lem:kerneldecay} absorbs, at the kernel level, uniformly in $K$. The finite-$K$ statements of this Supplement are exact for the retained-band model Eq.~\eqref{s:eq:familyrule}; the retained-band model approximates the physical device with the envelope-controlled error just located, and nowhere else.

\section{Proof of Universality}
\label{s:sec:proof}

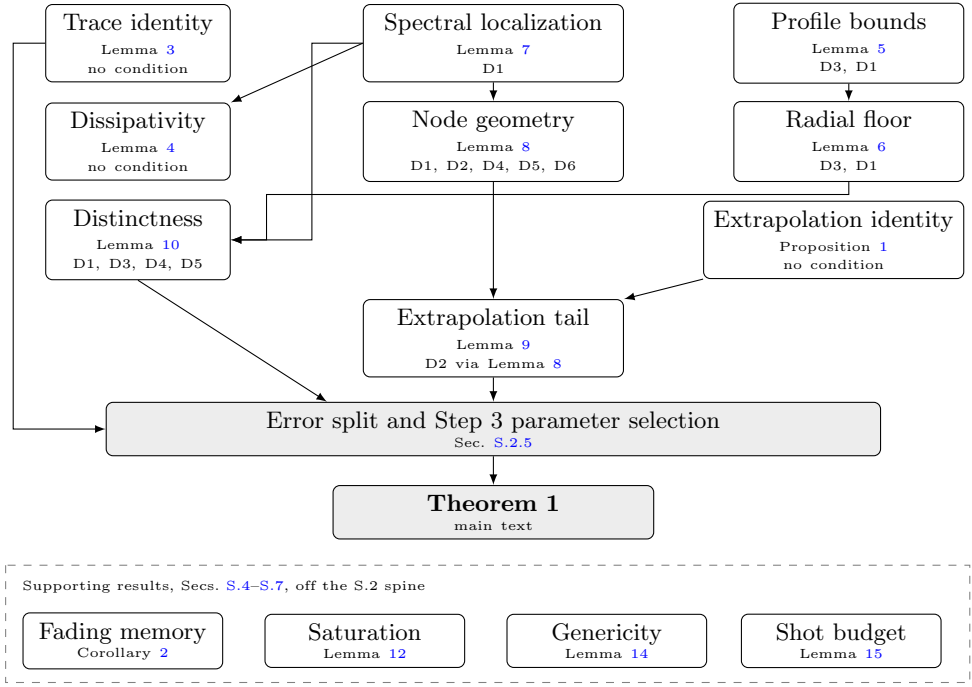
\begin{figure}[!t]
\centering
\begin{tikzpicture}[x=1mm,y=1mm,
  nd/.style={draw,rounded corners=0.9mm,align=center,inner sep=1.2mm,
             minimum height=9mm,font=\scriptsize},
  agg/.style={nd,fill=black!7,minimum height=7mm},
  sup/.style={nd,minimum height=6mm},
  ar/.style={-latex,thin}]

\node[nd,text width=22mm] (L3)  at (18,0)    {Trace identity\\{\tiny Lemma~\ref{s:lem:trace}}\\[-2pt]{\tiny no condition}};
\node[nd,text width=32mm] (L7)  at (65,0)    {Spectral localization\\{\tiny Lemma~\ref{s:lem:local}}\\[-2pt]{\tiny D1}};
\node[nd,text width=28mm] (L5)  at (112,0)   {Profile bounds\\{\tiny Lemma~\ref{s:lem:profile}}\\[-2pt]{\tiny D3, D1}};

\node[nd,text width=22mm] (L4)  at (18,-13)  {Dissipativity\\{\tiny Lemma~\ref{s:lem:strictdiss}}\\[-2pt]{\tiny no condition}};
\node[nd,text width=32mm] (L8)  at (65,-13)  {Node geometry\\{\tiny Lemma~\ref{s:lem:nodes}}\\[-2pt]{\tiny D1, D2, D4, D5, D6}};
\node[nd,text width=28mm] (L6)  at (112,-13) {Radial floor\\{\tiny Lemma~\ref{s:lem:radial}}\\[-2pt]{\tiny D3, D1}};

\node[nd,text width=22mm] (L10) at (18,-26)  {Distinctness\\{\tiny Lemma~\ref{s:lem:distinct}}\\[-2pt]{\tiny D1, D3, D4, D5}};
\node[nd,text width=32mm] (P1)  at (110,-26) {Extrapolation identity\\{\tiny Proposition~\ref{s:prop:extrapolation}}\\[-2pt]{\tiny no condition}};

\node[nd,text width=32mm] (L9)  at (65,-39)  {Extrapolation tail\\{\tiny Lemma~\ref{s:lem:tail}}\\[-2pt]{\tiny D2 via Lemma~\ref{s:lem:nodes}}};

\node[agg,text width=100mm] (S3) at (65,-51)
  {Error split and Step~3 parameter selection\\[-2pt]{\tiny Sec.~\ref{s:sec:errorbound}}};
\node[agg,text width=40mm] (TH) at (65,-62)
  {\textbf{Theorem 1}\\[-2pt]{\tiny main text}};

\draw[ar] (L7.west) -- (L4.north east);
\draw[ar] (L7) -- (L8);
\draw[ar] (L5) -- (L6);
\draw[ar] (L7.west) -| (41,-26) -- (L10.east);
\draw[ar] (L6.south) -- (112,-20) -- (35,-20) |- (L10.east);
\draw[ar] (L8) -- (L9);
\draw[ar] (P1.south west) -- (L9.north east);
\draw[ar] (L10.south) -- ([xshift=-22mm]S3.north);
\draw[ar] (L9) -- (S3);
\draw[ar] (L3.west) -- (1.5,0) -- (1.5,-51) -- (S3.west);
\draw[ar] (S3) -- (TH);

\draw[dashed,gray] (0.5,-69) rectangle (128,-84.5);
\node[anchor=north west,font=\tiny] at (1.5,-69.8)
  {Supporting results, Secs.~\ref{s:sec:fadingmemory}--\ref{s:sec:shotnoise}, off the S.2 spine};
\node[sup,text width=24mm] at (16,-79) {Fading memory\\[-2pt]{\tiny Corollary~\ref{s:cor:fmp}}};
\node[sup,text width=24mm] at (48,-79) {Saturation\\[-2pt]{\tiny Lemma~\ref{s:lem:saturation}}};
\node[sup,text width=24mm] at (80,-79) {Genericity\\[-2pt]{\tiny Lemma~\ref{s:lem:Q}}};
\node[sup,text width=24mm] at (111,-79) {Shot budget\\[-2pt]{\tiny Lemma~\ref{s:lem:shotnoise}}};
\end{tikzpicture}
\caption{Map of the universality argument. An arrow runs from a result to the
result that consumes it, and the graph is acyclic; the third line of each box
names the design conditions the result consumes, or records that it consumes
none. Theorem~1 of the main text is closed by the parameter selection of
Sec.~\ref{s:sec:errorbound}, which consumes the distinctness margins, the
extrapolation-tail bound, and the mean decay rate the trace identity pins. Two
results carry no design condition and are load bearing for that reason:
Lemma~\ref{s:lem:trace} is a rank argument on the generator, and
Proposition~\ref{s:prop:extrapolation} is an algebraic identity about Vandermonde
nodes. Four numbered lemmas are absent because no path from them reaches
Theorem~1: Lemmas~\ref{s:lem:profilegen} and~\ref{s:lem:envtail} support the
finite-$K$ family of Sec.~\ref{s:sec:family}, Lemma~\ref{s:lem:kerneldecay}
supports Corollary~\ref{s:cor:fmp}, and Lemma~\ref{s:lem:sector} supports
Lemma~\ref{s:lem:saturation}. Sec.~\ref{s:sec:errorbound} states the same
dependency order in prose; this figure is that paragraph drawn, and adds no claim
to it.}
\label{s:fig:depmap}
\end{figure}

We prove that the class $\mathcal{C}$ of Theorem~1 of the main text is universal. Figure~\ref{s:fig:depmap} maps the argument before it begins: which result feeds which, and which design condition each one consumes. The output is
\begin{align}
    \label{s:eq:output}
    \hat{y}_k &= \sum_{n=0}^N\int_0^{T_\mathrm{off}}W_n(t)\,\mathrm{Tr}[q_\mathrm{geo}^n e^{\mathcal{L}_0 t}\rho_{\zeta}(t_k)]\dd t, \qquad  
    \mathcal{L}_0\varrho = \left[\mathbf{\Gamma}_g\cdot a\varrho, a^\dagger\right] + \left[a, \varrho a^\dagger\cdot\mathbf{\Gamma}_g^\dagger\right],
\end{align}
with $q_\mathrm{geo}=\mathrm{Re}[v_\mathrm{geo}^*\cdot a]$ the measured quadrature of the geometric leakage channel and $t_k=kT$. The profile $v_\mathrm{geo}$ follows from the mode functions themselves. For the standing waves $u_n(x)=\sqrt{2/\ell}\sin(k_nx)$ of the terminated line, the atom sits at an interior point $x=\ell_a$ and samples $u_n(\ell_a)\propto\sin(\omega_n\tau/2)$---the profiled emission channel $\alpha$---while the outcoupling port sits at the boundary, where $u_n$ vanishes and the leakage is governed by $|\partial_xu_n|^2=(2/\ell)k_n^2$, identical across modes up to the slowly varying $k_n^2$. Hence $v_\mathrm{geo}$ is frequency-flat over the retained band to relative error $\varepsilon_v\approx B/\omega_0$, the same budget as the flat-coupling axiom A2; the two profiles differ because one samples the mode function and the other its boundary derivative. The homodyne local oscillator and the window functions $W_n(t)$ carry all spectral freedom of the readout, acting on the measurement record only. Expanded in a Volterra series,  
\begin{align*}
    \hat{y}_k = \sum_{n=0}^N\sum_{m_1,\dots,m_n\geq1}\hat{h}_n(m_1,\dots,m_n)\,u_{k-m_1}\cdots u_{k-m_n}.
\end{align*}
The goal is to show the weights $W_n(t)$ can realize any kernel $\hat{h}_n$ with $n\leq N$, $m_i\leq M$; if $M,N$ grow with the mode number, arbitrarily accurate approximation follows by Theorem~1 of Cuchiero \emph{et al.}~\cite{Cuchiero}, whose hypotheses Sec.~\ref{s:sec:cuchiero} verifies directly, so the citation supplies precedent rather than a load-bearing step. The Volterra route also certifies that a reservoir matching kernels up to $(N',M')\geq(N,M)$ reproduces every output of a smaller one and can match strictly more.

\subsection*{Roadmap and notation}
This subsection maps the proof. Secs.~S.2.2--S.2.4 derive the exact Volterra expansion and reduce kernel prescription to reduction modulo the node polynomial. Sec.~\ref{s:sec:errorbound} is the quantitative heart: it opens with the trace identity, which pins the mean decay rate $\lambda_\star=(\gamma+\gamma_g)/2K$ the design spends, and continues with the error bound, the designed operating point, the parameter selection, and the scaling ledger. Sec.~\ref{s:sec:cuchiero} verifies the density hypotheses.  

Table~\ref{s:tab:notation} collects the recurring symbols. Each is also defined at first use; the table exists so that no reader must reconstruct a definition from context.
\begin{table}[h]
\centering
\small
\setlength{\tabcolsep}{4pt}
\renewcommand{\arraystretch}{1.15}
\begin{tabular}{llp{8.6cm}}
\hline
Symbol & First use & Meaning\\
\hline
$K$ & S.1.1 & number of retained comb modes of the finite-$K$ description\\  
$M,N$ & S.2 intro & memory depth and Volterra order of the matched kernel block\\  
$T$ & Eq.~\eqref{s:eq:output} & readout period (one on/off drive-and-measure cycle)\\  
$\gamma,\gamma_g$ & S.1 & outcoupling (readout) and drive-channel rates; $\rho=\gamma_g/\gamma$\\  
$v_\mathrm{geo}$ & Eq.~\eqref{s:eq:output} & readout leakage profile, fixed by the outcoupling geometry, frequency-flat over the retained band ($v_{\mathrm{geo},k}=1/\sqrt K$)\\  
$\alpha$ & S.1 & emission profile of the drive channel, $\alpha_k\propto\sin(\omega_k\tau/2)$, normalized\\  
$\mathbf{E}$, $e_0$ & S.2.5 & Hermitian dressing $\tfrac{\gamma}{2}v_\mathrm{geo}v_\mathrm{geo}^*+\tfrac{\gamma_g}{2}\alpha\alpha^*$ and its operator norm\\  
$e_1$, $\bar e$ & S.2.5 & maximal off-diagonal row sum of $\mathbf{E}$ (Eq.~\eqref{s:eq:e1}, $K$-independent) and $\bar e=\max\{e_0,e_1\}$\\  
$\lambda_k$, $v_{R,k}$, $v_{L,k}$ & S.2.3 & eigenvalues and right/left eigenvectors of $\mathbf{\Gamma}_g$\\  
$\omega_k$, $\delta_k$, $\Delta_0$ & S.2.5 & comb frequencies, their detunings from $\omega_0$, and the comb spacing\\  
$g_0$, $\beta$ & S.2.5 & graded-defect amplitude and base, $\beta=4N+1$ (condition D5, Eq.~\eqref{s:eq:D5})\\  
$\lambda_\star$, $\bar r$ & S.2.5 & mean decay rate $(\gamma+\gamma_g)/2K$ and node radius $e^{-\lambda_\star T}$\\  
$x_k$ & S.2.4 & spectral nodes $e^{-\lambda_kT}$ of the Vandermonde construction\\  
$\bar\varepsilon$, $\bar\varepsilon_\mathrm{target}$ & S.2.5 & node-placement error and its assigned budget $\min\{\epsilon/(4eAM),1/4\}$\\  
$s_\alpha$ & S.2.5 & peak-to-mean ratio $K\max_k\alpha_k^2$ of the emission profile\\  
$A(M,N)$ & Eq.~\eqref{s:eq:termii} & target-dependent tail prefactor (kernel sup-norms and input bound)\\  
$S_\mathrm{out}$ & S.2.5 & extrapolation tail $\sum_{m>M}\Vert c_m\Vert_1$\\  
$h_K$, $h_\infty$ & S.3.1 & finite-comb and continuum input--output kernels (distinct objects; only $h_\infty$ obeys the delay equation)\\  
$c_0$, $D(c_0)$ & S.3.1 & loop-stability margin and the delay-stability condition\\  
$T_P$ & S.3.1 & revival time $2\pi/\Delta_0$ of the finite comb\\  
$C_h$, $C_\psi$ & S.3.1 & kernel envelope constant and band-envelope exponential moment\\  
$u_{\max}$ & S.2.2 & uniform input bound defining $\mathcal{K}_{u_{\max}}$\\  
$\eta$ & S.1 & input coupling amplitude of the drive channel, $\eta=\gamma_g/2g$\\  
$c_\mathrm{on}$ & S.3.2 & on-window injection factor $\int_0^{T_\mathrm{on}}e^{-\mathbf{\Gamma}_g s}\dd s$ acting on $\alpha$, entering $\zeta_1$\\  
\hline
\end{tabular}
\caption{Recurring notation of the universality proof, with the subsection of first use.}
\label{s:tab:notation}
\end{table}

\subsection{Normal ordering and Heisenberg evolution}
For a single mode with $q=\mathrm{Re}[a]$, the normal-ordered and ordinary exponentials are related through Baker--Campbell--Hausdorff by
\begin{align*}
    :e^{qt}: = e^{\frac12 a^\dagger t}e^{\frac12 a t}, \qquad
    e^{qt} = :e^{qt}:\,e^{\frac18 t^2}
    \;\Rightarrow\; q^n = \sum_{m=0}^{\lfloor n/2\rfloor}\frac{n!}{8^m(n-2m)!\,m!}:q^{n-2m}:.
\end{align*}
Hence any finite $\sum_{n=0}^N W_n q^n$ equals $\sum_{n=0}^N\widetilde{W}_n :q^n:$ with $\widetilde{W}_n=\sum_{m=0}^{\lfloor(N-n)/2\rfloor}\frac{(n+2m)!}{8^m n!\,m!}W_{n+2m}$, a triangular and hence invertible relabeling; the highest weights are unchanged and each lower one is introduced once, so the map $W_n\leftrightarrow\widetilde{W}_n$ is a bijection for finite $N$. Under $\mathcal{L}_0$ in the Heisenberg picture, $a\mapsto -\mathbf{\Gamma}_g\cdot a$ and $a^\dagger\mapsto -a^\dagger\cdot\mathbf{\Gamma}_g^\dagger$ act independently on each factor of a normal-ordered product, so
\begin{align*}
    (a^\dagger)^m a^n\,e^{\mathcal{L}_0 t} = \big(a^\dagger\cdot e^{-\mathbf{\Gamma}_g^\dagger t}\big)^m\big(e^{-\mathbf{\Gamma}_g t}\cdot a\big)^n,
\end{align*}
and therefore
\begin{align}
    \label{s:eq:return}
    \sum_{n=0}^N\int_0^{T_\mathrm{off}} W_n(t)\,q_\mathrm{geo}^n e^{\mathcal{L}_0 t}\dd t
    = \sum_{n=0}^N\int_0^{T_\mathrm{off}}\widetilde{W}_n(t)\,:\!\big(\mathrm{Re}[v_\mathrm{geo}^*\cdot e^{-\mathbf{\Gamma}_g t}\cdot a]\big)^n\!:\dd t.
\end{align}

\subsection{Coherent driven trajectory and Volterra kernels}
A note on notation: we write $\rho_{\zeta}(t)$ for the \emph{driven coherent-trajectory state} (the unique attractor of the time-dependent generator under the input Eq.~\eqref{s:eq:input}) and reserve $\rho_{\mathrm st}$ for the $u=0$ stationary state of $\mathcal{L}_0$. The Gaussian generator Eq.~\eqref{s:eq:LBDNMarkov} admits a coherent driven trajectory $\rho_{\zeta}(t)=\ket{\zeta(t)}\bra{\zeta(t)}$, where $\zeta(t)$ denotes the coherent mode-amplitude vector of the trajectory: substituting this ansatz reduces $\dot\rho_{\zeta}=\mathcal{L}(t)\rho_{\zeta}$ to  
\begin{align}
    \dot\zeta(t) = -\mathbf{\Gamma}_g\cdot\zeta(t) - i\eta u(t)\alpha,\qquad
    \zeta(t) = -i\eta\int_{-\infty}^t e^{-\mathbf{\Gamma}_g(t-\tau')}\cdot\alpha\,u(\tau')\dd\tau'.
\end{align}
With the input Eq.~\eqref{s:eq:input}, and with $t_k$ denoting the end of the $k$-th on-window (equivalently the start of the $k$-th measurement window), $\zeta(t_k)=\sum_{n\geq1}u_{k-n}\zeta_n$ where  
\begin{align}
    \label{s:eq:zetan}
    \zeta_n = -i\eta\left(\frac{1-e^{-\mathbf{\Gamma}_g T_\mathrm{on}}}{\mathbf{\Gamma}_g}\right)\cdot e^{-\mathbf{\Gamma}_g(n-1)T}\cdot\alpha.
\end{align}
Using Eq.~\eqref{s:eq:return} and $\langle\zeta|(a^\dagger\cdots)(a\cdots)|\zeta\rangle=(\zeta^*\cdots)(\zeta\cdots)$, the output collapses to
\begin{align}
    \hat{y}_k = \sum_{n=0}^N\int_0^{T_\mathrm{off}}\widetilde{W}_n(t)\big(\mathrm{Re}[v_\mathrm{geo}^*\cdot e^{-\mathbf{\Gamma}_g t}\cdot\zeta(t_k)]\big)^n\dd t,
\end{align}
so $\widetilde{W}_n(t)$ tunes the order-$n$ kernels, with
\begin{align}
    \label{s:eq:kernel}
    \hat{h}_n(m_1,\dots,m_n)=\int_0^{T_\mathrm{off}}\widetilde{W}_n(t)\prod_{i=1}^n\mathrm{Re}[v_\mathrm{geo}^*\cdot e^{-\mathbf{\Gamma}_g t}\cdot\zeta_{m_i}]\dd t.  
\end{align}

\subsection{Selecting eigenmodes by time integration}
\label{s:sec:selection}
We first record explicitly the spectral assumption used throughout this section.
\begin{assumption}[Finite-$K$ spectral resolution]
\label{s:ass:spectral}
At each finite mode number $K$, the matrix $\mathbf{\Gamma}_g$ is diagonalizable, with the biorthogonal spectral resolution
$\mathbf{\Gamma}_g=\sum_{k=1}^{K} v_{R,k}\,\lambda_k\, v_{L,k}^*$, $v_{L,k}^*\cdot v_{R,k'}=\delta_{kk'}$.  
\end{assumption}
We fix the normalisation once, since two conventions are in play. The resolvent construction of Lemma~\ref{s:lem:local} returns $\tilde v_{R,k}=e_k+a_k$ and $\tilde v_{L,k}=e_k+b_k$ with $k$-th component unity, so $(a_k)_k=(b_k)_k=0$ and $\tilde v_{L,k}^*\tilde v_{R,k}=1+b_k^*a_k$ with $|b_k^*a_k|\leq\Vert a_k\Vert\Vert b_k\Vert$: the duality defect is second order in the localisation radius, because both corrections are orthogonal to $e_k$. We therefore set $v_{R,k}\equiv\tilde v_{R,k}$ and $v_{L,k}\equiv\tilde v_{L,k}/(1+b_k^*a_k)$, which is biorthogonal exactly, and every overlap bound stated for the tilde vectors transfers with that single bounded factor, which the kernel construction absorbs into the prescribed moment. 

This holds automatically under the conditions of Theorem~1: the non-resonance condition implies in particular that the eigenvalues are pairwise distinct at each finite $K$ (the two-index case of Lemma~\ref{s:lem:distinct}(a,b) at designed operating points; at finite $K$ it also follows from the disjoint localization disks of Lemma~\ref{s:lem:local}), and a finite matrix with distinct eigenvalues is diagonalizable, its left/right eigenvector systems admitting the stated biorthogonal normalization. We emphasize that the resolution is invoked \emph{only at finite $K$}, where the sum is finite and no convergence question arises. No infinite-mode spectral decomposition of the (non-normal) generator is used anywhere in the proof: the kernel-matching construction below operates at finite $K$, and the many-mode limit enters solely through the scalar input--output kernel $h_K(t)$ and its uniform decay (Lemma~\ref{s:lem:kerneldecay}). Questions about the convergence of $\sum_k v_{R,k}\lambda_k v_{L,k}^*$ as $K\to\infty$---equivalently, whether the eigenvector system of the non-normal limit operator forms a Riesz basis---therefore never arise; the proof rests on the kernel precisely because operator-level arguments in the many-mode limit fail (Lemma~\ref{s:lem:trace}). The family of finite-$K$ generators to which this assumption applies is defined, from a single physical device by a single rule, in Sec.~\ref{s:sec:family}; in particular, no relation between the spectral data at different $K$ is assumed anywhere (Remark~\ref{s:rem:nonesting}).

Expanding Eq.~\eqref{s:eq:kernel} in this basis, each term's time integral has the form $F(s)=\int_0^{T_\mathrm{off}}f(t)e^{-st}\dd t$, whose Taylor coefficients about $s=0$ are $(-1)^m\int_0^{T_\mathrm{off}}f(t)t^m\dd t$. Real $f$ enforces $F(s)^*=F(s^*)$. What is needed is not the prescription of Taylor coefficients at $s=0$ but interpolation at the finitely many prescribed points $\Lambda_j$: taking $\widetilde W_n$ in the span of $\{t^p\}_{p<J}$ with $J$ the number of points, the realized values are $Mc$ with $M_{jp}=\int_0^{T_\mathrm{off}}t^pe^{-\Lambda_jt}\dd t$, and prescribability is the nonsingularity of $M$.  

As $T_\mathrm{off}\to\infty$, $M_{jp}\to p!/\Lambda_j^{p+1}$ and $\det M$ tends to a Vandermonde in the reciprocals $1/\Lambda_j$, nonzero exactly when the $\Lambda_j$ are distinct and nonzero---supplied by the distinctness of Lemma~\ref{s:lem:distinct} and by $\lambda_k\neq0$ of Lemma~\ref{s:lem:strictdiss}. At finite $T_\mathrm{off}$ the correction is $\int_{T_\mathrm{off}}^\infty$, exponentially small once $\mathrm{Re}\,\Lambda_j T_\mathrm{off}\gg J$; nonsingularity is certified outright above that threshold. Below it nonsingularity is not lost but merely non-uniform, and a genericity argument of the same type used for the delay in Sec.~\ref{s:sec:genericity} settles it: each entry $\int_0^{T_\mathrm{off}}t^pe^{-\Lambda_jt}\dd t$ is entire in the window length, hence so is the determinant, which tends as $T_\mathrm{off}\to\infty$ to a Vandermonde in the reciprocals $1/\Lambda_j$, nonzero whenever the $\Lambda_j$ are distinct and nonzero. An analytic function on $(0,\infty)$ that is not identically zero has a discrete zero set, and the union over the finitely many conjugation-resolved point sets at fixed $(M,N,K)$ is again discrete, so prescribability holds at every window length outside a discrete exceptional set. In plain terms: over a short window two nearby exponentials can look alike and the matrix can degenerate, but degeneracy is a coincidence rather than a regime, so the bad windows are isolated points on the clock exactly as the bad mirror distances are isolated points on the ruler. The deposited evaluation at the operated $T_\mathrm{off}$ confirms that this operating point is not exceptional, and discharges no step of the argument.

We state plainly why the threshold is not quantified and the genericity argument is carried instead. Certifying nonsingularity outright would require $\mathrm{Re}\,\Lambda_jT_\mathrm{off}\gtrsim J$ with $J$ the number of conjugation-resolved sum points, which grows combinatorially in $(M,N)$ as $J\leq\sum_{n\leq N}2^n\binom{M+n-1}{n}$, while the floor available from Lemma~\ref{s:lem:strictdiss} is $\mathrm{Re}\,\Lambda_j\geq\gamma/18K$; the resulting requirement $T_\mathrm{off}\gtrsim18KJ/\gamma$ is incompatible with the readout period $T\geq T_\mathrm{off}$ that Step~3 selects, which grows only as $O(M\ln M+\ln\epsilon^{-1})$. The threshold route is therefore closed to us, and $T_\mathrm{off}$ is deliberately not a selected parameter of the theorem: the discrete exceptional set is what carries this step, exactly as it carries the delay in Sec.~\ref{s:sec:genericity}.

Defining $F_n(\Lambda)=\int_0^{T_\mathrm{off}}\widetilde{W}_n(t)e^{-\Lambda t}\dd t$, we only need to fix $F_n$ at the discrete points $\sum_{i=1}^j\lambda_{k_i}^*+\sum_{i=j+1}^n\lambda_{k_i}$. The spectral conditions $\sum_k\mathrm{Re}[\lambda_k]n_k\neq0$, $\sum_k\mathrm{Im}[\lambda_k]n_k\neq0$ (for $\sum_k n_k=0$, $\sum_k n_k^2>0$) guarantee these points are distinct, so real basis functions $F^\pm_{k_1\dots k_n}$ can be built that equal $\pm1,\pm i$ on the target point and $0$ elsewhere while respecting $F(s)^*=F(s^*)$. This yields  
\begin{align}
    \label{s:eq:kernel2}
    \hat{h}_n(m_1,\dots,m_n)=\sum_{k_1,\dots,k_n}\mathrm{Re}\!\left[w_{\mathrm{Ord}(k_1\dots k_n)}\prod_{i=1}^n\big(v_\mathrm{geo}^*\cdot v_{R,k_i}\big)\big(v_{L,k_i}^*\cdot\zeta_{m_i}\big)\right].
\end{align}
The overlap conditions $v_\mathrm{geo}^*\cdot v_{R,k}\neq0$ and $v_{L,k}^*\cdot\alpha\neq0$ ensure no factor vanishes.

\subsection{From kernels to weights: Vandermonde inversion}
By Eq.~\eqref{s:eq:zetan}, $v_{L,k}^*\cdot\zeta_m = e^{-\lambda_k(m-1)T}(v_{L,k}^*\cdot\zeta_1)$; every $\mathrm{Re}\,\lambda_k>0$ by Lemma~\ref{s:lem:strictdiss}, so $|e^{-\lambda_k T_\mathrm{on}}|<1$ and $1-e^{-\lambda_k T_\mathrm{on}}\neq0$. Absorbing the $k$-dependent factors into $\mathcal{W}$, Eq.~\eqref{s:eq:kernel2} becomes  
\begin{align}
    \hat{h}_n(m_1,\dots,m_n)=\sum_{k_1,\dots,k_n}\mathrm{Re}\!\left[\mathcal{W}_{\mathrm{Ord}(k_1\dots k_n)}\prod_{i=1}^n V_{k_i m_i}\right],\quad V_{km}=x_k^{m-1},\; x_k=e^{-\lambda_k T}.  
\end{align}
$V$ is a Vandermonde matrix in the nodes $x_k=e^{-\lambda_k T}$. Injectivity of $x_k$ requires that distinct eigenvalues not be aliased by the exponential, i.e.\ $\mathrm{Im}[(\lambda_k-\lambda_{k'})]T\notin2\pi\mathbb{Z}$ for $k\neq k'$; this is exactly the non-resonance condition applied to the two-index tuple $n_k=+1,\,n_{k'}=-1$ (which has $\sum_k n_k=0$, $\sum_k n_k^2=2>0$), so distinctness of the complex eigenvalue sums guarantees it. Hence all $x_{k'}-x_k\neq0$, so $\det V=\prod_{k<k'}(x_{k'}-x_k)\neq0$ and $V$ is invertible when exactly $M$ modes participate. Inverting,
\begin{align*}
    \mathcal{W}_{\mathrm{Ord}(k_1\dots k_n)}&=\sum_{m_1'\leq\dots\leq m_n'}\mathcal{W}'_{m_1'\dots m_n'}\Big(\sum_{\mathcal{P}_n}\prod_{i=1}^n V^{-1}_{\mathcal{P}_n(m')_i k_i}\Big)\\
    &\;\Rightarrow\;\; \hat{h}_n(m_1,\dots,m_n)=\mathcal{W}'_{m_1\dots m_n},
\end{align*}
so all kernels of order $n\leq N$ and memory $m_i\leq M$ are fixed by the weights. This requires the device to support at least $M$ modes; no infinite-mode limit is taken, here or anywhere in the proof.
\begin{remark}[Exact matching persists for $K>M$]
\label{s:rem:KgtM}
For $K>M$ we do not resort to a pseudoinverse (which would yield only least-squares matching): the eigenmode-selection step of Sec.~S.2.3 allows the weights to be chosen so that the basis functions $F$ vanish on the eigenvalue sums of any $K-M$ unused modes, reducing the matching to a square Vandermonde system on the $M$ selected modes. This step consumes distinctness over the \emph{full} $K$-mode sum set---the sums involving unused modes must be resolved from those of the selected block---at the designed point this resolution is supplied by extending the defect hierarchy over every in-band mode (Remark~\ref{s:rem:embed}) together with Lemma~\ref{s:lem:radial}, whose prime $q$ is sized against $K_\mathrm{phys}$ by condition D3. The matching therefore remains exact for every $K\geq M$, and the strict-scaling corollary of the main text rests on this remark.
\end{remark}

\subsection{The error bound and the designed operating point}

\paragraph{Dependency order.} The conditions and lemmas of this section form an acyclic chain, and it
is worth displaying once. The physical data of Sec.~\ref{s:sec:family} fix the delay lock D3, which
feeds the profile bounds and the radial floor. The rate ratio D4 and the graded defects D5 are then
selected, and the comb spacing D1 last; its localization link D1a supports the spectral-localization
lemma, which in turn supports both the explicit strict-dissipativity floor and the node-geometry
lemma, and the latter with D2 supports the extrapolation-tail lemma. Distinctness draws on D4, D5 and
the radial floor, and feeds the selection and Vandermonde steps. Four citations inside this section run
forward in the text and not in the logic, and all four are recorded here rather than left for the
reader to notice: the explicit floor of Lemma~\ref{s:lem:strictdiss} uses the localization lemma
stated below it, and Lemmas~\ref{s:lem:strictdiss}, \ref{s:lem:profile} and \ref{s:lem:radial} each
cite an entry of the comb-spacing condition Eq.~\eqref{s:eq:D1}, which is displayed in full at the
localization lemma because that is where its first entry is consumed. The count is enforced by the
gate \texttt{check\_forward\_refs.py}; outside this section the only such citation is
Lemma~\ref{s:lem:saturation}'s use of Lemma~\ref{s:lem:sector}. In every case the cited statement is
proved without reference to the citing one, so no cycle exists. Step~3 selects in the order
$(M,N)\to q\to\bar\varepsilon_\mathrm{target}\to T\to\rho\to g_0\to\Delta_0$.
\label{s:sec:errorbound}
The mode-space generator obeys a conservation law that fixes the arithmetic of every rate in the proof, and we establish it before anything is built on top of it. The Hermitian part of $\mathbf{\Gamma}_g$ has rank at most two, so its trace is pinned by the two physical rates and cannot grow with the mode number. Two consequences follow immediately: the mean decay rate is exactly $\lambda_\star=(\gamma+\gamma_g)/2K$, which is the quantity the designed operating point of this section spends; and no gap $\mathrm{Re}[\lambda_k]\geq\varepsilon_D>0$ can hold uniformly in $K$, so spectral-gap hypotheses are unavailable to us as a matter of structure rather than of technique.
\begin{lemma}[Trace identity]
\label{s:lem:trace}
Let $\mathbf{\Gamma}_g = i\mathbf{\Omega} + \tfrac{\gamma}{2}(v_\mathrm{geo}\otimes v_\mathrm{geo}^*) + \tfrac{\gamma_g}{2}(\alpha\otimes\alpha^*)$ act on $\mathbb{C}^K$ with $\|v_\mathrm{geo}\|=\|\alpha\|=1$. Then
\begin{align}
    \sum_{k=1}^{K}\mathrm{Re}[\lambda_k] \;=\; \mathrm{Tr}\!\left[\tfrac{1}{2}(\mathbf{\Gamma}_g+\mathbf{\Gamma}_g^\dagger)\right] \;=\; \frac{\gamma+\gamma_g}{2},
\end{align}
independently of $K$. Consequently the mean of $\mathrm{Re}[\lambda_k]$ is exactly $\lambda_\star=(\gamma+\gamma_g)/2K$ and $\min_k \mathrm{Re}[\lambda_k] \leq \lambda_\star$, so no uniform gap can hold along the continuum limit: with the Hermitian part of rank at most two, the closing of the mode-space gap is structural.
\end{lemma}

\begin{lemma}[Strict dissipativity at finite $K$]
\label{s:lem:strictdiss}
Let $\mathbf{\Gamma}_g=i\mathbf{\Omega}+\mathbf{E}$ with $\mathbf{\Omega}=\mathrm{diag}(\delta_k)$ real and $\mathbf{E}=\tfrac{\gamma}{2}(v_\mathrm{geo}\otimes v_\mathrm{geo}^*)+\tfrac{\gamma_g}{2}(\alpha\otimes\alpha^*)$, with $\gamma>0$, $\gamma_g\geq0$, and the geometry-fixed flat coupler $v_{\mathrm{geo},k}=1/\sqrt K$ of Sec.~\ref{s:sec:family}. If the $\delta_k$ are pairwise distinct then every eigenvalue of $\mathbf{\Gamma}_g$ satisfies $\mathrm{Re}\,\lambda_k>0$. If in addition the detunings are separated, $|\delta_k-\delta_j|\geq\tfrac78\Delta_0$ for $j\neq k$ --- which conditions D1 and D5 supply on the designed comb, and which an undefected equidistant comb supplies with separation $\Delta_0$ --- and $\Delta_0\geq8\sqrt{K}\,\bar e$, then  
\begin{align}
    \label{s:eq:strictdiss}
    \mathrm{Re}\,\lambda_k\;\geq\;\frac{\gamma}{18K}\qquad\text{for every }k .
\end{align}
Consequently $\lambda_k\neq0$ for every $k$; under the spacing hypothesis every conjugation-resolved eigenvalue sum of order $n\geq1$ obeys $\mathrm{Re}\,\Lambda^{(S)}\geq\gamma/18K$, while the unconditional half $\mathrm{Re}\,\lambda_k>0$---which is all that the resolvent existence of Sec.~\ref{s:sec:shotnoise} requires---holds without it.
\end{lemma}
\begin{proof}
The strategy is to read the real part of any eigenvalue off the Rayleigh quotient: the frequency comb contributes nothing to it, so the whole real part comes from the Hermitian dressing, which the trace identity bounds below uniformly in $K$. Let $\mathbf{\Gamma}_gx=\lambda x$ with $x\neq0$. Then $\lambda=x^*\mathbf{\Gamma}_gx/(x^*x)$, and both $x^*\mathbf{\Omega}x$ and $x^*\mathbf{E}x$ are real, $\mathbf{\Omega}$ and $\mathbf{E}$ being Hermitian, so
\begin{align}
    \label{s:eq:relambda}
    \mathrm{Re}\,\lambda\;=\;\frac{x^*\mathbf{E}x}{x^*x}\;=\;\frac{\tfrac{\gamma}{2}|v_\mathrm{geo}^*x|^2+\tfrac{\gamma_g}{2}|\alpha^*x|^2}{x^*x}\;\geq\;0 .
\end{align}
Suppose $\mathrm{Re}\,\lambda=0$. Both terms are nonnegative and $\gamma>0$, so $v_\mathrm{geo}^*x=0$ and $\mathbf{E}x=0$, whence $i\mathbf{\Omega}x=\lambda x$ and $x$ is an eigenvector of the real diagonal $\mathbf{\Omega}$. Its entries being pairwise distinct, every eigenspace is one-dimensional, so $x=e_k$ for some $k$ up to scale; but then $v_\mathrm{geo}^*x=1/\sqrt K\neq0$, a contradiction. Hence $\mathrm{Re}\,\lambda>0$.

The $\delta_k$ are pairwise distinct on the designed comb: $|\delta_k-\delta_j|\geq\Delta_0-2g_0$ for $k\neq j$, and the second entry of Eq.~\eqref{s:eq:D1} with $g_0=\bar\varepsilon_\mathrm{target}/8T_{\max}$, where $T_{\max}$ is the readout-period ceiling fixed in the design step below gives $\Delta_0/g_0\geq512(e_0T_{\max}/\bar\varepsilon_\mathrm{target})^2>2048$, since $\lambda_\star T_{\max}=\Lambda_{\max}\geq2\ln 2M$ with $\lambda_\star\leq e_0$ and $\bar\varepsilon_\mathrm{target}\leq1$. Here $\lambda_\star\leq e_0$ because $\mathbf{E}$ is positive semidefinite of rank at most two with $\mathrm{tr}\,\mathbf{E}=(\gamma+\gamma_g)/2$ (both factors being unit vectors), so $e_0=\Vert\mathbf{E}\Vert\geq\tfrac12\mathrm{tr}\,\mathbf{E}=(\gamma+\gamma_g)/4\geq(\gamma+\gamma_g)/2K=\lambda_\star$ for $K\geq2$.  

For the explicit floor, Lemma~\ref{s:lem:local} gives $\Vert v_{R,k}-e_k\Vert\leq4e_0/\Delta_0$, which under $\Delta_0\geq8\sqrt K\,\bar e\geq8\sqrt Ke_0$ is at most $1/(2\sqrt K)\leq1/2$. Hence $\Vert v_{R,k}\Vert\leq3/2$ and
\begin{align}
    |v_\mathrm{geo}^*v_{R,k}|\;\geq\;|v_{\mathrm{geo},k}|-\Vert v_\mathrm{geo}\Vert\Vert v_{R,k}-e_k\Vert\;\geq\;\frac{1}{\sqrt K}-\frac{1}{2\sqrt K}\;=\;\frac{1}{2\sqrt K},
\end{align}
so Eq.~\eqref{s:eq:relambda} at $x=v_{R,k}$, discarding the nonnegative drive term, gives $\mathrm{Re}\,\lambda_k\geq\tfrac{\gamma}{2}\cdot\tfrac{1}{4K}\cdot\tfrac{4}{9}=\gamma/18K$. Finally $\mathrm{Re}\,\Lambda^{(S)}=\sum_i\mathrm{Re}\,\lambda_{k_i}\geq n\gamma/18K$, conjugation leaving real parts unchanged.
\end{proof}

The identity is immediate from the displayed trace: $i\mathbf{\Omega}$ is anti-Hermitian and the two rank-one Hermitian terms contribute $\gamma/2$ and $\gamma_g/2$; every $\mathrm{Re}\,\lambda_k>0$ at finite $K$, with the explicit floor $\gamma/18K$, by Lemma~\ref{s:lem:strictdiss}, which uses only the flat coupler and the distinctness of the $\delta_k$ and is independent of the emission profile.

This subsection contains the quantitative heart of the proof, and we begin with a plain statement of its plan. The argument has three steps.
\begin{enumerate}
\item \emph{From the error to the node geometry of an ideal comb.} The distance between target and device output is split into a truncation error of the target itself and the response of the matched device beyond the matched depth---the tail. An algebraic identity shows that once the kernels are matched, the tail is fixed by the spectral nodes $x_k=e^{-\lambda_kT}$ alone (no weight norm ever enters), and for an ideal equidistant comb with a common decay rate the tail is computable in closed form and exponentially small: $\leq 2M\bar r^{\,M}$.
\item \emph{The obstruction and the designed repair.} The same perfect regularity that makes the ideal tail computable defeats the order-$n$ kernel selection (degenerate eigenvalue sums) and zeroes the radial separations the selection also needs. This step repairs both, introducing each design condition at the point where the proof consumes it---graded frequency defects for the angular coincidences, the drive-channel spread for the radial ones, spacing and timing conditions as the localization and node-geometry lemmas require them---and assembles the conditions into a summary table at the end.
\item \emph{Parameter selection.} With the tail bound and the distinctness margins in hand, the accuracy target $\epsilon$ is met by choosing, once and in order, the depth and order $(M,N)$, the readout period $T$, and the comb; every constant is explicit.
\end{enumerate}
A ledger of how every design parameter scales with $(M,N,\epsilon)$, the accuracy envelope of a device at fixed geometry, and the distinction between the designed certificate and a generically fabricated device follow at the end.

\subsubsection*{Step 1: from the error split to the ideal comb}
\paragraph{Splitting the error.}
We bound the error against the target's own depth-$(M,N)$ Boyd--Chua truncation, so that the two integers $(M,N)$ are chosen jointly and once, and no constant is required to stay fixed while a memory depth is enlarged. Write $y^{(M,N)}$ for the truncation of the target functional $y$ at Volterra order $N$ and memory depth $M$, and let $\hat y$ denote the realized device output whose weights match, exactly, every kernel of $y^{(M,N)}$ (this is the kernel-matching construction of Secs.~S.2.3--S.2.4, valid for $K\geq M$). Then
\begin{align}
    |y_k-\hat{y}(\{u_k\})| \;\leq\; \underbrace{|y_k-y^{(M,N)}_k|}_{\text{(i) target truncation}} \;+\; \underbrace{|y^{(M,N)}_k-\hat{y}(\{u_k\})|}_{\text{(ii) realized tail beyond depth }M}.  
\end{align}
Quantitatively, for a target with a convergent Volterra expansion, term (i) is bounded by an element $\delta_{MN}$ of a sequence decreasing in $M,N$, by the Boyd--Chua approximation theorem for fading-memory functionals \cite{Boyd:1985}; the bound is uniform in the reservoir, and $K\geq M$ modes are required only so that term (ii) can be realized. Term (ii) is the total weight of the realized kernels beyond the matched depth---the device's response does not stop at lag $M$ merely because the weights were fitted there---and the next move shows that this excess is determined by the spectrum alone.  

\paragraph{The extrapolation identity.}
The key structural fact: \emph{once the kernels are matched at lags $\leq M$, the realized kernels at all lags $>M$ are determined by the node set $\{x_k\}$ alone, independently of the weights that realize the matching}. In particular no norm of the matched weights enters the error bound. (Supplementary Fig.~\ref{s:fig:conditioning} documents why any route through that norm fails: for nodes confined to an arc, $\ln\Vert V^{-1}\Vert_\infty$ grows linearly in $M$.) This is the fact that makes the bound finite-dimensional and explicit, and we now prove it.

\begin{proposition}[Extrapolation identity]
\label{s:prop:extrapolation}
Let $x_1,\dots,x_K\in\mathbb{C}$ be distinct, let $V$ be the Vandermonde matrix $V_{k\mu}=x_k^{\mu-1}$ ($\mu=1..K$), and for each $m\geq1$ let $c_m\in\mathbb{C}^K$ be the unique solution of  
\begin{align}
    (Vc_m)_k \;=\; x_k^{\,m-1}\qquad\text{for every }k=1,\dots,K
\end{align}
(so that $c_{m}=e_m$, the $m$-th standard basis vector, for $m\leq K$). Then the feature functions $f_m(t)=\sum_k g_k(t)\,x_k^{m-1}$ satisfy, identically in $t$ and for every coefficient family $g_k(t)$,  
\begin{align}
    f_m(t)\;=\;\sum_{\mu=1}^{K}c_{m,\mu}\,f_\mu(t),\qquad m\geq1 .
\end{align}
Consequently any output functional built multilinearly from $\{f_m\}$ and matched at lags $\leq K$ has its order-$n$ kernels at arbitrary lags given by the tensor extrapolation of the matched block,
\begin{align}
    \label{s:eq:tensorext}
    \hat h_n(m_1,\dots,m_n)\;=\;\sum_{\mu_1,\dots,\mu_n=1}^{K}\Big(\prod_{i=1}^{n}c_{m_i,\mu_i}\Big)\,H_n(\mu_1,\dots,\mu_n),  
\end{align}
where $H_n(\vec\mu)$ are the (weight-dependent, complex) matched-block moments; the extrapolation coefficients $c$ depend on the spectrum only.  
\end{proposition}
\begin{proof}
The defining equation $(Vc_m)_k=x_k^{m-1}$ reads, written out, $\sum_\mu c_{m,\mu}x_k^{\mu-1}=x_k^{m-1}$ for every $k$: the monomial of degree $m-1$, evaluated on the node set, coincides with a fixed linear combination of the first $K$ monomials evaluated on the same nodes. (For $m\leq K$ this combination is trivially the monomial itself, hence $c_m=e_m$; for $m>K$ it is the reduction of a high power modulo the degree-$K$ polynomial vanishing on the nodes.) Multiply this scalar identity by $g_k(t)$ and sum over $k$: every term of $f_m$ is reproduced, giving the first display. For the tensor statement, expand the product $\prod_i f_{m_i}(t)$ using the first display factor by factor; multilinearity distributes the sums, and integrating against the (fixed) weights preserves the linear combination, yielding Eq.~\eqref{s:eq:tensorext}.  
\end{proof}

In our setting $f_m(t)=v_\mathrm{geo}^*e^{-\mathbf{\Gamma}_gt}\zeta_m=\sum_kg_k(t)x_k^{m-1}$ with $g_k(t)=-i\eta(v_\mathrm{geo}^*v_{R,k})(v_{L,k}^*\alpha)\lambda_k^{-1}(1-e^{-\lambda_kT_\mathrm{on}})e^{-\lambda_kt}$ and $x_k=e^{-\lambda_kT}$ (Eq.~\eqref{s:eq:zetan}), so Proposition~\ref{s:prop:extrapolation} applies verbatim. Because the output uses real weights and real parts, the matched block enters through the conjugation-resolved moments $H^{(S)}_n(\vec\mu)$, one for each conjugation pattern $S\subseteq\{1..n\}$ of the factors $f/\bar f$. The constructive weights of Secs.~S.2.3--S.2.4 prescribe these moments directly. Two statements must be kept apart here. \emph{Prescribability}---that the moments can be set to chosen real values at all---requires the eigenvalue sums to be resolved, which the design of Sec.~\ref{s:sec:design} supplies with proven margins and a generic operating point supplies with verified ones. The resulting \emph{bound} $|H^{(S)}_n(\vec\mu)|\leq 2^n\max_{\vec\mu}|h^{(M,N)}_n(\vec\mu)|$ is then combinatorial, counting the $2^n$ conjugation patterns across which the real target kernel is distributed (Lemma~\ref{s:lem:distinct}); it carries no design margin, and the size of the weights realising it, which does, never enters the error bound (Sec.~S.2.5, resource (iv)).

With Eq.~\eqref{s:eq:tensorext} in hand, term (ii) reduces to a property of the node set. Write $S_\mathrm{out}\equiv\sum_{m>M}\Vert c_m\Vert_1$ for the total extrapolation weight beyond the matched depth (note $\Vert c_\mu\Vert_1=1$ for $\mu\leq M$). Since $y^{(M,N)}$ has no kernels beyond depth $M$ while the realized functional carries exactly the extrapolated kernels,
\begin{align}
    \label{s:eq:termii}
    |y^{(M,N)}_k-\hat y(\{u_k\})|
    \;&\leq\;\sum_{n=1}^{N}u_{\max}^n\!\!\sum_{\vec m\notin[1,M]^n}\!\!|\hat h_n(\vec m)|\nonumber\\
    \;&\leq\;\underbrace{\Big[\sum_{n=1}^{N}n\,2^{2n}\,u_{\max}^n\,M^{n-1}\,\Vert h^{(M,N)}_n\Vert_\infty\Big]}_{=:A(M,N)}\;\cdot\;S_\mathrm{out},
\end{align}
where the second line uses the counting bound $(M+S_\mathrm{out})^n-M^n\leq n\,S_\mathrm{out}(M+S_\mathrm{out})^{n-1}\leq n\,S_\mathrm{out}(2M)^{n-1}$ for $S_\mathrm{out}\leq M$, together with the $2^n$ moment bound above. The hypothesis $S_\mathrm{out}\leq M$ is not yet available at this point; it is a condition on the node geometry that the designed operating point delivers, and the parameter selection of Step 3 verifies it explicitly ($S_\mathrm{out}\leq2/M$ at every selected operating point). The prefactor $A(M,N)$ is a property of the target's own truncation, through its kernel sup-norms and the input bound; it involves no weight norms and no spectral constants. Everything now rides on making $S_\mathrm{out}$ small, which is a pure question of where the nodes $x_k$ sit in the complex plane. For orientation on the size of $A(M,N)$: when the target's kernels decay geometrically in order, $\Vert h^{(M,N)}_n\Vert_\infty\leq C_h\rho_h^{\,n}$, the sum is bounded by explicit geometric terms, $A(M,N)\leq(C_h/M)\sum_{n\leq N}n\,(4u_{\max}\rho_h M)^{n}$: summable uniformly in $N$ when $4u_{\max}\rho_h M<1$ and dominated by its last term otherwise. In either case $A$ is explicit in the target data alone; it enters the readout period only logarithmically and the comb spacing linearly (Table~\ref{s:tab:ledger}).

\paragraph{The ideal equidistant comb: the tail is not the problem.}
Suppose first that the device could be operated so that its spectral nodes sit exactly at
\begin{align}
    x^0_k=\bar r\,\omega^{k},\qquad \omega=e^{-2\pi i/M},\qquad k=0,\dots,M-1,\quad K=M:
\end{align}
the $M$-th roots of unity, shrunk onto a common radius $\bar r<1$. (The design below achieves this up to a small placement error; here we take it exactly. Note that the common radius means all decay rates $\mathrm{Re}\,\lambda_k$ coincide, so the radial separations that Lemma~\ref{s:lem:distinct} will later require are exactly zero here: the ideal comb is an illustrative limit for the tail computation rather than an admissible operating point, and Step 2 repairs precisely this.) The Vandermonde matrix then factors into a discrete Fourier transform and a radial scaling,
\begin{align}
    V^0_{k\mu}=(x^0_k)^{\mu-1}=\bar r^{\,\mu-1}\omega^{k(\mu-1)}=(F D_r)_{k\mu},  
    \qquad F_{k\mu}=\omega^{k(\mu-1)},\quad D_r=\mathrm{diag}(\bar r^{\,\mu-1}),  
\end{align}
and since $FF^*=MI$, the inverse is explicit: $(V^0)^{-1}=D_r^{-1}F^*/M$. The extrapolation coefficients can now be computed in closed form. The right-hand side of the defining equation has entries $(x^0_k)^{m-1}=\bar r^{\,m-1}\omega^{k(m-1)}$, and applying $F^*$ gives
\begin{align}
    (F^*\,\xi^0_m)_\mu&=\sum_{k}\omega^{-k(\mu-1)}\,\bar r^{\,m-1}\omega^{k(m-1)}  
    =\bar r^{\,m-1}\sum_k\omega^{k(m-\mu)}\nonumber\\
    &=\bar r^{\,m-1}\,M\,\mathbf{1}\{m\equiv\mu\ (\mathrm{mod}\ M)\},
\end{align}
by the geometric-sum orthogonality of the roots of unity. Dividing by $M$ and undoing the radial scaling,
\begin{align}
    \label{s:eq:c0closed}
    c^0_{m,\mu}\;=\;\bar r^{\,m-\mu}\,\mathbf{1}\{m\equiv\mu\ (\mathrm{mod}\ M)\}.
\end{align}
Physically, Eq.~\eqref{s:eq:c0closed} says the device's response at any lag $m>M$ is the response at the lag $\mu\equiv m\ (\mathrm{mod}\ M)$ inside the matched window, damped by one factor of $\bar r$ for every additional readout period. The angular positions repeat with period $M$ because the nodes are roots of unity; the radius supplies pure geometric decay. Each column of the tail therefore has exactly one nonzero entry, and for $m=\mu+jM$ ($j\geq1$, $1\leq\mu\leq M$),
\begin{align}
    \Vert c^0_m\Vert_1=\bar r^{\,jM}
    \quad\Longrightarrow\quad
    \sum_{m>M}\Vert c^0_m\Vert_1=\sum_{j\geq1}M\,\bar r^{\,jM}=\frac{M\,\bar r^{\,M}}{1-\bar r^{\,M}}\;\leq\;2M\,\bar r^{\,M}
\end{align}
for $\bar r^{\,M}\leq1/2$, a condition the parameter selection of Step 3 guarantees with room to spare: the readout period there enforces $\bar r\leq1/(2M)$, whence $\bar r^{\,M}\leq(2M)^{-M}\leq1/2$. The tail of the ideal comb is exponentially small in $M$, with the exponent set by the common decay rate through $\bar r=e^{-\lambda_\star T}$: the tail is not the problem.

\subsubsection{Step 2: the obstruction and the designed repair}
\label{s:sec:design}
The ideal comb cannot finish the proof. This step first identifies what must change relative to the ideal configuration of Step 1, then introduces the modifications and locates the error each contributes, and only then derives the error bounds, stating each design condition at the point where a proof consumes it. The conditions are collected in Table~\ref{s:tab:design} at the end of the step.

Two features of the ideal comb must change. First, the kernel matching of Secs.~S.2.3--S.2.4 requires more than good node positions. To prescribe an order-$n$ kernel, the time-integration step of Sec.~\ref{s:sec:selection} must resolve the individual eigenvalue sums $\Lambda^{(S)}(\vec k)=\sum_{i\in S}\lambda_{k_i}+\sum_{i\notin S}\bar\lambda_{k_i}$: two kernel entries can be prescribed independently only if their sums are distinct points in the complex plane. For the perfectly equidistant comb, $\delta_k=\Delta_0 k$, these sums collide massively. The simplest collision already occurs at order $n=2$:
\begin{align}
    \delta_0+\delta_2\;=\;0+2\Delta_0\;=\;\Delta_0+\Delta_0\;=\;\delta_1+\delta_1 ,  
\end{align}
so the selection integral cannot distinguish the kernel entry at mode pair $(0,2)$ from the entry at $(1,1)$; and every arithmetic-progression relation among the $\delta_k$ produces another such degeneracy. The perfect regularity that made the tail computable in Step 1 is exactly the regularity that defeats the order-$n$ selection. Second, the common radius of the ideal nodes means all decay rates coincide, so eigenvalue sums that differ only in their mode multisets are radially degenerate as well---the failure already flagged parenthetically in Step 1. The design must break both kinds of coincidence without moving the nodes appreciably, since the node positions carry the tail bound.

The design uses two mechanisms, angular and radial, both acting far below the node-placement budget; we fix the recurring symbols first, since every bound below prices its contribution against them.

All symbols are fixed once: $(M,T)$ come from the parameter selection of Step 3; $K=M$; $\lambda_\star\equiv(\gamma+\gamma_g)/2K$ is the mean decay rate that the trace identity (Lemma~\ref{s:lem:trace}) forces; $\bar r\equiv e^{-\lambda_\star T}$; $\rho\equiv\gamma_g/\gamma$ is the drive-to-readout rate ratio; $\mathbf{E}=\tfrac{\gamma}{2}(v_\mathrm{geo}\otimes v_\mathrm{geo}^*)+\tfrac{\gamma_g}{2}(\alpha\otimes\alpha^*)$ is the Hermitian dressing with norm $e_0\equiv\Vert\mathbf{E}\Vert\leq(\gamma+\gamma_g)/2$; $s_\alpha\equiv K\max_k\alpha_k^2$ measures the peak-to-mean ratio of the emission profile ($s_\alpha\leq q^2$ under condition D3, Lemma~\ref{s:lem:profile}(iii)); $\bar\varepsilon_\mathrm{target}$ is the node-placement budget handed down from Step 3; and $\beta\equiv4N+1$.

Two norms of the dressing must be kept apart, because the localization argument is a row-wise statement while the perturbation bounds are operator-norm statements. Alongside the operator norm $e_0=\Vert\mathbf{E}\Vert$ we therefore record the maximal off-diagonal row sum
\begin{align}
    \label{s:eq:e1}
    e_1\;\equiv\;\max_k\sum_{j\neq k}|E_{kj}|\;\leq\;\tfrac{\gamma}{2}\,|v_{\mathrm{geo},k}|\,\Vert v_\mathrm{geo}\Vert_1+\tfrac{\gamma_g}{2}\,|\alpha_k|\,\Vert\alpha\Vert_1\;\leq\;\frac{\gamma+\gamma_g\sqrt{s_\alpha}}{2},  
\end{align}
where the middle inequality is general while the final bound is specific to the flat coupler, $|v_{\mathrm{geo},k}|=1/\sqrt K$ with $\Vert v_\mathrm{geo}\Vert_1=\sqrt K$ (the geometry-fixed profile of Sec.~\ref{s:sec:family}), together with $|\alpha_k|\leq\sqrt{s_\alpha/K}$ and $\Vert\alpha\Vert_1\leq\sqrt K\Vert\alpha\Vert_2=\sqrt K$ by Cauchy--Schwarz. The point of Eq.~\eqref{s:eq:e1} is that $e_1$ is bounded \emph{independently of $K$}: this is a consequence of the rank-two structure of $\mathbf{E}$ with normalized factors, and it is not implied by the operator norm (a general matrix of norm $e_0$ has row sums as large as $\sqrt{K-1}\,e_0$). We write $\bar e\equiv\max\{e_0,e_1\}\leq(\gamma+\gamma_g\max\{1,\sqrt{s_\alpha}\})/2$ for the constant that the spacing condition D1, stated below at its point of use, controls. We caution against reading Eq.~\eqref{s:eq:e1} as an ordering of the norms themselves: it orders the two \emph{bounds}. Although $s_\alpha\geq1$ always (normalization forces $\max_k\alpha_k^2\geq1/K$), the flat profile $\alpha_k^2=1/K$ has $e_0=(\gamma+\gamma_g)/2$ exactly while $e_1=(\gamma+\gamma_g)(1-1/K)/2<e_0$, and the verified operating points likewise have $e_0>e_1$ (numerical paragraph below); random dressings realize either ordering, so the maximum is removable in neither direction.

\emph{Angular mechanism (designed): condition D5.} Add to each comb frequency an exponentially graded defect,
\begin{align}
    \tag{D5}\label{s:eq:D5}
    \delta_k\;=\;\Delta_0\,k\;+\;g_0\,\beta^{-k},\qquad \beta\equiv 4N+1 ,\qquad
    g_0\;=\;\frac{\bar\varepsilon_\mathrm{target}}{8\,T_{\max}},
\end{align}
the amplitude $g_0$ being fixed by the requirement, priced below in Lemma~\ref{s:lem:nodes}, that the largest defect displace any node phase by at most $g_0T\leq\bar\varepsilon_\mathrm{target}/8$, one quarter of the placement budget's half. Note that $g_0$ is an absolute frequency scale, independent of $\Delta_0$: enlarging the spacing (condition D1 below) leaves the defect hierarchy, and hence the separations it guarantees, untouched. The defects act like digits of a number written in base $\beta$: an eigenvalue sum at order $n\leq N$ picks up the defect combination $g_0\sum_k n_k\beta^{-k}$, where the integer $n_k$ (the signed multiplicity of mode $k$ in the sum) obeys $|n_k|\leq n\leq N$. Two different signed multiplicity vectors differ, at their first differing digit $j$, by at least $\beta^{-j}$ minus the largest possible carry from all deeper digits; because $\beta=4N+1$ exceeds four times the largest digit, the carries can never add up to a full digit, and the two combinations stay separated by at least $g_0\beta^{-j}/2$. This is the standard uniqueness of signed-digit representations, and it converts one continuous knob $g_0$ into a separation guarantee for every one of the finitely many coincidences at once.

\emph{Radial mechanism (generic): conditions D3 and D4.} Sums that share the same signed multiplicity vector but involve different mode multisets have identical comb and defect contributions; they are separated instead by the diagonal decay rates $E_{kk}=\gamma/2K+(\gamma_g/2)\alpha_k^2$, whose mode dependence enters through the emission profile $\alpha_k$. This spread is supplied by the drive channel, and two conditions make it usable. The profile is not free: the standing-wave geometry of Sec.~\ref{s:sec:family} fixes $\tilde\alpha_k\propto\sin(\omega_k\tau/2)$ with $\tau=2\ell_a/v$ the atom--mirror delay, $\omega_k=(n_0+k)\Delta_0+g_0\beta^{-k}$ and $\Delta_0=\pi v/\ell$, so that $\Delta_0\tau/2\pi=\ell_a/\ell$. Two integers control it: the atom's position as a fraction of the quantization length, and the band placement $n_0$. Condition \emph{D3 (delay lock)} fixes both. Let $q$ be the smallest prime with $q\geq2K_\mathrm{phys}+1$, where $K_\mathrm{phys}\geq K$ is the number of in-band modes; by Bertrand's postulate $q\leq4K_\mathrm{phys}+2$. Then  
\begin{align}
    \tag{D3}\label{s:eq:D3}
    &\frac{\ell_a}{\ell}=\frac{p}{q},\quad\gcd(p,q)=1,\\ \nonumber
    &n_0\bmod q=\nu\ \text{ with }\ (2\nu\bmod q)\in\{1,\dots,q-2K_\mathrm{phys}+1\} .  
\end{align}
The residue set is nonempty since $q\geq2K_\mathrm{phys}+1$, and an admissible $\nu$ exists since $q$ is an odd prime, so $\nu\mapsto2\nu$ is a bijection modulo $q$; $p=1$ is always admissible. Condition D3 places the atom at a rational point of the standing wave and shifts the retained band by at most $q$ modes. It replaces a requirement that $\tau$ merely avoid a discrete set, which left the radial margin without a symbolic lower bound; Eq.~\eqref{s:eq:D3} supplies one (Lemmas~\ref{s:lem:profile} and \ref{s:lem:radial}). 
\begin{lemma}[Profile bounds under the delay lock]
\label{s:lem:profile}
Assume Eq.~\eqref{s:eq:D3} and $\psi_0\equiv g_0\tau/2\leq1/q$, which is entry (iv) of Eq.~\eqref{s:eq:D1}. Write $A_k=\pi(n_0+k)p/q$ and $\psi_k=g_0\beta^{-k}\tau/2$, so $\tilde\alpha_k=\sin(A_k+\psi_k)$ and $\alpha_k=\tilde\alpha_k/\sqrt D$ with $D=\sum_j\tilde\alpha_j^2$. Then for every $0\leq k\leq K-1$: (i) $(n_0+k)p\not\equiv0\pmod q$, hence $|\sin A_k|\geq\sin(\pi/q)\geq2/q$; (ii) $1/q\leq|\tilde\alpha_k|\leq1$ and $K/q^2\leq D\leq K$; (iii) $|\alpha_k|\geq1/(q\sqrt K)$ and $s_\alpha\leq q^2$.  
\end{lemma}
\begin{proof}
(i) $q$ is prime and $\gcd(p,q)=1$, so $(n_0+k)p\equiv0$ would force $\nu\equiv-k$, hence $2\nu\equiv-2k\pmod q$ with $0\leq2k\leq2K-2$. The residues $-2k\bmod q$ lie in $\{0\}\cup\{q-2K_\mathrm{phys}+2,\dots,q-1\}$, disjoint from the admissible set of Eq.~\eqref{s:eq:D3}. Hence $|\sin A_k|=|\sin(\pi m_k/q)|$ with $m_k\in\{1,\dots,q-1\}$, so $|\sin A_k|\geq\sin(\pi/q)\geq2/q$ by concavity of $\sin$ on $[0,\pi/2]$. (ii) $|\psi_k|\leq\psi_0\leq1/q$ and $|\sin(A_k+\psi_k)-\sin A_k|\leq|\psi_k|$ give $|\tilde\alpha_k|\geq1/q$; summing over $K$ modes bounds $D$. (iii) Immediate from (ii).  
\end{proof}

\begin{lemma}[Radial separation floor]
\label{s:lem:radial}
Assume Eq.~\eqref{s:eq:D3} and $\psi_0\leq(4N)^{-(q-2)}/(32N)$, entry (iv) of Eq.~\eqref{s:eq:D1}. Let $d\in\mathbb{Z}^K$ satisfy $d\neq0$, $\sum_kd_k=0$, $\sum_k|d_k|\leq2N$. Then  
\begin{align}
    \label{s:eq:radfloor}
    \Big|\sum_k d_k\alpha_k^2\Big|\;\geq\;\frac{1}{8K}(4N)^{-(q-2)} .
\end{align}
\end{lemma}
\begin{proof}
The strategy is to turn the radial separation into a statement about a trigonometric sum over the comb, and then to bound that sum below by an algebraic argument at the $q$-th roots of unity rather than by estimating term by term. Since $\sum_kd_k=0$, $\sum_kd_k\tilde\alpha_k^2=-\tfrac12\sum_kd_k\cos(2A_k+2\psi_k)$, and expanding the cosine,
\begin{align}
    \label{s:eq:radsplit}
    &\sum_kd_k\tilde\alpha_k^2=-\tfrac12\sum_kd_k\cos2A_k+\tfrac12\sum_kd_k\cos2A_k(1-\cos2\psi_k)\\ \nonumber
    &\hspace{6em}+\tfrac12\sum_kd_k\sin2A_k\sin2\psi_k,
\end{align}
the last two terms bounded by $2N\psi_0^2$ and $2N\psi_0$ using $\sum_k|d_k|\leq2N$.

Let $\zeta=e^{2\pi ip/q}$, a primitive $q$-th root of unity, $\nu=n_0\bmod q$, and $P_d(x)=\sum_{k=0}^{K-1}d_kx^k$. Then $\sum_kd_k\cos2A_k=\mathrm{Re}[\zeta^\nu P_d(\zeta)]=\tfrac12R(\zeta)$ with $R(x)\equiv x^\nu P_d(x)+x^{q-\nu}P_d(x^{q-1})$ reduced modulo $x^q-1$, an integer polynomial of degree $<q$. Its two summands contribute $d_k$ at exponents $(\nu+k)\bmod q$ and $(-\nu-k)\bmod q$ respectively; these sets are disjoint, since $\nu+k\equiv-\nu-k'$ would give $2\nu\equiv-(k+k')$ with $0\leq k+k'\leq2K-2$, excluded by Eq.~\eqref{s:eq:D3} exactly as in Lemma~\ref{s:lem:profile}(i). No cancellation occurs, so $R\neq0$, $R$ has at most $2K$ nonzero coefficients, and its coefficient sum is $2\sum_k|d_k|\leq4N$.  

$q$ being prime, $\Phi_q=1+x+\dots+x^{q-1}$ is irreducible over $\mathbb{Q}$ (Eisenstein after $x\mapsto x+1$) and is the minimal polynomial of $\zeta$. If $R(\zeta)=0$ then $\Phi_q\mid R$, and $\deg R\leq q-1$ forces $R=c\,\Phi_q$; but $R$ has at most $2K\leq q-1$ nonzero coefficients while every coefficient of $\Phi_q$ is $1$, so $c=0$ and $R=0$, a contradiction. Hence $R(\zeta)$ is a nonzero algebraic integer of $\mathbb{Q}(\zeta)$, of degree $q-1$ over $\mathbb{Q}$, with conjugates $R(\zeta^j)$ each bounded in modulus by the coefficient sum $4N$. Its norm being a nonzero rational integer,  
\begin{align}
    1\leq\Big|\prod_{j=1}^{q-1}R(\zeta^j)\Big|\leq|R(\zeta)|(4N)^{q-2},
\end{align}
so $|R(\zeta)|\geq(4N)^{-(q-2)}$ and the leading term of Eq.~\eqref{s:eq:radsplit} has modulus at least $\tfrac14(4N)^{-(q-2)}$. With $\psi_0\leq(4N)^{-(q-2)}/(32N)\leq1$ the remainders total at most $4N\psi_0\leq\tfrac18(4N)^{-(q-2)}$, so $|\sum_kd_k\tilde\alpha_k^2|\geq\tfrac18(4N)^{-(q-2)}$. Dividing by $D\leq K$ gives Eq.~\eqref{s:eq:radfloor}. This proves the lemma: the radial separation of the nodes is bounded below by a quantity fixed by the comb arithmetic alone, so the floor survives at every $K$ and does not degrade as modes are added.
\end{proof}

Condition \emph{D4 (weak drive channel)}: the rate ratio obeys
\begin{align}
    \tag{D4}\label{s:eq:D4}
    \rho\;=\;\frac{\gamma_g}{\gamma}\;\leq\;\min\Big\{\frac{\bar\varepsilon_\mathrm{target}}{8\,s_\alpha\Lambda_{\max}},\;\frac{1}{8Ns_\alpha}\Big\},\qquad \Lambda\equiv\lambda_\star T=\max\Big\{\ln 2M,\;\tfrac{1}{M}\ln\tfrac{8MA}{\epsilon}\Big\} ,
\end{align}
so that the very spread being used for radial separation displaces the node moduli by less than one quarter of the budget's half (priced in Lemma~\ref{s:lem:nodes}): the drive channel is kept strong enough to separate and weak enough not to misplace. The spread is required only to exceed a perturbative remainder that the comb spacing (D1's third entry, below) makes arbitrarily small.

With the two mechanisms in place, the remaining conditions are on the comb spacing (D1) and the readout timing (D2), each stated immediately before the lemma that consumes it. The conditions divide three ways. D2 and the choices of $T$, $N$, $M$, the local-oscillator spectrum and the window functions are \emph{measurement settings}. D1, D3, D5 and the band placement are \emph{geometric settings}: dials of one fixed machine. The output coupler is neither: its profile and its flatness $\varepsilon_v$ are fabricated data.

\begin{remark}[Embedding the design in a device with $K_\mathrm{phys}>M$ in-band modes]
\label{s:rem:embed}
If the physical operating point retains $K_\mathrm{phys}>M$ comb modes in the occupied band (Sec.~\ref{s:sec:family}), the kernel matching itself is unaffected: the readout-side selection of Remark~\ref{s:rem:KgtM} confines every realized kernel, at all orders and all lags, to the $M$ selected modes, so the extrapolation identity and Lemma~\ref{s:lem:tail}(b) apply verbatim to the selected block. What changes is the rate bookkeeping. The trace budget of Lemma~\ref{s:lem:trace} is now shared among $K_\mathrm{phys}$ modes, so $\lambda_\star=(\gamma+\gamma_g)/2K_\mathrm{phys}$, and the advertised decay $\bar r^{\,M}=e^{-M\lambda_\star T}$ carries the factor $M/K_\mathrm{phys}$ in the exponent relative to the $K=M$ statement $e^{-(\gamma+\gamma_g)T/2}$. The designed operating point therefore chooses the quantization length $\ell$ and the occupied band so that approximately $M$ comb modes lie in-band, consistent with the main text's statement that modes are added by geometry; when $K_\mathrm{phys}>M$ is operated instead, the rate constant is the one just displayed, reported as such.
One further bookkeeping item completes the embedding: the readout-side selection of Remark~\ref{s:rem:KgtM} must resolve the eigenvalue sums involving the $K_\mathrm{phys}-M$ unselected modes, and at the designed point this resolution is supplied by extending the graded hierarchy of D5 over every in-band mode, $\delta_k=\Delta_0k+g_0\beta^{-k}$ for $k=0,\dots,K_\mathrm{phys}-1$. The signed-digit separation argument of Lemma~\ref{s:lem:distinct} is agnostic to the digit count, so it applies verbatim: the guaranteed angular floor becomes $g_0\beta^{-(K_\mathrm{phys}-1)}/4$, and the defect-protection entry of D1 carries the exponent $K_\mathrm{phys}-1$ in place of $M-1$, with the scaling ledger of Table~\ref{s:tab:ledger} read accordingly.
\end{remark}

We now derive the error bounds, in the order the construction forces: that the spectrum sits where the design intends (spectral localization), that the nodes land within budget of the roots-of-unity configuration (node geometry), that the tail bound of Step 1 survives the small displacements (extrapolation tail), and that the eigenvalue sums are pairwise resolved (design distinctness).

The localization lemma is the first consumer of a condition on the comb spacing, and we state that condition here, in full, since its five entries are consumed at five identified points. Condition \emph{D1 (equidistant comb, weak dressing, defect protection)}: $\omega_k=\omega_0+\Delta_0k$, $k=0..M{-}1$, with
\begin{align}
\tag{D1}\label{s:eq:D1}
&\Delta_0\;\geq\;\max\Big\{8\bar e\ \text{\footnotesize(D1a)},\ \ 8\sqrt{K}\,\bar e\ \text{\footnotesize(D1b)},\ \ 8q\sqrt{K}\,\bar e\ \text{\footnotesize(D1c)},\ \ \frac{64\,e_0^2\,T_{\max}}{\bar\varepsilon_\mathrm{target}},\\ \nonumber
&\hspace{4.2em}\frac{384\,N\,e_0^2\,T\,\beta^{\,M-1}}{\bar\varepsilon_\mathrm{target}},\ \ 
32\pi N g_0(4N)^{q-2},\ \ \frac{512\,N\,K\,e_0^2}{\gamma_g}(4N)^{q-2}\Big\}.
\end{align}
The first entry keeps the dressing perturbative and is consumed twice. Its factor $8$ is what makes the localization radius satisfy $8e_0^2/\Delta_0\leq\Delta_0/8$, an estimate the resolvent chain of Lemma~\ref{s:lem:local} uses twice; its factor $q\sqrt K$ is what makes \emph{both} overlap conditions hold with an explicit margin. The two sides are not symmetric, and the asymmetry sets the entry. Writing $\Vert v_{R,k}-e_k\Vert,\Vert v_{L,k}-e_k\Vert\leq4e_0/\Delta_0$ (Lemma~\ref{s:lem:local}) and $\Vert v_\mathrm{geo}\Vert=\Vert\alpha\Vert=1$,
\begin{align}
\label{s:eq:overlaps}
|v_\mathrm{geo}^*\cdot v_{R,k}|\;&\geq\;|v_{\mathrm{geo},k}|-\frac{4e_0}{\Delta_0}\;=\;\frac{1}{\sqrt K}-\frac{4e_0}{\Delta_0},\\ \nonumber
|v_{L,k}^*\cdot\alpha|\;&\geq\;|\alpha_k|-\frac{4e_0}{\Delta_0}\;\geq\;\frac{1}{q\sqrt K}-\frac{4e_0}{\Delta_0},
\end{align}
the drive-side floor $|\alpha_k|\geq1/(q\sqrt K)$ coming from Lemma~\ref{s:lem:profile}(iii). The readout side, with its floor $1/\sqrt K$, would close already at $\Delta_0\geq8\sqrt K\,\bar e$, which leaves $1/(2\sqrt K)$; the drive side, whose floor is smaller by the factor $q$, requires $\Delta_0\geq8q\sqrt K\,e_0$ for the remainder to consume at most half of it, and then leaves $1/(2q\sqrt K)$. The entry is stated at $\bar e\geq e_0$ so that one expression serves both. The second entry keeps the dressing's node displacement inside the placement budget and is consumed by Lemma~\ref{s:lem:nodes}. The third entry protects the deepest graded separation against the dressing shift and is consumed by Lemma~\ref{s:lem:distinct}; which entry \emph{binds} is determined once, in the reach paragraph below, and it is the fifth. The third entry is the price of the constructive certificate and grows exponentially with $M$ (see the resource paragraph below). Because the exponential entries exceed the first by factors growing exponentially in $M$, the precise constant in the first entry---including its factor $q\leq4K_\mathrm{phys}+2$---is immaterial at every realized operating point: strengthening it to secure the drive-side margin cannot change which entry binds.

\begin{lemma}[Spectral localization]
\label{s:lem:local}
Assume the detunings are separated, $|\delta_k-\delta_j|\geq\tfrac78\Delta_0$ for $j\neq k$, and that the dressing is weak, $\Delta_0\geq8\bar e$. Conditions D1 and D5 together imply the first and D1 the second, and at an undefected equidistant comb the first holds with separation $\Delta_0$ exactly, so the lemma applies at designed and generic operating points alike. Then the eigenvalues of $\mathbf{\Gamma}_g=i\mathbf{\Omega}+\mathbf{E}$ satisfy, for each $k$,
\begin{align}
    \big|\lambda_k-i\delta_k-E_{kk}\big|\;\leq\;\frac{8e_0^2}{\Delta_0},
\end{align}
each disk containing exactly one eigenvalue; moreover $\mathrm{Re}\,\lambda_k\geq0$ always, \emph{both} eigenvector systems satisfy $\Vert v_{R,k}-e_k\Vert\leq4e_0/\Delta_0$ and $\Vert v_{L,k}-e_k\Vert\leq4e_0/\Delta_0$, and the diagonal decay rates are $E_{kk}=\gamma/2K+(\gamma_g/2)\alpha_k^2$, with mean value exactly $\lambda_\star$.
\end{lemma}
The proof introduces seven quantities local to it: the intended disk centre $c_k$, the disk $D_k$ and its Schur function $s_k$, the dressing blocks $E_k$ and $E_{jk}$, the $k$-th column $F_k$ of $\mathbf{E}$, and the comb part $\Omega_k$. None appears outside this proof.  
\begin{proof}
Throughout, write $c_k\equiv i\delta_k+E_{kk}$ for the intended centre of the $k$-th disk, and note first that the comb frequencies are separated even after the graded defects are switched on: for $j\neq k$, $|\delta_k-\delta_j|\geq\Delta_0-2g_0\geq\tfrac{15}{16}\Delta_0$, since $g_0=\bar\varepsilon_\mathrm{target}/8T_{\max}$ (D5) while the second entry of Eq.~\eqref{s:eq:D1} forces $\Delta_0\geq64e_0^2T_{\max}/\bar\varepsilon_\mathrm{target}$, whence $\Delta_0/g_0\geq512(e_0T_{\max}/\bar\varepsilon_\mathrm{target})^2\geq2048\geq32$; the last step uses $e_0T_{\max}\geq\lambda_\star T_{\max}=\Lambda_{\max}\geq2\ln2M\geq2\ln4$ and $\bar\varepsilon_\mathrm{target}\leq1$, valid at every operating point of Step 3. We suppress the $\tfrac{15}{16}$ below and use the weaker $|\delta_k-\delta_j|\geq\tfrac78\Delta_0$, which is all the chain needs.

\emph{First-order picture.} That each eigenvalue lies near its comb frequency is, at first order, Gershgorin's theorem, and the relevant radius is the off-diagonal \emph{row sum} rather than the largest entry. By Eq.~\eqref{s:eq:e1} that row sum is at most $e_1$, uniformly in $K$; since the centres $c_k$ are separated by at least $\tfrac78\Delta_0\geq7\bar e\geq7e_1$ under D1, the $K$ Gershgorin disks of radius $e_1$ are pairwise disjoint, and by the standard homotopy argument (deform the off-diagonal part to zero; eigenvalues move continuously and cannot leave a disjoint disk) each contains exactly one eigenvalue. The $K$ disks therefore account for the whole spectrum. We stress that the $K$-independence of $e_1$ is supplied by the rank-two structure of $\mathbf{E}$ with normalized factors and not by $\Vert\mathbf{E}\Vert$ alone: bounding each entry by $e_0$ would give a row sum as large as $\sqrt{K-1}\,e_0$, which is useless here. This picture gives radius $e_1$; the design needs the sharper radius $8e_0^2/\Delta_0$, and the rest of the proof supplies it.

\emph{$\mathrm{Re}\,\lambda\geq0$.} The numerical range of $\mathbf{\Gamma}_g$ has real part $W(\mathbf{E})\subseteq[0,e_0]$ and contains the spectrum.

\emph{Sharper localization.} Fix $k$, set $\kappa\equiv8e_0^2/\Delta_0$, and let $\bar D_k=\{z:|z-c_k|\leq\kappa\}$ be the \emph{closed} disk. By the first entry of Eq.~\eqref{s:eq:D1}, $e_0\leq\bar e\leq\Delta_0/8$, hence  
\begin{align}
    \label{s:eq:kappabound}
    \kappa\;=\;\frac{8e_0^2}{\Delta_0}\;\leq\;\frac{8e_0}{\Delta_0}\cdot\frac{\Delta_0}{8}\;=\;e_0\;\leq\;\frac{\Delta_0}{8},
\end{align}
which is the estimate the chain below uses twice. (This is where the constant $8$ rather than $4$ in D1 is needed: with $\Delta_0\geq4e_0$ one obtains only $\kappa\leq\Delta_0/2$, and the separation estimate that follows fails.)

Let $\hat A_k$ denote the $(K{-}1)$-dimensional block of $\mathbf{\Gamma}_g$ with the $k$-th row and column removed, $i\hat\Omega_k$ its diagonal comb part, $\hat E_k$ its dressing part ($\Vert\hat E_k\Vert\leq e_0$), and $F_k$ the $k$-th column of $\mathbf{E}$ with its diagonal entry removed ($\Vert F_k\Vert\leq e_0$). We first establish invertibility of $\hat A_k-z$ on all of $\bar D_k$, since this is what licenses the Schur complement, and only then form it.  

For $z\in\bar D_k$ and $j\neq k$,
\begin{align}
    |i\delta_j-z|\;\geq\;|\delta_k-\delta_j|-|E_{kk}|-|z-c_k|\;\geq\;\tfrac78\Delta_0-e_0-\kappa\;\geq\;\tfrac58\Delta_0,
\end{align}
since each subtrahend is at most $\Delta_0/8$, by $|E_{kk}|\leq e_0\leq\Delta_0/8$ and Eq.~\eqref{s:eq:kappabound}. Hence $i\hat\Omega_k-z$ is invertible with $\Vert(i\hat\Omega_k-z)^{-1}\Vert\leq8/(5\Delta_0)\leq2/\Delta_0$, and
\begin{align}
    \Vert(i\hat\Omega_k-z)^{-1}\hat E_k\Vert\;\leq\;\frac{2e_0}{\Delta_0}\;\leq\;\frac14\;<\;1 .
\end{align}
Writing $\hat A_k-z=(i\hat\Omega_k-z)\big(I+(i\hat\Omega_k-z)^{-1}\hat E_k\big)$, the second factor is invertible by Neumann series for \emph{every} $z\in\bar D_k$, so $\hat A_k-z$ is invertible on the closed disk with
\begin{align}
    \Vert(\hat A_k-z)^{-1}\Vert\;\leq\;\frac{2/\Delta_0}{1-1/4}\;=\;\frac{8}{3\Delta_0}\;\leq\;\frac{4}{\Delta_0}.
\end{align}
This is the point at issue: the resolvent bound is established on the whole closed disk and not merely on its boundary, which is what both the Schur complement and the holomorphy hypothesis of Rouch\'e's theorem require.

With invertibility in hand, the Schur complement on the $k$-th coordinate says that $z\in\bar D_k$ is an eigenvalue of $\mathbf{\Gamma}_g$ if and only if
\begin{align}
    s_k(z)\;\equiv\;c_k-z+F_k^*(\hat A_k-z)^{-1}F_k\;=\;0 ,  
\end{align}
and $s_k$ is holomorphic on $\bar D_k$ because the resolvent is. The coupling correction obeys
\begin{align}
    \big|F_k^*(\hat A_k-z)^{-1}F_k\big|\;\leq\;\frac{4e_0^2}{\Delta_0}\;=\;\frac{\kappa}{2}\;<\;\kappa ,
\end{align}
so on the boundary circle $|z-c_k|=\kappa$ we have $|s_k(z)-(c_k-z)|\leq\kappa/2<\kappa=|c_k-z|$, with a factor-of-two margin. By Rouch\'e's theorem $s_k$ has exactly as many zeros in the open disk as the linear function $c_k-z$, namely one. Since the $K$ disks $\bar D_k$ are pairwise disjoint (their radii $\kappa\leq\Delta_0/8$ against centre separation $\geq\tfrac78\Delta_0$) and each carries one eigenvalue, they account for the full spectrum, which is the ``exactly one'' claim of the statement.

The right-eigenvector bound follows from $v_{R,k}\propto e_k-(\hat A_k-\lambda_k)^{-1}F_k$ with the same resolvent bound, $\lambda_k\in\bar D_k$ being covered by it. The left eigenvectors are the right eigenvectors of $\mathbf{\Gamma}_g^\dagger=-i\mathbf{\Omega}+\mathbf{E}$, since $\mathbf{E}$ is Hermitian. That matrix has the same diagonal separation $|{-\delta_k}+\delta_j|=|\delta_k-\delta_j|$, the same dressing operator norm $e_0$, and the same off-diagonal row-sum bound $e_1$---Eq.~\eqref{s:eq:e1} is invariant under conjugate transpose, $\sum_{j\neq k}|E_{jk}|=\sum_{j\neq k}|E_{kj}|$ for Hermitian $\mathbf{E}$---so every step of this proof applies to it verbatim, giving $\Vert v_{L,k}-e_k\Vert\leq4e_0/\Delta_0$. Finally $E_{kk}=\tfrac{\gamma}{2}v_{\mathrm{geo},k}^2+\tfrac{\gamma_g}{2}\alpha_k^2=\gamma/2K+(\gamma_g/2)\alpha_k^2$ by the geometry-fixed coupler of Sec.~\ref{s:sec:family}, and averaging over $k$ using $\sum_k\alpha_k^2=1$ gives $\lambda_\star$.  
\end{proof}

The node-geometry lemma is where the readout timing enters, and it consumes two conditions that we state here in displayed form. Condition \emph{D2 (readout-timing condition)}:
\begin{align}
    \tag{D2}\label{s:eq:D2}
    \Delta_0T\;=\;\frac{2\pi(nM+1)}{M},\qquad n\in\mathbb{N},
\end{align}
so that $\delta_kT\equiv2\pi k/M\pmod{2\pi}$ for the undressed comb: the node angles are exactly the $M$-th roots of unity, with the comb wrapping the circle $n$ times between readouts. Because consecutive grid values of $T$ differ by $2\pi/\Delta_0$, the grid always contains a point within one revival time above any required minimum, so this is a timing condition on the readout, satisfiable at fixed comb spacing, and it restricts no device parameter.

The fourth and last contribution to the node budget is the readout coupler's own flatness. The geometry-fixed profile $v_{\mathrm{geo},k}=1/\sqrt K$ of Sec.~\ref{s:sec:family} is flat only to relative error $\varepsilon_v$ (Sec.~\ref{s:sec:proof}), so the diagonal decay rate carries a mode-dependent part $\tfrac{\gamma}{2K}\nu_k$ with $|\nu_k|\leq\varepsilon_v$, displacing each node modulus by $\lambda_\star T\varepsilon_v/(1+\rho)$. Condition \emph{D6 (coupler flatness)} caps that displacement at the same quarter of the budget's half that D1, D4 and D5 cap the other three:
\begin{align}
    \tag{D6}\label{s:eq:D6}
    \varepsilon_v\;\leq\;\frac{\bar\varepsilon_\mathrm{target}}{8\,\Lambda_{\max}},\qquad \Lambda_{\max}\equiv\lambda_\star T_{\max} .
\end{align}
Eq.~\eqref{s:eq:D6} is Eq.~\eqref{s:eq:D4}'s first entry with $s_\alpha$ replaced by unity and $\rho$ by $\varepsilon_v$: the two channels enter the node budget through the same product of a spread with the dimensionless readout period, and are capped identically. Unlike D1--D5, D6 is not a setting. Its quantity $\varepsilon_v\approx B/\omega_0$ is fixed once the band and the transition frequency are, so Eq.~\eqref{s:eq:D6} is a feasibility test evaluated at the operating point rather than a dial turned to meet it.

\begin{lemma}[Node geometry]
\label{s:lem:nodes}
Let $M\geq2$. Under D1--D5 the nodes $x_k=e^{-\lambda_kT}$ satisfy $x_k=\bar r\,\omega^k(1+\varepsilon_k)$ with $\omega=e^{-2\pi i/M}$, $\bar r=e^{-\lambda_\star T}$ (up to one global phase absorbed into the mode labeling), and
\begin{align}
    \label{s:eq:epsbar}
    |\varepsilon_k|\;\leq\;\bar\varepsilon\;&\equiv\;2\Big(\underbrace{\rho\,s_\alpha\lambda_\star T/(1+\rho)}_{\text{drive-channel spread}}\;+\;\underbrace{\frac{8e_0^2T}{\Delta_0}}_{\text{dressing shift}}\\ \nonumber
    &\qquad\quad+\;\underbrace{g_0T}_{\text{graded defects}}\;+\;\underbrace{\lambda_\star T\,\varepsilon_v/(1+\rho)}_{\text{coupler flatness}}\Big)\;\leq\;\bar\varepsilon_\mathrm{target},
\end{align}
provided the bracket is $\leq1/2$; the final inequality holds because D4, D1, D5 and D6 cap the four contributions at $\bar\varepsilon_\mathrm{target}/8$ each. The fourth term is the readout coupler's own spread: the outcoupling profile is flat only to relative error $\varepsilon_v$ (Sec.~\ref{s:sec:proof}), so $E_{kk}$ inherits a mode-dependent part $\tfrac{\gamma}{2K}\nu_k$ with $|\nu_k|\leq\varepsilon_v$, and D6 caps its node displacement alongside the other three. The nodes are pairwise distinct for $\bar\varepsilon<\sin(\pi/M)$.
\end{lemma}
\begin{proof}
The strategy is to factor the node into the four independent contributions the design controls, bound each against the budget entry that governs it, and collect them. Write each node as a product of the four factors the design controls:
\begin{align}
    x_k\;=\;e^{-\lambda_kT}
    \;=\;\underbrace{e^{-i\Delta_0kT}}_{\text{comb angle}}\;
    \underbrace{e^{-ig_0\beta^{-k}T}}_{\text{defect phase}}\;
    \underbrace{e^{-E_{kk}T}}_{\text{diagonal decay}}\;
    \underbrace{e^{-R_kT}}_{\text{dressing remainder}},  
\end{align}
with $|R_k|\leq8e_0^2/\Delta_0$ by Lemma~\ref{s:lem:local}. Take the factors in turn. The comb angle is exact: by D2, $\Delta_0kT\equiv2\pi k/M\pmod{2\pi}$, so $e^{-i\Delta_0kT}=\omega^k$. The defect phase deviates from unity by at most $g_0\beta^{-k}T\leq g_0T$. The diagonal decay is $e^{-E_{kk}T}=\bar r\,e^{-(E_{kk}-\lambda_\star)T}$, and the exponent obeys $|E_{kk}-\lambda_\star|T\leq(\gamma_g/2)\max_k|\alpha_k^2-1/K|\,T\leq(\gamma_g/2)(s_\alpha/K)T=\rho\,s_\alpha\,\lambda_\star T/(1+\rho)$, which condition D4 caps at $\bar\varepsilon_\mathrm{target}/8$ through the readout-period ceiling $T_{\max}$. The dressing remainder deviates by at most $|R_k|T\leq8e_0^2T/\Delta_0$. Combining the three small factors with the elementary bound $|e^{w}-1|\leq2|w|$ for $|w|\leq1/2$ gives Eq.~\eqref{s:eq:epsbar}. Distinctness: adjacent ideal nodes are separated by $2\bar r\sin(\pi/M)$, which exceeds the total displacement $2\bar r\bar\varepsilon$.
\end{proof}

The tail beyond the matched depth is controlled at any operating point, and
controlled far more sharply at the designed one. Both bounds are the same
computation --- reduction modulo the node polynomial --- and differ only in
the estimate available for it, so we state them together.

\begin{lemma}[Extrapolation tail, generic and designed]
\label{s:lem:tail}\label{s:lem:tailgen}
Let $x_1,\dots,x_M\in\mathbb{C}$ be the nodes and let $c_m$ be the  
extrapolation coefficients of Proposition~\ref{s:prop:extrapolation}. As
there, the nodes are pairwise distinct; this is what makes $c_m$ well
defined, since Proposition~\ref{s:prop:extrapolation} fixes $c_m$ by a
Vandermonde solve. Write $S_\mathrm{out}$ for the total realized kernel weight beyond the matched
depth, defined in Eq.~\eqref{s:eq:Sout} below as a sum of moduli, where
$\Vert\cdot\Vert_1$ denotes the sum of moduli of the entries of a coefficient
vector; and let
\begin{align}
    \label{s:eq:rstar}
    r_\star\;\equiv\;\max_k|x_k|
\end{align}
denote the spectral radius of the node set. Then:
\begin{enumerate}
\item[(a)] \emph{(Generic operating point.)} If $r_\star<1/(M+1)$, then
\begin{align}
    \label{s:eq:Soutgen}
    S_\mathrm{out}\;\leq\;M(1+r_\star)^M\sum_{j\geq1} r_\star^{\,j}\,
        \frac{(M+j)^{M+j}}{j^{\,j}M^M}
    \;\leq\; e\,M(M+1)\,r_\star\,\big(1+O(Mr_\star)\big).
\end{align}
In particular, with $x_k=e^{-\lambda_kT}$ and $T$ the readout period,
$S_\mathrm{out}$ decays as $e^{-\min_k\mathrm{Re}\,\lambda_k\,T}$.

\item[(b)] \emph{(Designed operating point.)} If in addition the nodes take
the designed form $x_k=\bar r\,\omega^k(1+\varepsilon_k)$ with
$\omega=e^{-2\pi i/M}$ the primitive $M$-th root of unity, $\bar r$ the
common radius, and $|\varepsilon_k|\leq\bar\varepsilon\leq1/4$, and if
$\bar r\leq1/(2M)$ and $M\geq2$, then
\begin{align}
    \label{s:eq:Sout}
    S_\mathrm{out}\;=\;\sum_{m>M}\Vert c_m\Vert_1\;\leq\;2M\bar r^{\,M}\;+\;e\,M\,\bar\varepsilon .
\end{align}
\end{enumerate}
The hypotheses nest: those of (b) imply that of (a), since
$r_\star\leq\bar r(1+\bar\varepsilon)\leq\tfrac{5}{8M}<1/(M+1)$ for
$M\geq2$, the last step being $5(M+1)<8M\iff M>5/3$. The conclusions do not
nest, and (b) is not a corollary of (a): Eq.~\eqref{s:eq:Sout} decays
exponentially in $M$ where Eq.~\eqref{s:eq:Soutgen} decays only linearly in
$r_\star$, and the two are reached by different arguments.
\end{lemma}

The proof introduces seven quantities local to it: the reduction operator $\mathcal{R}$ and the smear operator $\mathcal{E}$, the polynomial coefficients $a_i$ and $b_i$, the standard basis vectors $e_i$ and $e_j$, and the reduced polynomial $p_m$. None appears outside this proof.  
\begin{proof}
Both parts compute the same object. By
Proposition~\ref{s:prop:extrapolation}, $c_m$ is the coefficient vector of
$x^{m-1}$ reduced modulo the node polynomial $\Pi(x)=\prod_k(x-x_k)$, so  
$S_\mathrm{out}$ asks how fast repeated reduction shrinks a monomial. Part
(a) estimates this by a contour integral, which needs nothing of the node
configuration beyond its radius. Part (b) exploits the fact that at the
designed point $\Pi$ is a perturbed binomial, for which reduction is a single
explicit instruction. Throughout, $\Vert\cdot\Vert_1$ on a polynomial means
the sum of moduli of its coefficients.

\emph{(a) Generic.} Write $P(x)\equiv\Pi(x)=\sum_{i=0}^Ma_ix^i$ and let
$p_m(x)=x^{m-1}\bmod P(x)$, so that $c_m$ is the coefficient vector of $p_m$;
equivalently $p_m$ interpolates the entire function $z\mapsto z^{m-1}$ at the
nodes. No inverse of the Vandermonde matrix is formed. The idea is that
interpolation at nodes confined to a small disk cannot produce large
coefficients, and a contour integral is the cleanest way to say so. By the
Hermite formula, for any circle $\Gamma=\{|z|=R\}$ enclosing every node,  
\begin{align}
    p_m(x)=\frac{1}{2\pi i}\oint_\Gamma z^{m-1}\,
        \frac{P(z)-P(x)}{(z-x)P(z)}\dd z .
\end{align}
Expanding $\frac{P(z)-P(x)}{z-x}=\sum_{\mu=0}^{M-1}x^\mu\sum_{i>\mu}a_iz^{i-1-\mu}$
and taking $R<1$, so that $|z|^{i-1-\mu}\leq1$ for every $i>\mu$, its
coefficient $\ell^1$ norm in $x$ is at most $M\sum_i|a_i|\leq M(1+r_\star)^M$,
using $|a_i|\leq\binom{M}{i}r_\star^{\,M-i}$. On $\Gamma$ the denominator is
bounded below, $|P(z)|\geq\prod_k(R-|x_k|)\geq(R-r_\star)^M$. Hence for every
$R\in(r_\star,1)$,
\begin{align}
    \Vert c_m\Vert_1\;\leq\;M(1+r_\star)^M\,\frac{R^{\,m}}{(R-r_\star)^M}.
\end{align}
The contour radius is a free parameter here, its freedom being exactly the
statement that the estimate holds for every $R\in(r_\star,1)$, so we may
optimise it per term rather than once. Minimising $R^{M+j}/(R-r_\star)^M$
over $R$ gives $R_\star=(M+j)r_\star/j$, admissible ($R_\star<1$) for every
$j\geq1$ when $(M+1)r_\star<1$, and substituting yields the summand of
Eq.~\eqref{s:eq:Soutgen}. The $j=1$ term is
$M(1+r_\star)^Mr_\star(M+1)^{M+1}/M^M\leq e\,M(M+1)r_\star(1+r_\star)^M$, and
the remaining terms are smaller by factors $O(Mr_\star)$.

\emph{(b) Designed.} Everything follows from what the node polynomial is at
the designed point. The $M$ numbers $\bar r\omega^k$ are precisely the roots
of $x^M-\bar r^{\,M}$, which is monic of degree $M$, so at
$\bar\varepsilon=0$ the node polynomial \emph{is} that binomial and reduction
is a single instruction: replace $x^M$ by $\bar r^{\,M}$. Away from
$\bar\varepsilon=0$ write
\begin{align}
    \label{s:eq:delta}
    \Pi(x)\;=\;x^M-\bar r^{\,M}+\delta(x),\qquad \deg\delta\leq M-1,  
\end{align}
so that $\delta$ measures, and is the only thing that measures, how far the
design pushes the node polynomial off that binomial.

\emph{Step 1: the size of $\delta$.} The coefficients of $\delta$ are the
elementary symmetric functions $e_j(\{x_k\})$, and these vanish
\emph{identically} at $\bar\varepsilon=0$ for $1\leq j\leq M-1$: that
vanishing is the whole content of ``roots of unity at a common radius''.
Using $|\prod_{k\in S}(1+\varepsilon_k)-1|\leq(1+\bar\varepsilon)^{j}-1\leq
j\bar\varepsilon(1+\bar\varepsilon)^{j-1}$ on each of the $\binom Mj$ subsets
of size $j$, and $\binom Mj j=M\binom{M-1}{j-1}$,
\begin{align}
    \label{s:eq:deltanorm}
    D\;\equiv\;\Vert\delta\Vert_1\;\leq\;M\bar\varepsilon\,\bar r\,
        \big(1+\bar r(1+\bar\varepsilon)\big)^{M-1}\;\leq\;
        \tfrac e2\,\bar\varepsilon ,
\end{align}
the last step using $\bar r(1+\bar\varepsilon)\leq1/M$ and $M\bar r\leq1/2$,
so that $(1+1/M)^{M-1}<e$. The sum is dominated by its first term
$M\bar r\bar\varepsilon$: the radius discounts the $j$-th symmetric function
by $\bar r^{\,j}$, and $M\bar r\leq1/2$ by the readout period.

\emph{Step 2: the reduction operator does not amplify.} Let $\mathcal{R}$
denote the operator ``multiply by $x$, then reduce modulo $\Pi$'', written
with a script letter to keep it distinct from the readout period $T$. In the
monomial basis $\mathcal{R}$ is the companion matrix of $\Pi$: it sends
$e_i\mapsto e_{i+1}$ for $i<M-1$, columns of $\ell^1$ norm exactly one, and
sends $e_{M-1}$ to the coefficient vector of $x^M\bmod\Pi=\bar r^{\,M}-\delta(x)$,
of $\ell^1$ norm at most $\bar r^{\,M}+D$. Hence
$\Vert\mathcal{R}\Vert_{1\to1}=\max\{1,\bar r^{\,M}+D\}=1$, the maximum being
over columns. Multiplying by $x$ only relabels coefficients except at the top,
where it costs one reduction, and one reduction costs less than it saves.

\emph{Step 3: $M$ reductions are a scalar plus a controlled remainder.} For
$\deg p\leq M-1$ the product $\bar r^{\,M}p$ is already reduced, so
\begin{align}
    \mathcal{R}^Mp\;=\;(x^Mp)\bmod\Pi\;=\;\big((\bar r^{\,M}-\delta)p\big)\bmod\Pi
    \;=\;\bar r^{\,M}p-(\delta p\bmod\Pi),
\end{align}
that is, $\mathcal{R}^M=\bar r^{\,M}I+\mathcal{E}$ with
$\mathcal{E}p=-(\delta p\bmod\Pi)$, the script $\mathcal{E}$ again chosen to
avoid the comb spacing's $\Delta$. Writing $\delta p=\sum_ib_ix^i$ gives  
$\Vert b\Vert_1\leq D\Vert p\Vert_1$ by submultiplicativity of $\ell^1$ under
convolution, and $(\delta p)\bmod\Pi=\sum_ib_i\mathcal{R}^i(e_0)$ with
$\Vert\mathcal{R}^i(e_0)\Vert_1\leq1$ by Step 2, so
$\Vert\mathcal{E}\Vert_{1\to1}\leq D$. At the ideal point $M$ reductions
return the coefficient vector unchanged up to the factor $\bar r^{\,M}$; off
it, each full turn also smears the answer across residue classes by $\delta$,
and the whole error analysis is the size of that smear.

\emph{Step 4: summation.} Write $m-1=aM+s$ with $0\leq s<M$. Then
$c_m=\mathcal{R}^{m-1}e_0=(\mathcal{R}^M)^a\mathcal{R}^se_0$, so
$\Vert c_m\Vert_1\leq\Vert\mathcal{R}^M\Vert^a\Vert\mathcal{R}^se_0\Vert_1
\leq(\bar r^{\,M}+D)^a$. Every $m>M$ has $a\geq1$, and each $a$ admits exactly
$M$ values of $s$, whence
\begin{align}
    S_\mathrm{out}\;\leq\;M\sum_{a\geq1}(\bar r^{\,M}+D)^a
    \;=\;\frac{M(\bar r^{\,M}+D)}{1-(\bar r^{\,M}+D)}
    \;\leq\;2M\big(\bar r^{\,M}+D\big)
\end{align}
whenever $\bar r^{\,M}+D\leq1/2$, which $\bar r^{\,M}\leq(2M)^{-M}\leq1/16$
and $D\leq\tfrac e2\bar\varepsilon\leq7/16$ guarantee at
$\bar\varepsilon\leq1/4$. Substituting Eq.~\eqref{s:eq:deltanorm} gives
Eq.~\eqref{s:eq:Sout}. Both sums are sums of moduli throughout, over $\mu$
inside $\Vert c_m\Vert_1$ and over $m$ outside, as Eq.~\eqref{s:eq:Sout}
requires; no $\ell^2$ quantity is used, and none is available to be traded
for one.

\emph{Global phase.} If the ideal nodes carry a common phase $e^{i\theta}$,
the binomial becomes $x^M-(\bar re^{i\theta})^M$ and every step is unchanged,
$|\bar re^{i\theta}|=\bar r$. This proves both parts of the lemma: the extrapolation tail is bounded by the node radius alone in the generic case, and by the designed geometry in the second, in each case without any hypothesis on the weights.
\end{proof}

What the design buys is now a comparison inside one statement. At the
designed point the tail decays as $\bar r^{\,M}=e^{-M\lambda_\star T}=
e^{-(\gamma+\gamma_g)T/2}$, the full trace budget of
Lemma~\ref{s:lem:trace} spent over the memory span; at a generic point it
decays as $e^{-\min_k\mathrm{Re}\,\lambda_k\,T}$, the slowest mode's share
alone. The design equalises the decay rates, converting the \emph{minimum}
into the \emph{mean}. The ratio of the two exponents is bounded on both sides
by results that use no design condition: $\min_k\mathrm{Re}\,\lambda_k\leq
\lambda_\star$ by Lemma~\ref{s:lem:trace}, and $\min_k\mathrm{Re}\,\lambda_k
\geq\gamma/18K$ by Lemma~\ref{s:lem:strictdiss}, whence
\begin{align}
    \label{s:eq:penalty}
    M\;\leq\;\frac{M\lambda_\star}{\min_k\mathrm{Re}\,\lambda_k}\;\leq\;
    9M(1+\rho_{\gamma}),\qquad \rho_\gamma=\gamma_g/\gamma .
\end{align}
The certificate's comb spacing therefore purchases a readout period shorter by
a factor in this band, and nothing else: universality itself, with an explicit
rate, is available at any admissible operating point through part~(a).

\begin{lemma}[Design distinctness for the order-$n$ selection]
\label{s:lem:distinct}
Under D1, D4, and D5, consider the conjugation-resolved eigenvalue-sum points $\Lambda^{(S)}(\vec k)=\sum_{i\in S}\lambda_{k_i}+\sum_{i\notin S}\bar\lambda_{k_i}$ over all orders $n\leq N$, mode multisets $\vec k$, and conjugation patterns $S$. Classify each point by its signed multiplicity vector $\mathbf{n}\in\mathbb{Z}^K$ (entry $n_k$ counts appearances of mode $k$ in $S$ minus appearances outside $S$; $|n_k|\leq N$) and its total multiplicity vector $\mathbf{m}\in\mathbb{N}^K$. Then:
\begin{enumerate}
\item[(a)] Points with distinct signed vectors $\mathbf{n}\neq\mathbf{n}'$ have imaginary parts separated by at least $g_0\beta^{-(M-1)}/4$.
\item[(b)] Points with equal signed vectors but distinct total-multiplicity vectors $\mathbf{m}\neq\mathbf{m}'$ are separated in their real parts: by at least $\gamma/4K$ if the total orders differ, and otherwise by the designed drive-channel margin $\varepsilon_\mathrm{rad}\equiv\min|\tfrac{\gamma_g}{2}\sum_k(m_k-m'_k)\alpha_k^2|\geq\tfrac{\gamma_g}{16K}(4N)^{-(q-2)}>0$, with $q$ the prime of condition D3.
\item[(c)] The only remaining coincidences are the exact complement pairs $S\leftrightarrow S^c$ at equal multisets, which satisfy $\Lambda^{(S)}=\overline{\Lambda^{(S^c)}}$ identically; the reality constraint $F(s)^*=F(s^*)$ of Sec.~S.2.3 prescribes these jointly.
\end{enumerate}
Consequently the constructive weights can prescribe every conjugation-resolved moment $H^{(S)}_n(\vec\mu)$ to a chosen real value, and choosing these values as the symmetrized target combinations yields $|H^{(S)}_n(\vec\mu)|\leq2^n\max|h^{(M,N)}_n|$.
\end{lemma}
\begin{proof}
The strategy is to classify every eigenvalue-sum point by its multiplicity vector, so that two points coincide only when their vectors do, and then to show the design separates the finitely many vectors that survive. Every eigenvalue decomposes, by Lemma~\ref{s:lem:local}, as $\lambda_k=i\delta_k+E_{kk}+r_k$ with $|r_k|\leq8e_0^2/\Delta_0$. A sum of $n\leq N$ eigenvalues or conjugates therefore has
\begin{align}
    \mathrm{Im}\,\Lambda^{(S)}=\sum_k n_k\,\delta_k+\mathrm{Im}\,R,
    \qquad
    \mathrm{Re}\,\Lambda^{(S)}=\sum_k m_k\,E_{kk}+\mathrm{Re}\,R,
    \qquad |R|\leq N\,\frac{8e_0^2}{\Delta_0}.
\end{align}
(a) For two points with $\mathbf{n}\neq\mathbf{n}'$, write the digit-difference vector $\mathbf{d}=\mathbf{n}-\mathbf{n}'$, $|d_k|\leq2N$, $\mathbf{d}\neq0$. The unperturbed separation is
\begin{align}
    \Big|\sum_k d_k\delta_k\Big|=\Big|\Delta_0\sum_k d_k\,k\;+\;g_0\sum_k d_k\beta^{-k}\Big|.
\end{align}
If the comb part is nonzero, $|\Delta_0\sum d_kk|\geq\Delta_0$, which dwarfs everything else. If the comb part vanishes, let $j$ be the smallest index with $d_j\neq0$; then  
\begin{align}
    \Big|g_0\sum_{k\geq j}d_k\beta^{-k}\Big|\;\geq\;g_0\beta^{-j}\Big(1-\frac{2N}{\beta-1}\Big)\;=\;g_0\beta^{-j}\Big(1-\frac{2N}{4N}\Big)\;=\;\frac{g_0\beta^{-j}}{2}\;\geq\;\frac{g_0\beta^{-(M-1)}}{2},
\end{align}
where the parenthesis bounds the worst-case carry from all deeper digits by a geometric series: this is the uniqueness of signed-digit representations in base $\beta=4N+1$ with digits bounded by $2N$. The dressing remainders shift each of the two points by at most $N\cdot8e_0^2/\Delta_0$, and the third entry of Eq.~\eqref{s:eq:D1} guarantees $2N\cdot8e_0^2/\Delta_0\leq g_0\beta^{-(M-1)}/4$, so half the unperturbed separation survives.

(b) With $\mathbf{n}=\mathbf{n}'$ the imaginary parts agree to within the remainders, and the real parts differ by $\sum_k(m_k-m'_k)E_{kk}=(\gamma/2K)(n-n')+(\gamma_g/2)\sum_k(m_k-m'_k)\alpha_k^2$. If the total orders differ, the first term contributes at least $\gamma/2K$ while the second is at most $2N\gamma_g s_\alpha/2K=\rho Ns_\alpha\,\gamma/K$, which D4 makes smaller than $\gamma/4K$; the remainders are smaller still by Eq.~\eqref{s:eq:D1}. If the total orders agree, the separation is the pure drive-channel expression $\tfrac{\gamma_g}{2}\sum_k(m_k-m'_k)\alpha_k^2$ with $\sum_k(m_k-m'_k)=0$. Writing $d=\mathbf{m}-\mathbf{m}'$, we have $d\neq0$, $\sum_kd_k=0$ and $\sum_k|d_k|\leq2N$, so Lemma~\ref{s:lem:radial} bounds the expression below by $\varepsilon_\mathrm{rad}=\tfrac{\gamma_g}{16K}(4N)^{-(q-2)}$. The dressing remainders shift the two points by at most $2N\cdot8e_0^2/\Delta_0$ in total, which the fifth entry of Eq.~\eqref{s:eq:D1} makes at most $\varepsilon_\mathrm{rad}/2$; half the separation survives.

(c) Complement pairs at equal multisets are exact conjugates by inspection, and the joint prescription in the $F^{\pm}$ basis of Sec.~S.2.3 is compatible with the reality constraint; distinct-or-conjugate is exactly what that construction requires. This proves the lemma: at the designed operating point the eigenvalue-sum points of every order $n\leq N$ are distinct or exact conjugates, which is precisely the input the order-$n$ selection of Sec.~S.2.3 requires.
\end{proof}

The two separations are both design outputs: the angular floor $g_0\beta^{-(M-1)}/4$ from the graded defects, and the radial floor $\varepsilon_\mathrm{rad}$ from the delay lock. Neither is a property measured at an operating point.

\begin{table}[h]
\centering
\small
\setlength{\tabcolsep}{4pt}
\renewcommand{\arraystretch}{1.15}
\begin{tabular}{lp{7.6cm}p{3.3cm}}
\hline
Cond. & Statement and role & Consumed by\\
\hline
D1 & Comb spacing $\Delta_0$ bounded below by the five-entry maximum of Eq.~\eqref{s:eq:D1}: keeps the dressing perturbative; caps its node displacement; protects the deepest graded separation; locks the delay; protects the radial separation (the last is the entry that binds) & Lemmas~\ref{s:lem:local}, \ref{s:lem:nodes}, \ref{s:lem:distinct}\\
D2 & Readout period $T$ on the grid $\Delta_0T=2\pi(nM{+}1)/M$: undressed node angles exactly the $M$-th roots of unity; a measurement setting only & Lemma~\ref{s:lem:nodes}\\
D3 & Atom at a rational point $\ell_a/\ell=p/q$ of the standing wave, $q$ the smallest prime $\geq2K_\mathrm{phys}+1$, with band residue $n_0\bmod q$ admissible: fixes the emission profile, bounds $s_\alpha$, and floors the radial separation & Lemmas~\ref{s:lem:profile}, \ref{s:lem:radial}, \ref{s:lem:distinct}\\
D4 & Rate ratio $\rho$ bounded by Eq.~\eqref{s:eq:D4}: the drive-channel spread separates radially without misplacing the nodes & Lemmas~\ref{s:lem:nodes}, \ref{s:lem:distinct}\\
D5 & Graded defects of Eq.~\eqref{s:eq:D5}, base $\beta=4N+1$: supplies $\mathbb{Q}$-independence of the detunings, which no comb polynomial in $k$ provides beyond order $\deg+1$, breaking every angular coincidence with margin $g_0\beta^{-(M-1)}/4$. Unlike D1--D4 this is a property of the fabricated resonator spectrum rather than a runtime setting & Lemmas~\ref{s:lem:nodes}, \ref{s:lem:distinct}\\
D6 & Coupler flatness $\varepsilon_v\leq\bar\varepsilon_\mathrm{target}/8\Lambda_{\max}$, Eq.~\eqref{s:eq:D6}: caps the readout coupler's own node displacement. Physical data of the fabricated coupler rather than a setting; verified at the operating point & Lemma~\ref{s:lem:nodes}\\
\hline
\multicolumn{3}{p{12.1cm}}{\footnotesize The output coupler's \emph{profile} is geometry-fixed physical data (Sec.~\ref{s:sec:family}) and no condition sets it. Its \emph{flatness} $\varepsilon_v$ is likewise physical data, but it is not unconstrained: D6 is the feasibility test that data must pass rather than a dial. D1--D5 are settings; D6 is a check.}\\
\hline
\end{tabular}
\caption{The designed operating point, assembled. Each condition is stated in Step 2 at its point of use; the table exists so the complete set can be audited at a glance.}
\label{s:tab:design}
\end{table}

\subsubsection*{Step 3: parameter selection, and the theorem's quantifiers}
The proof closes by explicit parameter selection. Given a target with a convergent Volterra expansion and an accuracy $\epsilon>0$:
\begin{enumerate}
\item Choose $(M,N)$ so that the Boyd--Chua truncation error obeys $\delta_{MN}\leq\epsilon/2$ (term (i)); this fixes $A=A(M,N)$ from the target's own kernels. Without loss of generality take $M\geq2$: enlarging $M$ only decreases $\delta_{MN}$.
\item Set $K=M$, let $q$ be the smallest prime $\geq2K_\mathrm{phys}+1$, and fix the delay and band placement by condition D3. Set the node-placement budget
\begin{align}
    \label{s:eq:epstarget}
    \bar\varepsilon_\mathrm{target}=\min\Big\{\frac{\epsilon}{4e\,A\,M},\;\frac14\Big\}
\end{align}
and the drive ratio $\rho$ by D4, whose right-hand side depends only on $\bar\varepsilon_\mathrm{target}$, $N$, $M$ and the bound $s_\alpha\leq q^2$ of Lemma~\ref{s:lem:profile}. Then set $\lambda_\star=(\gamma+\gamma_g)/2K$.
\item Choose the readout period $T$ on the D2 grid so that
\begin{align}
    \label{s:eq:rbar}
    \bar r=e^{-\lambda_\star T}\leq\min\Big\{\frac{1}{2M},\;\Big(\frac{\epsilon}{8MA}\Big)^{1/M}\Big\},
    \qquad \bar r^{\,M}=e^{-(\gamma+\gamma_g)T/2},
\end{align}
i.e.
\begin{align}
T\;&\geq\;\lambda_\star^{-1}\max\Big\{\ln 2M,\;\tfrac{1}{M}\ln\tfrac{8MA}{\epsilon}\Big\}
\;=\;\frac{2}{\gamma+\gamma_g}\max\Big\{M\ln 2M,\;\ln\tfrac{8MA}{\epsilon}\Big\}\nonumber\\
&=\;O\!\Big(\frac{M\ln M+\ln\epsilon^{-1}}{\gamma+\gamma_g}\Big),
\label{s:eq:Tselect}
\end{align}
which the D2 grid permits at any comb spacing. The middle equality substitutes $\lambda_\star=(\gamma+\gamma_g)/2M$ at $K=M$, and it is worth pausing on because the factor is easy to count twice: the $1/\lambda_\star$ is already carried inside the displayed numerator, so the clock grows as $\Theta(M\ln M)$ in the memory depth and not as $\Theta(M^2\ln M)$. The rate itself is the check. At the binding branch $(\gamma+\gamma_g)T/2=M\ln 2M$, so $\bar r^{\,M}=e^{-(\gamma+\gamma_g)T/2}=(2M)^{-M}$, which is exactly the bound Step~1 uses at Eq.~\eqref{s:eq:c0closed}; the discarded reading $T=O(M\ln M)/\lambda_\star$ would give $(\gamma+\gamma_g)T/2=M^2\ln 2M$ and hence $\bar r^{\,M}=(2M)^{-M^2}$, contradicting it. The trace budget rather than the per-mode rate, sets the clock.
\item Choose the defect amplitude $g_0$ by D5 and then the comb spacing $\Delta_0$ by Eq.~\eqref{s:eq:D1}.
\end{enumerate}
Then Lemma~\ref{s:lem:nodes} gives $\bar\varepsilon\leq\bar\varepsilon_\mathrm{target}$, whose two entries discharge the two standing hypotheses: $\bar\varepsilon\leq1/4$ is the condition of Lemma~\ref{s:lem:tail}(b), and with $\bar r\leq1/(2M)$ at $M\geq2$ it gives $S_\mathrm{out}\leq2M\bar r^{\,M}+eM\bar\varepsilon\leq M$, the counting hypothesis of Eq.~\eqref{s:eq:termii}; while $\bar\varepsilon\leq\epsilon/(4eAM)$ gives $S_\mathrm{out}\leq\epsilon/(2A)$, so term (ii) $\leq\epsilon/2$ and the total error is at most $\epsilon$. Every constant is explicit and no limit other than the finite construction is taken.

Write $T_{\min}=\Lambda_{\min}/\lambda_\star$ for the floor of item~3, with $\Lambda_{\min}=\max\{\ln2M,\tfrac1M\ln(8MA/\epsilon)\}$ dimensionless, and set $T_{\max}\equiv2T_{\min}$, $\Lambda_{\max}\equiv\lambda_\star T_{\max}$. Conditions D1's second and third entries, D4, D5 and D6 are asserted at $T_{\max}$; the remaining entries are $T$-free once $g_0$ is fixed, so every condition asserted at the ceiling holds at every admissible $T\leq T_{\max}$. That the grid reaches into the window is then a one-line consequence rather than a hypothesis: the D2 spacing is $2\pi/\Delta_0$, and the second entry of Eq.~\eqref{s:eq:D1} at $T_{\max}$ gives $\Delta_0T_{\min}\geq128(e_0T_{\min})^2/\bar\varepsilon_\mathrm{target}\geq128>2\pi$, using $e_0T_{\min}\geq\lambda_\star T_{\min}=\Lambda_{\min}\geq\ln4>1$ and $\bar\varepsilon_\mathrm{target}\leq1$, so a grid point lies in $[T_{\min},T_{\max}]$.

The selection order is acyclic: $(M,N)\to q\to\bar\varepsilon_\mathrm{target}\to\Lambda_{\min}\to\Lambda_{\max}\to[\text{check D6 against }\varepsilon_v]\to\rho\to\lambda_\star\to T_{\min}\to T_{\max}\to g_0\to\Delta_0\to\text{grid}\to T$. The bracketed step is the only one that is not a selection: Eq.~\eqref{s:eq:D6} is evaluated against fabricated data and either passes or fails, and if it fails the accuracy is unreachable on that hardware rather than at a different setting.
$\qquad\blacksquare$

\subsubsection*{Scaling of the design parameters}
\begin{table}[h]
\centering
\small
\setlength{\tabcolsep}{4pt}
\renewcommand{\arraystretch}{1.15}
\begin{tabular}{@{}p{2.6cm}p{2.7cm}p{4.5cm}p{2.4cm}@{}}
\hline
Quantity & Formula & Growing $M$ & Shrinking $\epsilon$\\
\hline
Readout period $T$ & Eq.~\eqref{s:eq:Tselect} & $\Theta(M\ln M)$ & $+\,\Theta(\ln\epsilon^{-1})$\\
Node budget $\bar\varepsilon_\mathrm{target}$ & Eq.~\eqref{s:eq:epstarget} & $\Theta\big(1/(AM)\big)$ & $\Theta(\epsilon)$ (small $\epsilon$)\\
Drive ratio $\rho$ & Eq.~\eqref{s:eq:D4}, first entry & $\Theta\big(1/(AM^{3}\ln M)\big)$ & $\Theta(\epsilon/\ln\epsilon^{-1})$\\
Defect amplitude $g_0$ & Eq.~\eqref{s:eq:D5} & $\Theta\big(1/(AM^{2}\ln M)\big)$ & $\Theta(\epsilon/\ln\epsilon^{-1})$\\
Comb spacing $\Delta_0$ & Eq.~\eqref{s:eq:D1}, fifth entry & $\Theta\big(A\,M^{4}\ln M\,(4N)^{q-2}\big)$ & $\Theta(\epsilon^{-1}\ln\epsilon^{-1})$\\
Node radius $\bar r^{\,M}$ & Eq.~\eqref{s:eq:rbar} & exponentially small & $\Theta(\epsilon)$\\
\hline
\end{tabular}
\caption{Scaling ledger of the designed operating point. The two regimes are growing $M$ at fixed $\epsilon$ and shrinking $\epsilon$ at fixed $(M,N)$. $A=A(M,N)$ is the target-dependent prefactor of Eq.~\eqref{s:eq:termii}; $\beta=4N+1$; $q$ is the least prime $\geq2M{+}1$, and the drive-ratio row uses $s_\alpha\leq q^2$. Every formula column cites the equation it is read off, and no entry is asymptotic guesswork. In the readout-period row, $\lambda_\star\propto1/M$ at the designed point $K=M$ is already absorbed in Eq.~\eqref{s:eq:Tselect}, whence $\Theta(M\ln M)$; the defect-amplitude row inherits that exponent through $g_0=\bar\varepsilon_\mathrm{target}/8T_{\max}$. The exponential factor $\beta^M$ in the comb spacing is the defect-protection entry of Eq.~\eqref{s:eq:D1}; it is a property of the designed certificate, and the paragraph on designed versus generic operating points below explains what replaces it on a fabricated device. At $K_\mathrm{phys}>M$ in-band modes the exponent reads $K_\mathrm{phys}-1$ (Remark~\ref{s:rem:embed}).}
\label{s:tab:ledger}
\end{table}

\paragraph{Reading the ledger.} Each growth column is a product of factors already fixed above, and we write the chain out so that no exponent lives only inside a table cell. Take the accuracy fixed and the depth growing, on the branch $\bar\varepsilon_\mathrm{target}=\epsilon/4eAM$ of Eq.~\eqref{s:eq:epstarget}, which binds once $A>\epsilon/(eM)$; the Volterra prefactor carries $M^{n-1}$ at order $n$, so this is every target of order $N\geq2$. The budget is then $\Theta(1/(AM))$. The clock follows from Eq.~\eqref{s:eq:Tselect} as $\Theta(M\ln M)$, so the defect amplitude $g_0=\bar\varepsilon_\mathrm{target}/8T_{\max}$ is $\Theta\big(1/(AM^{2}\ln M)\big)$. The drive ratio divides the budget by the profile's peak-to-mean ratio and by the dimensionless clock, $\rho=\bar\varepsilon_\mathrm{target}/8s_\alpha\Lambda_{\max}$ with $s_\alpha\leq q^2=\Theta(M^2)$ under the delay lock D3 and $\Lambda_{\max}=\Theta(\ln M)$, giving $\Theta\big(1/(AM^{3}\ln M)\big)$.

The comb spacing is then forced, and it is worth seeing why it cannot be read independently of the row above it. Its binding entry is $\Delta_0\geq512NKe_0^2(4N)^{q-2}/\gamma_g$ with $\gamma_g=\rho\gamma$, so $\Delta_0\propto K/\rho$ at fixed $N$ and fixed $e_0=\Theta(\gamma)$: the spacing is the mode count divided by the drive ratio, and its $M$-exponent is therefore exactly one more than the exponent of $1/\rho$. With $1/\rho=\Theta(AM^{3}\ln M)$ this is $\Theta\big(AM^{4}\ln M\,(4N)^{q-2}\big)$, the entry tabulated. The physical reading is that the design pays twice for the same radial margin: the drive channel must be weak enough not to misplace the nodes (D4), which shrinks the very separation D5 and D1 must then protect, and the comb must be opened far enough to protect it. Every polynomial factor here is dominated by $(4N)^{q-2}$ with $q=\Theta(M)$, so the exponential entry is what a fabricated device actually feels; the polynomial prefactor matters only as an audit trail.

With the depth fixed and the accuracy tightening, only two factors move: the budget is $\Theta(\epsilon)$ and the clock acquires $+\,\Theta(\ln\epsilon^{-1})$. Both $g_0$ and $\rho$ divide a budget of order $\epsilon$ by a clock of order $\ln\epsilon^{-1}$, so both are $\Theta(\epsilon/\ln\epsilon^{-1})$; the comb spacing, proportional to $1/\rho$, is $\Theta(\epsilon^{-1}\ln\epsilon^{-1})$; and $\bar r^{\,M}=\Theta(\epsilon)$ by Eq.~\eqref{s:eq:rbar}. Buying a decimal place costs a factor of ten in the comb spacing and a logarithm in the clock, and nothing in the mode count.

\section{Reach, Scaling, and the Operating Class}
\label{s:sec:reach-section}

This section collects the results that follow from the construction of Sec.~\ref{s:sec:proof} without being part of the proof of Theorem~1: what a fixed geometry can reach, how the design parameters scale, the weaker operating class on which the guarantee still holds, the physical resources the certificate spends, the scaling corollary, and the density step.

\subsection*{The accuracy envelope of a fixed geometry}
The selection of Step 3 can be inverted, which answers the operational question: with the hardware fabricated and the geometry already set, which accuracies are reachable by measurement settings alone?

\begin{corollary}[Fixed-geometry envelope]
\label{s:cor:envelope}  
Fix the fabricated hardware and the geometric settings: a comb of $K_\mathrm{phys}$ in-band modes at spacing $\Delta_0$ with defect amplitude $g_0$ and drive ratio $\rho$ (satisfying D3--D5 at level $\bar\varepsilon_0\equiv6\max\{\rho s_\alpha\ln 2K_\mathrm{phys},\,g_0T\}$ for the periods used). An accuracy $\epsilon$ is then achievable by measurement settings alone (a choice of $M\leq K_\mathrm{phys}$, $N$, $T$ on the D2 grid, local-oscillator spectrum, and window functions) whenever there exist $(M,N)$ with
\begin{align}
    \delta_{MN}\;\leq\;\frac{\epsilon}{2},
    \qquad
    \bar\varepsilon\big(\Delta_0,\rho,g_0,\varepsilon_v;M,T\big)\;\leq\;\min\Big\{\frac{\epsilon}{4eA(M,N)M},\,\frac14\Big\},
    \nonumber\\
    2M\bar r^{\,M}\;\leq\;\frac{\epsilon}{8A(M,N)},
\end{align}
with $\bar\varepsilon$ the explicit four-term expression of Eq.~\eqref{s:eq:epsbar} evaluated at the fixed geometry, and with the distinctness margins of Lemma~\ref{s:lem:distinct} (the graded floor $g_0\beta^{-(M-1)}/4$ and the radial floor $\varepsilon_\mathrm{rad}$ of Eq.~\eqref{s:eq:radfloor}, the prime $q$ of condition D3 belonging to the fixed geometry) exceeding the dressing remainders at that geometry. The infimum of such $\epsilon$ is the accuracy envelope $\epsilon_{\min}$ of the fixed geometry. Below the envelope, two resources must move together: the geometric dial (larger quantization length $\ell$, refining $\Delta_0$ and raising $K_\mathrm{phys}$) and the retained bandwidth. They are coupled: refining $\Delta_0$ at fixed $B=K\Delta_0$ lowers the spacing, while the first entry of Eq.~\eqref{s:eq:D1} demands $\Delta_0\geq8q\sqrt K\,\bar e$ with $q\approx2K$, so that $B\gtrsim16K^{5/2}\bar e$: the bandwidth must grow at least as $K^{5/2}$ for the certificate to survive rising mode number, on top of the exponential growth of the fifth entry. Subject to that, the strict-scaling corollary of the main text guarantees the move is never wasted.  
\end{corollary}
\begin{proof}
Immediate from Step 3 read backwards: at fixed geometry the only selected quantities are $(M,N,T)$ and the record-side freedoms, $T$ ranges over the D2 grid whose spacing $2\pi/\Delta_0$ is available at fixed $\Delta_0$, and the three displayed inequalities are precisely the conditions under which steps 1, 3, and 4 close with the geometric quantities held fixed. The third inequality is satisfiable at fixed $\lambda_\star$ by taking $T$ large on the grid, so the envelope is set by the second inequality and by $M\leq K_\mathrm{phys}$.
\end{proof}

Two consequences of the corollary deserve emphasis. First, within the envelope, the retained modes are selected by the readout: the record-side machinery of Remark~\ref{s:rem:KgtM} confines the realized kernels to the $M$ modes the measurement addresses, so no physical mode-selection mechanism is needed, and the question of how $K$ modes are singled out from the one device has a one-word answer: measurement. Second, the envelope is a statement about a setting of the device, never about the device: every accuracy is reachable on the same fabricated hardware, and the envelope marks where the purchase switches from measurement settings to the geometric dial.

\subsection*{The operating class, and the theorem stated on it}
\label{s:sec:opclass}
The conditions the construction of this section supplies are sufficient, and the proof consumes less
than all of them. We therefore isolate what is consumed, so that the guarantee can be stated on
operating points a fabricated device can occupy.

\begin{definition}[Operating class]
\label{s:def:opclass}
A finite-$K$ member $\mathbf{\Gamma}_g^{(K)}$ of the family of Sec.~\ref{s:sec:family}, together with a
readout period $T$, a measurement window $T_\mathrm{off}$ and a readout order $N$, lies in
$\mathcal{A}_0(M,N)$ when
\begin{enumerate}
\item[(O1)] the detunings $\delta_k$ are pairwise distinct;
\item[(O2)] the eigenvalues $\lambda_k$ of $\mathbf{\Gamma}_g$ are pairwise distinct;
\item[(O3)] the overlaps are nonzero, with margins $\omega_{\min}\equiv\min_k|v_\mathrm{geo}^*v_{R,k}|>0$
and $a_{\min}\equiv\min_k|v_{L,k}^*\alpha|>0$;
\item[(O4)] the conjugation-resolved eigenvalue-sum points of order $n\leq N$, taken over all in-band
modes, are pairwise distinct or exactly conjugate, with minimal separation $\delta_\Lambda>0$, and the
drive channel is active, $\gamma_g>0$;
\item[(O5)] $T_\mathrm{off}$ lies outside the discrete exceptional set of Sec.~\ref{s:sec:selection};
\item[(O6)] $r_\ast\equiv\max_k|e^{-\lambda_kT}|<1/(M+1)$;
\item[(O7)] $\mathrm{Im}(\lambda_k-\lambda_{k'})T\notin2\pi\mathbb{Z}$ for $k\neq k'$.
\end{enumerate}
The subclass $\mathcal{A}_1(M,N)\subset\mathcal{A}_0(M,N)$ adds
\begin{enumerate}
\item[(O8)] the weak-dressing condition $\Delta_0\geq8\sqrt{K}\,\bar e$.
\end{enumerate}
\end{definition}

Conditions D1--D5 imply (O1)--(O4) and (O7) through Lemmas~\ref{s:lem:local} and~\ref{s:lem:distinct}, and
D1's coupler entry is (O8); (O6) is a condition on the readout period, and (O5) holds at every window
outside a discrete set. Every entry is an open condition, so $\mathcal{A}_0$ has nonempty interior.
Write $\lambda_{\min}\equiv\min_k\mathrm{Re}\,\lambda_k$, which is strictly positive on $\mathcal{A}_0$ by
Lemma~\ref{s:lem:strictdiss} and is bounded below by $\gamma/18K$ on $\mathcal{A}_1$.

\begin{theorem}[Universality with an explicit rate on the operating class]
\label{s:thm:opclass}
Let $y$ be a continuous fading-memory target with a convergent Volterra expansion on
$\mathcal{K}_{u_{\max}}$, let $\epsilon>0$, and choose $(M,N)$ so that the Boyd--Chua truncation error
obeys $\delta_{MN}\leq\epsilon/2$, fixing $A=A(M,N)$. Let the device be operated at any point of
the operating class $\mathcal{A}_0(M,N)$ of Definition~\ref{s:def:opclass} (whose conditions O1--O8 the proof consumes) with $K\geq M$. Then for every readout period satisfying (O6)--(O7) and
\begin{align}
    \label{s:eq:opclassrate}
    T\;\geq\;\frac{1}{\lambda_{\min}}\,\ln\frac{2e\,A\,M(M+1)}{\epsilon}
\end{align}
there exist weights $\{W_n(t)\}$ with $|y_k-\hat y_k|\leq\epsilon$ on $\mathcal{K}_{u_{\max}}$. On
$\mathcal{A}_1$ the same holds with $\lambda_{\min}$ replaced by its symbolic floor $\gamma/18K$.
\end{theorem}
\begin{proof}
Kernel matching to depth $M$ and order $N$ is the construction of Secs.~S.2.3--S.2.4, which consumes
(O2), (O3), (O4) and (O7) for the selection and the Vandermonde inversion, and (O5) for
prescribability at the operated window; for $K>M$ the readout-side selection of
Remark~\ref{s:rem:KgtM} confines the realized kernels to the $M$ selected modes and consumes (O4) over
all in-band modes. The extrapolation identity (Proposition~\ref{s:prop:extrapolation}) is algebraic and
consumes only (O2). Lemma~\ref{s:lem:tail}(a) then bounds the tail by
$S_\mathrm{out}\leq eM(M+1)r_\ast(1+O(Mr_\ast))$ under (O6), where $r_\ast=e^{-\lambda_{\min} T}$ is a maximum
over node moduli whose exponent carries the minimum over decay rates. Requiring
$A\,S_\mathrm{out}\leq\epsilon/2$ gives Eq.~\eqref{s:eq:opclassrate}, and adding the truncation error
gives $\epsilon$. No design condition is consumed at any step.
\end{proof}

In plain terms, the designed operating point of this section engineers the spectrum so that every
constant can be written down in advance; the class above asks only that the comb lines be distinct,
that the readout couple to every mode, that the eigenvalue sums not collide, and that the device be
watched long enough. The price of the weaker hypothesis is that the rate constant $\lambda_{\min}$ is a measured
property of the operating point rather than a formula in the physical rates: one must look at the
machine to know how long to watch it. Comparing the two at $K=M$, the designed period is
$\Theta(M\ln M/\gamma)$ and the class period is $\Theta(M\ln(AM^2/\epsilon)/\gamma)$, the same order in
the memory depth, so the designed point buys a bounded factor in the readout period rather than the
availability of the guarantee.

\subsection*{Designed and generic operating points}
The design D1--D5 is a constructive certificate: it exhibits, with every constant explicit, one operating point at which all hypotheses hold with proven margins. Its conditions are sufficient; we do not establish that any is necessary. They carry a stated price, the exponential defect-protection entry of Eq.~\eqref{s:eq:D1}, and the guarantee does not depend on paying it (Lemma~\ref{s:lem:tail}(a)). A generic fabricated device pays no such price: its eigenvalue sums are separated by its own spectral irregularity, with margins measured rather than prescribed, and Sec.~\ref{s:sec:genericity} verifies this by direct diagonalization at every operating point simulated here.

\subsection*{Physical resources of the design}
The construction spends seven resources, and we state them plainly, for the designed certificate. (i) \emph{Time}: $T=O(M\ln M+\ln\epsilon^{-1})/(\gamma+\gamma_g)$ per readout step, by Eq.~\eqref{s:eq:Tselect}. (ii) \emph{Bandwidth}: $B=K\Delta_0$ with $\Delta_0$ carrying the exponential defect-protection and radial-protection factors of Eq.~\eqref{s:eq:D1}; the flat-coupling budget $\varepsilon_\mathrm{flat}(B)$ of the Discussion is charged accordingly. A generic operating point replaces this entry by its verified margins and pays only the polynomial second entry of Eq.~\eqref{s:eq:D1}. (iii) \emph{Timing precision}: the D2 condition must hold to a phase accuracy of order $\bar\varepsilon_\mathrm{target}$, which at spacing $\Delta_0$ means a relative timing precision of order $\bar\varepsilon_\mathrm{target}/(\Delta_0T)$, which falls exponentially in $M$ through $\Delta_0$, so the certificate's clock is as far beyond laboratory reach as its comb spacing; growing precision requirements on settings are the standard price of universal-approximation constructions (classical weight precision grows likewise), and we state the requirement rather than leave it implicit. (iv) \emph{Shots}: at the designed operating point the graded separations are exponentially small in $M$, so the constructive weights there, while never entering the error bound, are exponentially large, and realizing them at fixed output precision costs shots accordingly; the theorem is an approximation statement, and its sample-complexity accounting is Sec.~\ref{s:sec:shotnoise}'s. (v) \emph{Rate ordering}: condition D4 drives the drive-channel rate far below the mean mode decay, by the two entries of D4, so the designed certificate sits outside the adiabatic-slaving ordering $\gamma_g\gg\lambda_\star$ under which the linear-transducer limit is derived from the physical atom (Methods).

The certificate's requirement $s\leq10^{-2}$ is not in conflict with this: it is imposed at the designed point, where D4 caps $\rho$ near $10^{-4}$, so the admissible drive is far smaller than the simulated one and the saturation bound is slack by orders of magnitude.

Which entry supplies that floor must be said, because the entries do not scale alike. The defect-protection entry gives $768\,M^{2}\ln(2M)\,5^{M-1}$, about $2.1\times10^{4}$ at $M=2$. Both expressions are read off Eq.~\eqref{s:eq:D1} by the same substitutions, and we give them once so the reader can check either: $e_0=\gamma/2$ because the Hermitian part has rank two with unit vectors (Lemma~\ref{s:lem:trace}); $T=\Lambda/\lambda_\star=2M\Lambda/\gamma$ at $K=M$; $\gamma_g=\rho\gamma$ with $\rho$ at its D4 ceiling $\bar\varepsilon_\mathrm{target}/(8s_\alpha\Lambda_{\max})$ and $s_\alpha\leq q^2$ from D3; $\Lambda_{\max}=2\Lambda_{\min}$; and $B=K\Delta_0=M\Delta_0$. The third entry $384Ne_0^2T\beta^{M-1}/\bar\varepsilon_\mathrm{target}$ then gives $B/\gamma\geq192M^2\Lambda\beta^{M-1}/\bar\varepsilon_\mathrm{target}$, and the fifth $512NKe_0^2(4N)^{q-2}/\gamma_g$ gives $B/\gamma\geq2^{\,2q+6}M^2q^2\Lambda_{\max}/\bar\varepsilon_\mathrm{target}$; at $N=1$, $\beta=5$ and $\bar\varepsilon_\mathrm{target}=1/4$ these are the two displayed forms. Their ratio grows as $(16/5)^M$ when $q$ is taken as $2M{+}1$, and jumps with $q$ at the primes. The \emph{radial}-protection entry, the fifth, gives more: with $\rho$ at its D4 cap, at $s_\alpha\leq q^2$ and with the band constraint $M\Delta_0\leq B$, it reads
\begin{align}
\label{s:eq:reach}
B/\gamma\;\geq\;\frac{2^{\,2q+7}}{\bar\varepsilon_\mathrm{target}}\,M^{2}q^{2}\,\Lambda_{\min},\qquad \Lambda=\lambda_\star T,\qquad q\ \text{the least prime}\ \geq2K_\mathrm{phys}+1 ,
\end{align}
including the factor-two margin taken in the pair selection of Step~3. The prefactor is a single power of two because every factor entering it is one: $512$ from the fifth entry of Eq.~\eqref{s:eq:D1}, $e_0^2=\gamma^2/4$, the $8$ of the D4 ceiling on $\rho$, and $\bar\varepsilon_\mathrm{target}=1/4$; at that budget the prefactor reads $2^{\,2q+9}$. At the minimal readout period $\Lambda=\ln 2M$ this is about $7.3\times10^{7}$ at $M=2$, $6.6\times10^{9}$ at $M=3$, and $8.6\times10^{12}$ at $M=4$ (deposited script \texttt{resources\_v36.py}), and it degrades by $\Lambda/\ln2M$ when the accuracy branch of Step~3 binds. Since $q\geq2M+1$, the fifth entry exceeds the third by a factor growing like $(16/5)^{M}$, so it is the fifth entry that binds and Eq.~\eqref{s:eq:reach} that states the reach. The platform reading of the main text follows: the certified construction fits no depth $M\geq2$ at the superconducting example ($B/\gamma\approx10^{2}$), and $M=2$--$3$ only for kHz-class linewidths under GHz-flat coupling. What this floor prices is the comparison following Lemma~\ref{s:lem:tail}: a readout period shorter by the factor of Eq.~\eqref{s:eq:penalty} rather than the availability of the guarantee, which Lemma~\ref{s:lem:tail}(a) supplies at any admissible operating point.

Condition D6 carries a cost of the same kind, and we state it because it is the entry that decides what is buildable. Since $\varepsilon_v\leq\bar\varepsilon_\mathrm{target}/8\Lambda_{\max}$ by Eq.~\eqref{s:eq:D6} and the geometry gives $\varepsilon_v\approx B/\omega_0$ with $\omega_0$ the transition frequency, the two combine into a carrier-to-linewidth requirement
\begin{align}
\label{s:eq:d6cost}
\frac{\omega_0}{\gamma}\;\geq\;\frac{B}{\gamma}\cdot\frac{8\Lambda_{\max}}{\bar\varepsilon_\mathrm{target}}\;=\;\frac{2^{\,2q+11}}{\bar\varepsilon_\mathrm{target}^{2}}\,M^{2}q^{2}\Lambda^{2}
\end{align}
which at $\bar\varepsilon_\mathrm{target}=1/4$ reads $2^{\,2q+15}M^{2}q^{2}\Lambda^{2}$, about $6.4\times10^{9}$ at $M=2$. The relaxation of the node budget from $1/(4M^{3/2})$ to $1/4$, which the remainder estimate of Lemma~\ref{s:lem:tail}(b) permits, improves this by a factor $8$ at $M=2$ and by $4M^{3/2}$ in general; at the retired budget the same expression gives $5.2\times10^{10}$. Even relaxed, the requirement is severe: it is the coupler's flatness, rather than the comb spacing, that sets the hardest demand on the fabricated device, and no platform named in this work meets it at $M=2$.

\subsection*{Check of the designed operating point}
The design checks of Sec.~\ref{s:sec:design} are inequalities between explicit expressions and are evaluated by the deposited scripts at operating points constructed from Table~\ref{s:tab:design}. They check the construction; no lemma of this Supplement rests on them.

\begin{figure}[h]
    \centering
    \includegraphics[width=.92\linewidth]{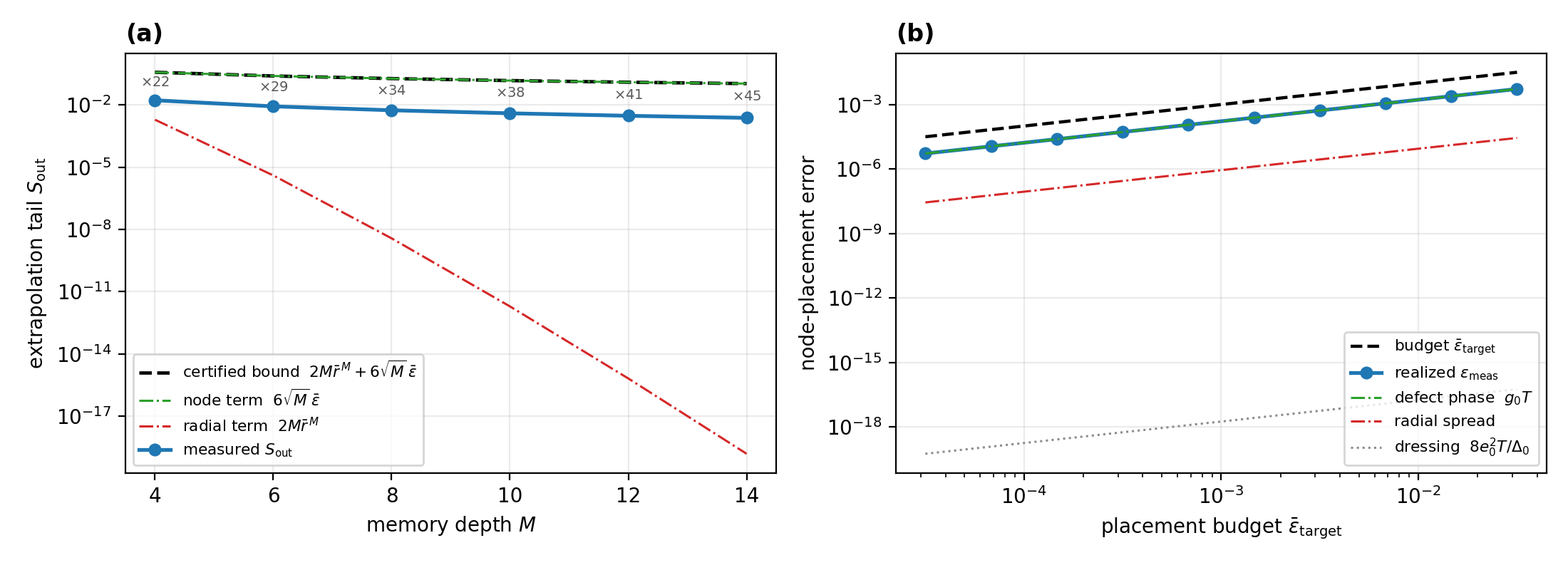}
    \caption{Verification of the designed operating point (geometry-fixed coupler). \textbf{a}, Measured extrapolation tail $\sum_{m>M}\Vert c_m\Vert_1$, computed directly from the node set at the extreme admissible modulus $\bar\varepsilon=1/4$ and $\bar r=1/(2M)$, against the bound of Lemma~\ref{s:lem:tail}(b) (dashed) and its two constituents. The node term dominates the bound over the whole range while the radial term $2M\bar r^{\,M}$ is already negligible, and the measurement sits below the bound by the factors annotated, rising monotonically from $22$ at $M=4$ to $45$ at $M=14$. The plotted value is the worst case over the exhaustive deterministic phase family $\theta_k=2\pi jk/M$, $j=0,\dots,M-1$, rather than a random sample: random draws under-sample the $M$-torus non-uniformly in $M$ and produce a spurious non-monotonicity. \textbf{b}, Realized node-placement error against the assigned budget $\bar\varepsilon_\mathrm{target}$ at $M=6$, $N=2$, with the three contributions of Eq.~\eqref{s:eq:epsbar} resolved: the placement is realized at $\tfrac16$ of budget with unit slope over three decades, consumed almost entirely by the graded-defect phase $g_0T$, with the radial spread two orders below it and the dressing displacement $8e_0^2T/\Delta_0$ some fifteen further orders down. Two remarks on scope. Both panels run well past the depth at which the certified operating point is physically reachable---the bandwidth requirement of Eq.~\eqref{s:eq:reach} already exceeds any demonstrated platform at $M=4$---and every design inequality of Eq.~\eqref{s:eq:D1} is asserted and holds at each point plotted; the construction is therefore verified over a range where the certificate itself is not realizable, which is what makes the margins in \textbf{a} and \textbf{b} informative rather than merely reassuring. And both margins are one-sided: the bounds are conservative, the conservatism grows with $M$, and none of it is fitted.}
    \label{s:fig:designverif}
\end{figure}

\begin{figure}[h]
    \centering
    \includegraphics[width=\linewidth]{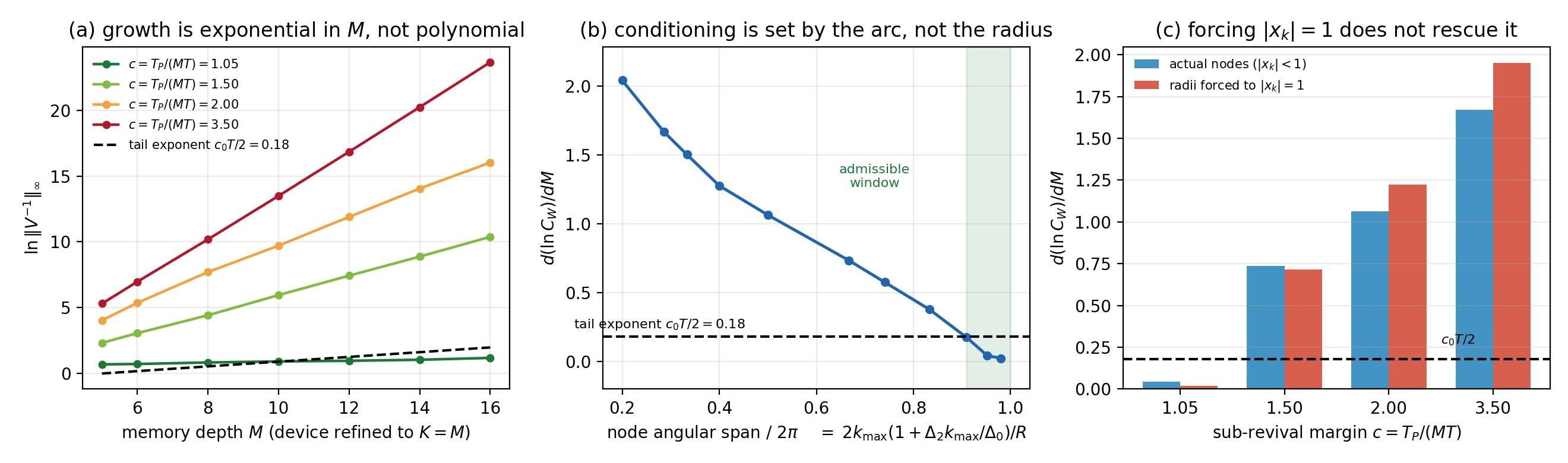}
    \caption{Why weight-norm routes fail: motivation for the weight-independent proof. \textbf{a}, For nodes confined to an arc (the geometry forced when a sub-revival condition $T_P>MT$ caps the angular span), $\ln\Vert V^{-1}\Vert_\infty$ grows linearly in $M$ ($R^2>\NumAPJJJ$): any bound routed through the matched-weight norm inherits exponential growth. \textbf{b}, The growth rate is set by the angular span rather than the radius. \textbf{c}, Forcing $|x_k|=1$ does not repair it. The proof of this subsection avoids the weight norm entirely (Proposition~\ref{s:prop:extrapolation}), and the design removes the sub-revival constraint that created the arc.}
    \label{s:fig:conditioning}
\end{figure}

\subsection{Capability containment and strict scaling: proof of the corollary}
\label{s:sec:strictscaling}
The strict-scaling corollary of the main text asserts that enlarging the accessible mode number never decreases approximation capability, and strictly enlarges the span of exactly matchable kernel blocks. We prove the two halves separately.

\begin{proposition}[Capability containment and strict growth of the span]
\label{s:prop:scaling}  
Let $\mathcal{R}(M,N)$ denote the set of target-accuracy pairs $(y,\epsilon)$ the $(M,N)$-device
achieves, and let $\mathcal{S}(M,N)$ denote the set of kernel blocks
$\{h_n(\vec m):n\leq N,\ \vec m\in[1,M]^n\}$ it realizes exactly. Let $M\leq M'$ and $N\leq N'$. Then

\emph{(i) Containment.} $\mathcal{R}(M,N)\subseteq\mathcal{R}(M',N')$.

\emph{(ii) Strict growth of the span.} $\mathcal{S}(M,N)$ is the full block space of symmetric
kernels, of dimension $\sum_{n\leq N}\binom{M+n-1}{n}$, the kernels being identified only through
their symmetric part, as the inversion of Sec.~S.2.4 over ordered indices makes explicit. Hence
$\mathcal{S}(M,N)\subsetneq\mathcal{S}(M',N')$ strictly whenever $(M',N')\geq(M,N)$ componentwise
with at least one strict inequality.
\end{proposition}
\begin{proof}
\emph{(i).} Let $(y,\epsilon)\in\mathcal{R}(M,N)$: the parameter selection of Step~3 chose $(M,N)$ from
the target so that the truncation error satisfies $\delta_{MN}\leq\epsilon/2$, and met the remaining
budget through Eq.~\eqref{s:eq:termii} by driving the extrapolation tail below $\epsilon/(2A(M,N))$.
We exhibit settings of the $(M',N')$-device achieving the same $\epsilon$ in three steps. The
intuition is that a larger device can be told to imitate a smaller one by being handed the smaller
one's kernels and zeros everywhere else, and that the zeros cost nothing.

\emph{Step A (zero-padded matching).} Operate the $(M',N')$-device at any point at which its nodes are
pairwise distinct and its conjugation-resolved eigenvalue sums of order $n\leq N'$ are resolved; the
designed point of this section is one such, and a generically fabricated device is another, by
Sec.~\ref{s:sec:genericity}. By Lemma~\ref{s:lem:nodes} its $M'$ nodes are pairwise distinct, and by
Lemma~\ref{s:lem:distinct} at base $\beta=4N'+1$ every conjugation-resolved eigenvalue-sum point of
order $n\leq N'$ needed by the matching is resolved, so the inversion of Sec.~\ref{s:sec:proof}
prescribes \emph{any} values on the $(N',M')$ block. Prescribe the target's truncated kernels
$h^{(M,N)}_n(\vec m)$ on the sub-block $n\leq N$, $\vec m\in[1,M]^n$, and zero on the remainder. This
is an admissible choice of the weights $\{W_n(t)\}$; nothing else about the device changes.

\emph{Step B (zeros extrapolate to zero).} By Proposition~\ref{s:prop:extrapolation} the realized
kernels at all lags are the tensor extrapolation Eq.~\eqref{s:eq:tensorext} of the matched-block
moments, and Eq.~\eqref{s:eq:tensorext} is linear in $H_n$ separately at each order. At orders
$n\in(N,N']$ the matched block is identically zero, hence $H_n\equiv0$, hence the realized kernels
vanish at \emph{every} lag: the padding introduces no spurious response at the padded orders. At
orders $n\leq N$ the realized functional agrees with the truncated target on $[1,M']^n$, the values on
$[1,M']^n\setminus[1,M]^n$ being the target's own zeros, and deviates from it only at lags
$\vec m\notin[1,M']^n$.

\emph{Step C (tail budget).} The deviation of Step~B is controlled by Eq.~\eqref{s:eq:termii} applied
to the $(M',N')$ matching. The zero-padded block has the target's own kernel sup-norms, zeros not
increasing a supremum, so the prefactor is
$A'\equiv\sum_{n=1}^{N}n\,2^{2n}u_{\max}^{n}M'^{\,n-1}\Vert h^{(M,N)}_n\Vert_\infty<\infty$, the sum
terminating at $N$ because the padded orders carry no realized content by Step~B; the tail factor is
the $M'$-node $S_\mathrm{out}'\leq2M'\bar r'^{\,M'}+e\,M'\bar\varepsilon'$ of
Lemma~\ref{s:lem:tail}(b). By the trace identity $\bar r'^{\,M'}=e^{-(\gamma+\gamma_g)T'/2}$ with $T'$
free, and $\bar\varepsilon'$ is driven below any tolerance by the comb spacing
(Lemma~\ref{s:lem:nodes}); the parameter selection of Step~3, run verbatim at $(M',N')$ with budget
$\epsilon/(2A')$, therefore delivers $S_\mathrm{out}'\leq\epsilon/(2A')$, with the same quantifier
order and no circularity. Since $\delta_{MN}\leq\epsilon/2$ is a property of the target alone and is
unchanged, the total error is at most $\epsilon$.

\emph{(ii).} Surjectivity of the matching onto the full block space is the inversion itself: node
distinctness (Lemma~\ref{s:lem:nodes}) makes the order-one prescription invertible, and design
distinctness (Lemma~\ref{s:lem:distinct}) makes the order-$n$ prescription well posed for all
$n\leq N$, so every block value is independently prescribable. The dimension count and the strict
inclusion follow directly, every block supported on the enlarged index set and not on the smaller one
being realizable at $(M',N')$ and not indexable at $(M,N)$. No auxiliary hypothesis is required. This establishes both halves of the proposition: the capability sets are nested as the memory depth and Volterra order grow, and the nesting is strict, since each enlargement of the index set carries a kernel block the smaller device cannot index at all.
\end{proof}

\begin{remark}[Sharper prefactor]
Tracking the support of the nonzero matched entries (contained in $[1,M]^n$) through the counting bound of Eq.~\eqref{s:eq:termii} replaces $M'^{\,n-1}$ by $M^{\,n-1}$ in $A'$, recovering the original $A(M,N)$; the loose constant above is used in the proof because it requires no support bookkeeping, and either constant is finite and target-determined, which is all Step C consumes.
\end{remark}

Two cautions delimit what has been proved, and they are the reason (i) and (ii) are separate
statements rather than one. Containment is a statement about \emph{achievable accuracies} rather than
about exact reproduction of output functionals: distinct members of the family have distinct spectra,
no spectral nesting being assumed (Sec.~\ref{s:sec:family}), so their extrapolation tails differ, and
the $(M',N')$-device reproduces the $(M,N)$-device's outputs to every tolerance rather than
identically. This is the operative sense in which scaling is never wasted, and the only sense the
theorem's own error metric defines. Strictness, meanwhile, is proved at the level of the exactly
matchable span: a capability-level separation, a target and an $\epsilon$ achievable at $(M',N')$ and
provably \emph{not} achievable at $(M,N)$ over all of the smaller device's admissible settings, would
require a lower bound we do not claim. Since deeper truncation strictly reduces $\delta_{MN}$ for any
target with content at the added lags, the span statement is what the strict-containment claim of the
abstract asserts.

\subsection{Verification of the hypotheses of the density theorem}
\label{s:sec:cuchiero}
The passage from finitely many matched kernels to approximation of the full functional invokes Theorem~1 of Cuchiero \emph{et al.}~\cite{Cuchiero}. We verify its hypotheses in our setting. The input space is $\mathcal{K}_{u_{\max}}=[-u_{\max},u_{\max}]^{\mathbb{Z}_-}$---inputs uniformly bounded by $u_{\max}$, the bound already used in Lemma~\ref{s:lem:sector}; the symbol $B$ is reserved for the bandwidth---compact in the product topology by Tychonoff's theorem. The target is a continuous fading-memory functional on $\mathcal{K}_{u_{\max}}$, i.e.\ continuous with respect to a weighted norm $\|u\|_w=\sup_m w_m|u_{-m}|$ with $w_m\downarrow0$; on $\mathcal{K}_{u_{\max}}$ this topology coincides with the product topology, so the target lies in $C(\mathcal{K}_{u_{\max}})$. The finite Volterra polynomials $u\mapsto\prod_i u_{-m_i}$ form a point-separating subalgebra of $C(\mathcal{K}_{u_{\max}})$ containing the constants: two distinct sequences differ at some lag $m$, separated by the degree-one monomial $u\mapsto u_{-m}$. By Stone--Weierstrass this algebra is dense in $C(\mathcal{K}_{u_{\max}})$, which is the density hypothesis of the cited theorem; the kernel-matching construction of Secs.~S.2.3--S.2.4 realizes every element of the algebra exactly (for $K\geq M$), which is its realizability hypothesis. The quantification is over the family: exact realization of the degree-$(N,M)$ monomial block needs $K\geq M$ modes, so density is attained along the nested family of operating points as $M$ grows---the family-level, rate-free sense in which merely continuous targets are reached, exactly as Theorem~1's statement records. We emphasize that Stone--Weierstrass enters here only to characterize the closure of the \emph{target class}; the strict-scaling corollary of the main text continues to follow from kernel containment, which density arguments do not provide.  

Theorem~1 adds three physical statements to the abstract density results it invokes. Boyd--Chua \cite{Boyd:1985} and Cuchiero \emph{et al.}~\cite{Cuchiero} establish that \emph{some} functional with rich enough kernels approximates any fading-memory target; they say nothing about which hardware realizes those kernels, at what rate, or with what scaling. Here: (i) the atom--mirror filter's own kernels are provably rich enough to match any target up to $(N,M)$ under generic conditions on one mirror distance; (ii) the convergence rate is the trace budget of the fixed hardware; and (iii) enlarging the accessible mode number never reduces capability and strictly enlarges the matchable span.

\section{Fading Memory of the Device}
\label{s:sec:fadingmemory}

The device has fading memory, and the fact has two statements of different strength. We give the elementary one first, because it follows from the kernel bound alone and because it is all that is needed at a fixed operating point, and then prove the stronger statement that survives the many-mode limit.

\paragraph{The elementary statement, at fixed mode number.} In the linear-transducer limit the reduced generator $\mathcal{L}_0$ is quadratic with a c-number drive, so a constant drive relaxes to a coherent steady state with amplitude $\mathbf{\Gamma}_g^{-1}\alpha$, and the input--output kernel is the finite sum of exponentials $h_K(t)=v_\mathrm{geo}^*e^{-\mathbf{\Gamma}_gt}\alpha=\sum_k(v_\mathrm{geo}^*v_{R,k})(v_{L,k}^*\alpha)e^{-\lambda_kt}$. Every $\mathrm{Re}\,\lambda_k>0$ at finite $K$ (Lemma~\ref{s:lem:strictdiss}, which needs only the flat coupler and the distinctness of the $\delta_k$), so $|h_K(t)|\leq C(K)e^{-\varepsilon_D(K)t}$ with $\varepsilon_D(K)=\min_k\mathrm{Re}\,\lambda_k>0$: the kernel decays, the influence of any input recedes, and Corollary~\ref{s:cor:fmp} follows at that $K$. Nothing more is required for any single operating point, and no delay equation is involved.

\paragraph{Why the elementary statement is not enough.} Its rate is not uniform in the mode number. By the trace identity of Lemma~\ref{s:lem:trace}, $\varepsilon_D(K)\leq(\gamma+\gamma_g)/2K$, so the elementary bound degrades as $1/K$ and says nothing about the family as a whole---even though the physical loop that produces the memory is unchanged as the comb is refined. The discrepancy is not an artefact of a loose bound: the slow eigenmodes are the weakly coupled ones, and they carry almost none of the input--output response. The physically meaningful statement is therefore about the \emph{kernel} rather than the spectrum, and it is uniform: on any sub-revival horizon the kernel decays at the loop rate $c_0$ set by delay stability, with constants independent of $K$. That is the content of the remainder of this section.

\subsection{Uniform kernel decay under delay stability}
\label{s:sec:kerneldecay}
\begin{lemma}[Uniform kernel decay under delay stability]
\label{s:lem:kerneldecay}
Two kernels appear in this lemma, and we define them separately because they obey different equations. The finite-comb kernel is the matrix element
\begin{align}
    h_K(t) \;=\; v_\mathrm{geo}^*\cdot e^{-\mathbf{\Gamma}_g t}\cdot\alpha ,
\end{align}
the input--output impulse response of the $K$-mode Gaussian-limit generator, drawn from the nested family whose coupling density carries the fixed band envelope of Theorem~1 (flat with smooth Gaussian rolloff). The continuum kernel $h_\infty(t)$ is its unbandlimited limit. It is $h_\infty$, and only $h_\infty$, that obeys the delay-differential equation of the reduced linear transducer: at any finite $K$ the comb kernel $h_K$ satisfies no delay equation (it is a finite sum of exponentials, and its revival echoes are the signature of the difference).

The physical difference between the two kernels is the mirror's apparent distance. In $h_\infty$ the atom faces a genuine continuum: the emitted pulse reflects once, returns after the delay $\tau$, and then escapes forever out the open end, and this one-echo causal structure is exactly what a delay-differential equation encodes. In $h_K$ the field lives on a discrete comb, and a discrete comb is a periodic system: $K$ equally spaced tones cannot dephase forever and rephase after the revival time $T_P=2\pi/\Delta_0$. Physically, finite mode spacing means finite quantization length, so the pulse has a second boundary to return from: $h_K$ shows the fast loop transient, decays as if the field were gone, and partially reconstructs at $t\approx T_P$, the revival echo, which is the arrival of radiation a true continuum would have carried to infinity. The two kernels therefore agree on the sub-revival window, where the finite comb does not yet know it is finite, and differ precisely at and beyond $T_P$; this is why $h_K$ obeys no delay equation and why the uniform bound below is honestly restricted to horizons $T_H<T_P$. The characteristic equation of the continuum delay dynamics is
\begin{align}
    \label{s:eq:charroot}
    s + \frac{\gamma}{2}\Big(1 + r\,e^{i\phi}e^{-s\tau}\Big) = 0, \qquad r=\gamma_R/\gamma\leq1,\quad \phi=\pi-\omega_0\tau .  
\end{align}
Here $\gamma_R$ is the emission rate into the returning mirror channel as seen at the atom, round-trip losses included, so that $r=\gamma_R/\gamma\leq1$ is the returned fraction of the total decay, $r=1$ the lossless ideal mirror, and $r$ the quantity the feedback transmissivity $\eta$ of Sec.~\ref{s:sec:robustness} attenuates (see the paragraph on detection versus feedback loss in Sec.~\ref{s:sec:shotnoise}); and $\phi=\pi-\omega_0\tau$ is the round-trip phase, the propagation phase $\omega_0\tau$ at the atomic transition frequency plus the $\pi$ of the mirror reflection. (The characteristic equation describes the undriven loop, so its phase is evaluated at the atomic frequency; the main text's interaction-picture phase $\pi-\omega_L\tau$ is the same quantity in the frame rotating at the drive, the two coinciding on resonance and differing by the detuning contribution $\Delta\tau$ off it.) Consistency is internal: $\phi=\pi$ corresponds to $\sin(\omega_0\tau/2)=0$, the atom at a field node, which is exactly the dark-state set excluded by the overlap conditions, while $\phi=0$ is the antinode of maximal collective decay, matching the static anchors of Sec.~\ref{s:sec:robustness}. Define the \emph{delay-stability condition} $D(c_0)$: all roots of Eq.~\eqref{s:eq:charroot} satisfy $\mathrm{Re}[s]\leq -c_0<0$. Let $T_P=2\pi/\Delta_0$ be the revival time of a comb of spacing $\Delta_0$. If $D(c_0)$ holds, then for every horizon $T_H<T_P$ there exist $C_h(c_0,T_P)<\infty$ and $K_0$, both independent of $K$, such that for all $K\geq K_0$,
\begin{align}
    |h_K(t)| \;\leq\; C_h\, e^{-c_0 t/2} \qquad \text{for all } 0\leq t\leq T_H ,
\end{align}
i.e.\ on any sub-revival horizon the kernel decay is uniform in the mode number. Refining the comb ($\Delta_0\downarrow0$) raises $T_P$ and hence extends the horizon on which the uniform bound holds; in particular, whenever the comb is refined so that $T_P$ exceeds the memory horizon $MT$ used by the readout, the bound holds with $K$-independent constants over the entire range of lags that enter the output.
\end{lemma}
\begin{proof}
We first derive the delay equation for $h_\infty$, since the rest of the proof rests on it. In the continuum the linear transducer couples to the semi-infinite waveguide through the standing-wave density $\tilde g(\omega)^2=(\gamma/\pi)\sin^2(\omega\tau/2)$ of Sec.~\ref{s:sec:exactderiv}. Writing $\sin^2(\omega\tau/2)=\tfrac12(1-\cos\omega\tau)$ and passing to the frame rotating at $\omega_0$ under the flat-band axiom A2, Wigner--Weisskopf elimination of the field gives the memory kernel as the Fourier transform of the density,
\begin{align*}
\int\dd\nu\,\tfrac{\gamma}{2\pi}\big(1-\cos((\nu+\omega_0)\tau)\big)e^{-i\nu t}\;=\;\gamma\,\delta(t)-\tfrac{\gamma}{2}e^{-i\omega_0\tau}\delta(t-\tau)-\tfrac{\gamma}{2}e^{i\omega_0\tau}\delta(t+\tau).
\end{align*}
The rate appearing here is the atom's mirror-modified emission into the guided field; in the mode-space generator it is the profiled channel $\alpha_k\propto\sin(\omega_k\tau/2)$ that carries this structure, the frequency-flat readout channel being delta-correlated and contributing no delay. The causal contributions (the local term counting one half, the advanced term discarded) yield $\dot h_\infty(t)=-\tfrac{\gamma}{2}h_\infty(t)+\tfrac{\gamma}{2}e^{-i\omega_0\tau}h_\infty(t-\tau)$ for $t>\tau$, with the memoryless segment on $[0,\tau]$; since $e^{-i\omega_0\tau}=-e^{i\phi}$ at the round-trip phase $\phi=\pi-\omega_0\tau$ (the $\pi$ being the mirror reflection), and a lossy return scales the delayed term by $r=\gamma_R/\gamma$, this is $\dot h_\infty(t) = -\tfrac{\gamma}{2}h_\infty(t) - \tfrac{\gamma}{2} r\, e^{i\phi} h_\infty(t-\tau)$, whose characteristic equation is Eq.~\eqref{s:eq:charroot}. Solutions of a linear autonomous delay equation admit the spectral bound $|h_\infty(t)|\leq C e^{(\sigma_0+\epsilon)t}$ for any $\epsilon>0$, with $\sigma_0$ the supremum of real parts of characteristic roots \cite{HaleVerduynLunel}; $D(c_0)$ gives $\sigma_0\leq-c_0$, hence the bound for $h_\infty$ at rate $c_0$ (take $\epsilon=c_0/2$).  

Uniformity in $K$ requires relating the finite-comb kernel to $h_\infty$, and the relation is not a plain Riemann approximation. Two steps are involved, and we keep them separate. First, the fixed band envelope of the theorem's nested-family setup acts on the continuum kernel by convolution: writing $\psi$ for the (inverse) Fourier transform of the fixed envelope shape,
\begin{align}
    \label{s:eq:envconv}
    h^{(\mathrm{env})}_\infty \;=\; \psi * h_\infty, \qquad\text{i.e.}\qquad h^{(\mathrm{env})}_\infty(t) \;=\; \int \psi(s)\,h_\infty(t-s)\,\dd s ,
\end{align}
with $\psi$ fixed once and independent of $K$ because the envelope shape is. Second, for a comb of spacing $\Delta_0$ the Poisson summation formula relates the finite-comb \emph{memory} kernel $C_K$ to periodic repetitions of the envelope-shaped one, $C_K(t)=\sum_{p\in\mathbb{Z}}C^{(\mathrm{env})}_\infty(t-pT_P)$; the repetitions are disjoint on $[0,T_P)$, so $C_K\equiv C^{(\mathrm{env})}_\infty$ there, and since the response solves a causal Volterra equation with a unique solution, the responses agree exactly on the same window. The response is not itself a periodization of $h_\infty$---the map from memory kernel to response is a resolvent rather than a linear one---and the display below is written for the memory kernel, with the echo picture used only as an illustration of what happens beyond $T_P$:
\begin{align}
    \label{s:eq:poisson}
    h_K(t) \;=\; \sum_{p\geq0} h^{(\mathrm{env})}_\infty\big(t - p\,T_P\big)\,\mathbf{1}(t-pT_P), \qquad T_P=\frac{2\pi}{\Delta_0},
\end{align}
where the $p\geq1$ terms are \emph{revival echoes} of the finite comb. Each echo is a copy of the decaying kernel restarted at $t=pT_P$. Equation~\eqref{s:eq:poisson} is exact for the equidistant comb; which is the case the lemma claims, together with the designed comb of Sec.~\ref{s:sec:design}, whose defects satisfy $|\omega_k-\omega_0-\Delta_0k|\leq g_0$. The quadratically chirped comb $\Delta_0+2\Delta_2|k|$ has varying local spacing, the repetitions are not exact, and it lies outside the scope of this statement. Eq.~\eqref{s:eq:poisson} is written here for the idealized comb, in which the repetitions are exactly disjoint; with the smeared envelope the echoes acquire tails that leak marginally across $t=pT_P$, and we account for that leakage explicitly below.  

The band envelope is what makes the $p=0$ term decay uniformly in $K$, and it does so without any appeal to an ideal (brick-wall) band cutoff---whose filter would have divergent $L^1$ norm (the sinc kernel), and whose one-sided transform decays only as $1/|\omega|$, so that a naive band-limitation constant would grow like $\log B_K$ rather than staying $K$-uniform. Instead, band-limitation with the fixed envelope is the convolution Eq.~\eqref{s:eq:envconv}, whose kernel $\psi$ possesses a finite exponential moment,
\begin{align}
    \label{s:eq:envmoment}
    \int |\psi(s)|\,e^{c_0|s|}\,\dd s \;\equiv\; C_\psi \;<\;\infty, \qquad \text{$K$-independently.}
\end{align}
This is where the band-edge shape enters, and the sufficient condition is exactly Eq.~\eqref{s:eq:envmoment}: the envelope must possess a finite exponential moment at rate $c_0$, equivalently its frequency profile must extend analytically to a horizontal strip of half-width $c_0$. A merely smooth rolloff need not satisfy this, but the condition is far from Gaussian-specific: an exponential edge of width $\sigma_B$ satisfies it whenever $\sigma_B>c_0$, while a Lorentzian edge fails. The Gaussian edge of fixed width $\sigma_B$ used here gives a Gaussian $\psi$ of width $1/\sigma_B$ in time, for which every exponential moment is finite and $C_\psi=e^{c_0^2/(2\sigma_B^2)}$ up to normalization. The constant therefore depends only on the ratio $c_0/\sigma_B$ of the loop-stability rate to the fixed band-edge width---a single physical number, $K$-independent because $\sigma_B$ is fixed once with the envelope shape. We note that $\sigma_B$ is the \emph{band-edge} scale and is logically distinct from the bandwidth $B$ over which the flat-coupling budget $\varepsilon_\mathrm{flat}(B)$ of the Discussion is assessed: the envelope constrains how the density rolls off at the edges, $\varepsilon_\mathrm{flat}$ how flat it is across the interior.

The band-limitation step then collapses to a three-line convolution bound. Using $|h_\infty(t-s)|\leq C\,e^{-c_0(t-s)}\mathbf{1}(t-s)$ from delay stability $D(c_0)$ in Eq.~\eqref{s:eq:envconv},
\begin{align}
    |h^{(\mathrm{env})}_\infty(t)|
    \;&\leq\; \int |\psi(s)|\,|h_\infty(t-s)|\,\dd s\nonumber\\
    \;&\leq\; C\,e^{-c_0 t}\!\int_{s<t}\!|\psi(s)|\,e^{c_0 s}\,\dd s \;+\; (\text{leakage from } s>t)\nonumber\\
    \;&\leq\; C_h\, e^{-c_0 t/2},
\end{align}
where the first integral is bounded by $C_\psi$ via the exponential moment Eq.~\eqref{s:eq:envmoment} (it absorbs the $e^{c_0 s}$), and the leakage from the causal tail $s>t$---the region where $\mathbf{1}(t-s)$ has cut off $h_\infty$---is Gaussian-small in $t$ because $\psi$ inherits the Gaussian rolloff, hence absorbed into $C_h$. On a horizon $t\leq T_H<T_P$ only the $p=0$ term of Eq.~\eqref{s:eq:poisson} is present ($\theta(t-pT_P)=0$ for $p\geq1$), so $|h_K(t)|=|h^{(\mathrm{env})}_\infty(t)|\leq C_h\,e^{-c_0 t/2}$ with $C_h$ built from $C_\psi$ and $C$ alone. The constants $C_h$ and $K_0$ therefore depend on $(c_0,T_P)$ and the fixed envelope but not on $K$, which is the uniformity claimed; the pointwise exponential envelope is now \emph{derived} from Eq.~\eqref{s:eq:envmoment} rather than asserted, so no separate ripple-crossing estimate is needed. Beyond one revival ($t\gtrsim T_P$) the echoes reconstruct the kernel and no fading-memory bound of this form can hold at fixed spacing---the honest content of the trace identity (Lemma~\ref{s:lem:trace}). The uniform bound therefore holds only on horizons below the revival time. The echo leakage noted after Eq.~\eqref{s:eq:poisson} is Gaussian-small in $B_K(T_P-T_H)$ and is absorbed into $C_h$, so the sub-revival bound is unaffected. Whether a given operating point lies below its own revival is a joint statement about the readout period and the comb spacing; we do not establish it for the simulations reported here, and the uniform-in-$K$ constants of this lemma are correspondingly not claimed at those operating points.
\end{proof}
\begin{remark}[The contraction and sub-revival requirements cannot both be met]
\label{s:rem:jointunsat}
Two requirements bear on the readout period from opposite sides, and they do not overlap. The
extrapolation-tail estimate of Lemma~\ref{s:lem:tail}(a) requires the contraction
$\max_k|e^{-\lambda_kT}|<1/(M+1)$, hence $T>\ln(M+1)/\lambda_{\min}$ with $\lambda_{\min}\equiv\min_k\mathrm{Re}\,\lambda_k$
the slowest decay rate of the retained block; the uniform bound of Lemma~\ref{s:lem:kerneldecay}
requires the memory horizon to lie below the revival, $MT<T_P$. The trace identity
(Lemma~\ref{s:lem:trace}) caps $\lambda_{\min}\leq\lambda_\star=(\gamma+\gamma_g)/2K$, and a comb of $K$ modes
filling a band $B$ has $T_P=2\pi K/B$. Chaining the three,
\begin{align}
    \label{s:eq:jointunsat}
    \frac{2KM\ln(M+1)}{\gamma+\gamma_g}\;<\;MT\;<\;\frac{2\pi K}{B}
    \qquad\Longleftrightarrow\qquad
    B\,M\ln(M+1)\;<\;\pi(\gamma+\gamma_g),
\end{align}
in which the mode number cancels identically: refining the comb raises the revival time and lowers the
slowest decay rate in the same proportion, so the two effects annihilate. Axiom A2 places the band far
above both rates, so the right-hand side of Eq.~\eqref{s:eq:jointunsat} is smaller than the left for
every $M\geq1$, and no operating point of the finite comb meets both requirements. In plain terms, the
device must be watched long enough for its slowest mode to decay, and that time grows with the mode
number in exactly the proportion that the comb's rephasing time does; adding modes buys both sides
equally and never closes the gap. No step of the approximation bound of Sec.~\ref{s:sec:errorbound} is
affected, since none of it uses fading memory; what Eq.~\eqref{s:eq:jointunsat} withdraws is the
possibility of extending the uniform-in-$K$ constants of Lemma~\ref{s:lem:kerneldecay} over the lags the
readout uses.
\end{remark}

\paragraph{Relation to the approximation theorem.} The uniform decay proven here characterizes the device: under delay stability the physical input--output kernel decays at the loop rate on sub-revival horizons, uniformly in the mode number, which is the fading-memory property of Corollary~\ref{s:cor:fmp} and the measurable signature of the feedback loop. It is a separate statement from the approximation bound of Sec.~\ref{s:sec:errorbound}, which is finite-dimensional and spectral and in which neither $C_h$ nor $c_0$ appears. The two are complementary: the theorem certifies what a chosen operating point can approximate, and this section certifies that the machine itself forgets.

\paragraph{The constant $C_h$ at the operating comb.} The lemma asserts existence of a finite, $K$-independent $C_h$ on the sub-revival horizon; the constant itself should be evaluated at the operating comb, and we report its measured behaviour. On a sub-revival window $T_H=0.5\,T_P$, with the comb refined as the theorem prescribes ($R=T_P/T$ growing with $K$), the equidistant comb gives $C_h=0.9251,\,0.9231,\,0.9226,\,0.9223,\,0.9222$ at $K=6,10,14,20,28$: stable to four significant figures, a direct confirmation of the claimed $K$-uniformity. The equidistant comb is not the worst case: at $K=14$ and $R=21$ on $T_H=0.8\,T_P$, moderate chirps raise $C_h$ to between $1.8$ and $3.7$ times its undispersed value, while chirped combs place their echo reconstruction at or beyond $T_P$ rather than before it. The margin to the revival also matters: at $T_H=0.8\,T_P$ the equidistant $C_h$ itself begins to grow with $K$ ($0.92\to1.73\to5.30$ at $K=14,20,28$) as the horizon approaches the revival, so the sub-revival window should be taken with room to spare rather than pressed against $T_P$; we accordingly operate, and recommend, $T_H\leq0.5\,T_P$. Nothing in the error bound of Sec.~\ref{s:sec:errorbound} depends on $C_h$.

\paragraph{Exponent bookkeeping.} Three rates appear, and we track the factors of two once. Delay stability $D(c_0)$ places all characteristic roots at $\mathrm{Re}[s]\leq-c_0$, so the continuum kernel decays at rate $c_0$ (up to any $\epsilon>0$ in the spectral bound). Lemma~\ref{s:lem:kerneldecay} states its conclusion at the halved rate $c_0/2$: the slack absorbs the $\epsilon$ of that spectral bound (we take $\epsilon=c_0/2$) and the envelope's exponential-moment constant $C_\psi$, leaving a clean pointwise bound $C_h e^{-c_0t/2}$. The fading-memory corollary (Cor.~\ref{s:cor:fmp}) halves once more, to $c_0/4$: the weighted-norm statement splits the per-lag decay factor into one half used to sum the geometric tail into the Lipschitz constant $L$ and one half retained as the weight sequence $w_m=e^{-c_0(m-1)T/4}$. No step is tight in these constants, and no qualitative claim depends on the halvings.

\paragraph{Measured envelope constants.} The uniform bound is an analytic statement about the idealized many-mode nested family, proven above with no appeal to numerics. Independently, the finite-$K$ devices actually simulated are verified to be members of $\mathcal{C}$ by direct diagonalization (Sec.~\ref{s:sec:genericity}), and on those devices the kernel envelope can be read off directly: the fitted decay rates are $c_\mathrm{eff}=0.15\gamma$ ($K=6$) and $0.09\gamma$ ($K=14$), and the measured envelope constants $\sup_{t\leq T_P}|h_K(t)|\,e^{+c_\mathrm{eff}t/2}$ are $0.56$ and $0.63$ respectively: finite, $K$-stable over the sub-revival horizon, and comfortably below the bound the lemma proves at the operating scale. These constants are a consistency check on the simulated devices rather than an ingredient of the proof. The fixed envelope disturbs nothing else in the construction: it is positive on the occupied band, so the overlap conditions $v_\mathrm{geo}^*\cdot v_{R,k}\neq0$, $v_{L,k}^*\cdot\alpha\neq0$ still hold; it is $\tau$-independent, so the almost-every-$\tau$ genericity of Sec.~\ref{s:sec:genericity} is untouched; it preserves the normalization $\|\alpha\|=\|v_\mathrm{geo}\|=1$, so the trace identity (Lemma~\ref{s:lem:trace}) is unchanged; and it is a property of the many-mode family only, so the finite-$K$ Vandermonde construction of Secs.~S.2.3--S.2.4 is unaffected.

\begin{figure}[h]
    \centering
    \includegraphics[width=.72\linewidth]{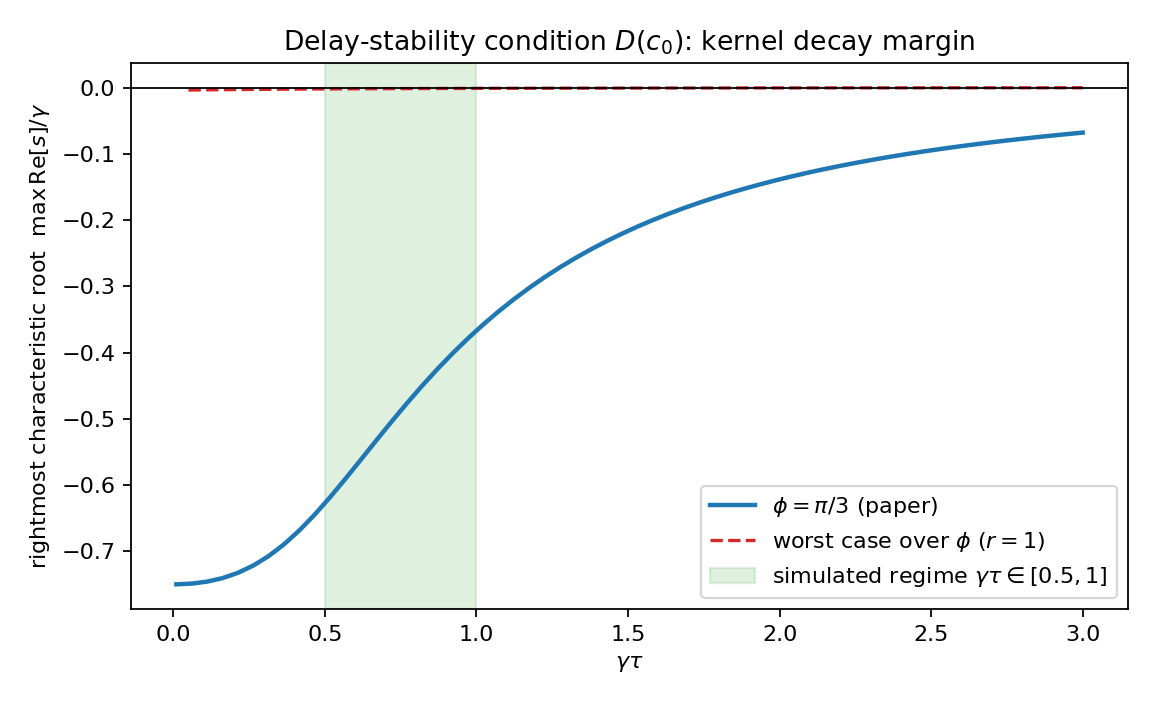}
    \caption{The delay-stability condition $D(c_0)$. Rightmost characteristic root of Eq.~\eqref{s:eq:charroot} versus $\gamma\tau$ at the simulated phase $\phi=\pi/3$ (solid) and in the worst case over $\phi$ at $r=1$ (dashed). The simulated regime $\gamma\tau\in[\NumAPF,1]$ (shaded) has margin $c_0\geq0.36\gamma$.}
    \label{s:fig:delayroots}
\end{figure}

\paragraph{Where $D(c_0)$ holds.} Solving Eq.~\eqref{s:eq:charroot} via the Lambert $W$ function, $D(c_0)$ fails only at the exact anti-node resonance $\phi=\pi\,(\mathrm{mod}\,2\pi)$ with $r=1$, where $s=0$ is a root: the dark-state configuration at which the atom decouples, the same measure-zero set excluded by the overlap conditions. Numerically (Supplementary Fig.~\ref{s:fig:delayroots}), at the simulated $\phi=\pi/3$ the margin is $c_0=0.62\gamma$ at $\gamma\tau=0.5$ and $c_0=0.36\gamma$ at $\gamma\tau=1.0$, and the worst case over all $\phi$ remains strictly stable for $\gamma\tau\leq3$. $D(c_0)$ is a stability requirement on the feedback loop, the natural quantum analogue of the echo-state condition of classical delay-line reservoir computing, and it replaces the mode-space gap in Theorem~1.

\subsection{Fading-memory property of the composed map}
\label{s:sec:fmp}
\begin{corollary}[FMP of the realized reservoir functional]
\label{s:cor:fmp}
Under condition $D(c_0)$, the composed input--output map $\hat y:\mathcal{K}_{u_{\max}}\to\mathbb{R}$ realized with matched weights satisfies, for all $u,v\in\mathcal{K}_{u_{\max}}$,
\begin{align}
    |\hat y(u)-\hat y(v)| \;\leq\; L \sup_{1\leq m\leq\lfloor T_P/T\rfloor} e^{-c_0(m-1)T/4}\,|u_{-m}-v_{-m}|,
\end{align}
with $L=L(u_{\max},N,C_W,C_h,c_0,T)<\infty$ uniform in $K\geq K_0$ for combs refined so that $T_P>MT$, where $C_W\equiv\max_{n\leq N}\sup_{t\in[0,T_\mathrm{off}]}|\widetilde W_n(t)|$ is a bound on the fixed matched weight functions. Hence $\hat y$ has the fading-memory property in the weighted-norm sense of the echo-state literature, with weight sequence $w_m=e^{-c_0(m-1)T/4}$, uniformly along the mode-number family, over every lag the matched weights use: those are supported on the last $M$ windows and $MT<T_P$ by hypothesis. Beyond the first revival the finite-$K$ kernel still decays, at the rate $\min_k\mathrm{Re}\,\lambda_k$, so the all-lag statement holds with the $K$-dependent weight $e^{-(\min_k\mathrm{Re}\,\lambda_k)(m-1)T}$; by Lemma~\ref{s:lem:trace} no $K$-uniform weight is available over all lags, and none is needed, since no step of the approximation bound uses fading memory.
\end{corollary}
\begin{remark}[Fading memory is the device's; truncation is the readout's]
\label{s:rem:fmp-vs-trunc}
Two distinct statements are easily conflated here, and we separate them. Corollary~\ref{s:cor:fmp} is an \emph{all-lag} statement: the weighted bound holds over every $m\geq1$, which is what the fading-memory property means in the echo-state sense, and it is a property of the \emph{device}---it follows from Lemma~\ref{s:lem:kerneldecay}, which bounds the physical kernel and knows nothing of any readout. The memory depth $M$, by contrast, belongs to the \emph{readout}: the matched weights are supported on the last $M$ on/off windows, and the resulting truncation is the approximation step quantified by term (ii) of Sec.~\ref{s:sec:errorbound}. The device does not ``have memory $M$''; it has fading memory at rate $c_0$, and the readout chooses to read $M$ lags of it. The sub-revival caveat attaches to the certification rather than to the property: it is over the horizon $MT$, kept below $T_P$ by comb refinement, that the uniform-in-$K$ constants of Lemma~\ref{s:lem:kerneldecay} are available.
\end{remark}
\begin{proof}
The output is a polynomial of degree $\leq N$ in the linear functionals $\ell_t(u)=\mathrm{Re}[v_\mathrm{geo}^*e^{-\mathbf{\Gamma}_g t}\zeta(u)]=\sum_{m\geq1}\mathrm{Re}[h_K(t+(m-1)T)\,c_{\mathrm{on}}]\,u_{-m}$ (with $c_\mathrm{on}$ the fixed on-window factor of Eq.~\eqref{s:eq:zetan}), whose coefficient at lag $m$ is bounded by $|c_\mathrm{on}|\,C_h e^{-c_0(m-1)T/2}$ by Lemma~\ref{s:lem:kerneldecay}, where $c_\mathrm{on}=\int_0^{T_\mathrm{on}}e^{-\mathbf{\Gamma}_g s}\dd s$ acting on $\alpha$ defines the fixed on-window injection factor entering $\zeta_1$ (Eq.~\eqref{s:eq:zetan}); $|c_\mathrm{on}|$ is a $K$-uniform constant by the same contractivity bound. On the bounded domain the polynomial is Lipschitz in each $\ell_t$ with constant set by $u_{\max}$, $N$, $C_W$; summing the geometric tail and splitting the decay factor gives the stated weighted-sup bound.  
\end{proof}

\section{Continuity of the Volterra Kernels in the Saturation Parameter}
\label{s:sec:saturation}

The main text states that the gap between the linear-transducer limit (in which Theorem~1 holds) and the fully saturable device (which the simulations run) is a quantified discrepancy of order the saturation parameter $s=(\varepsilon_\mathrm{INPUT}/\gamma_g)^2$. We record the perturbative statement here.

\begin{lemma}[Kernel continuity in $s$]
\label{s:lem:saturation}
Fix the reservoir parameters $(\mathbf{\Omega},\alpha,v_\mathrm{geo},\gamma,\tau,T_\mathrm{on},T_\mathrm{off})$ and bounded inputs $|u_k|\leq u_\mathrm{max}$. Let the full generator Eq.~\eqref{s:eq:LBDNfull} and the Gaussian-limit generator Eq.~\eqref{s:eq:LBDNMarkov} be run from a common initial state on the sector $\mathcal{E}(n_{\max})$ of Lemma~\ref{s:lem:sector}, driven by the same input word, over a comparison horizon of $M+1$ readout periods. Then the output $\hat y_k$ of the full generator has Volterra kernels at lags $m_i\leq M$ satisfying
\begin{align}
    \hat{h}^{\,\mathrm{full}}_n(m_1,\dots,m_n) = \hat{h}^{\,\mathrm{G}}_n(m_1,\dots,m_n) + O(s),
\end{align}
where $\hat{h}^{\,\mathrm{G}}_n$ are the kernels of the Gaussian-limit generator Eq.~\eqref{s:eq:LBDNMarkov}, with the explicit bound
\begin{align}
    \label{s:eq:satconst}
    \big|\hat h^{\,\mathrm{full}}_n(\vec m)-\hat h^{\,\mathrm{G}}_n(\vec m)\big|\;\leq\;|s|\;C_\mathrm{ext}(n,u_{\max})\,\Vert O\Vert_\infty\,C_Vg\sqrt{n_{\max}+1}\,(M{+}1)T ,
\end{align}
where $\Vert O\Vert_\infty$ bounds the readout observable on the sector and $C_\mathrm{ext}(n,u_{\max})$ is the coefficient-extraction constant of the input monomials. The constant depends on $u_\mathrm{max}$, $n$, $m_i$, the fixed reservoir parameters, the comparison horizon $(M{+}1)T$, and the sector size $n_{\max}$ of Lemma~\ref{s:lem:sector}; it is therefore a per-operating-point statement rather than a $K$-uniform one.
\end{lemma}
\begin{proof}
Work in the displaced frame of Lemma~\ref{s:lem:sector} and on the bounded-excitation sector $\mathcal{E}(n_{\max})$ there constructed, on which the physical trajectory is supported up to tails bounded by $e^{-n_{\max}/(2\beta_{\max}^2)}$, absorbed into the constant below. The sector is finite-dimensional ($K$ modes carrying at most $n_{\max}$ total excitations), so every operator on it is bounded; write the full generator there as $\mathcal{L}_\mathrm{full}=\mathcal{L}_\mathrm{G}+s\,\mathcal{V}$ with $\Vert\mathcal{V}\Vert\leq C_Vg\sqrt{n_{\max}+1}$ (Lemma~\ref{s:lem:sector}).

\emph{Analyticity of the one-period propagator.} The on/off drive is $T$-periodic, so the driven trajectory is a fixed point of the one-period propagator $\mathcal{P}_s$. Expanding $\mathcal{P}_s$ in a Dyson series in $s\mathcal{V}$ about the Gaussian propagation, the order-$n$ term is an $n$-fold time-ordered integral bounded in norm by $(|s|\,\Vert\mathcal{V}\Vert\,T)^n/n!$ times the uniformly bounded Gaussian segment norms, the piecewise-constant switching entering only through the segmentation of the integrals; the series converges absolutely for every $s$, and $\mathcal{P}_s$ is entire in $s$ on the sector.

\emph{Comparison of the two trajectories.} Write $\rho_\mathrm{full}(t)$ and $\rho_\mathrm{G}(t)$ for the states generated by $\mathcal{L}_\mathrm{full}$ and $\mathcal{L}_\mathrm{G}$ from the common initial state under the same input word, and $\mathcal{U}_\mathrm{full}(t,\tau')$ for the propagator of the full generator. Duhamel's formula is exact,
\begin{align}
    \rho_\mathrm{full}(t)-\rho_\mathrm{G}(t)\;=\;\int_0^{t}\mathcal{U}_\mathrm{full}(t,\tau')\,[s\mathcal{V}]\,\rho_\mathrm{G}(\tau')\,\dd\tau' .
\end{align}
Take trace norms. The propagator sits on the left of each product and is completely positive and trace preserving, so $\Vert\mathcal{U}_\mathrm{full}(t,\tau')\Vert_{1\to1}\leq1$ for every pair of times---a supremum over $\tau'$ rather than an average---while $\Vert\rho_\mathrm{G}(\tau')\Vert_1=1$ and $\Vert\mathcal{V}\Vert\leq C_Vg\sqrt{n_{\max}+1}$ by Lemma~\ref{s:lem:sector}. Hence
\begin{align}
    \label{s:eq:duhamel}
    \Vert\rho_\mathrm{full}(t)-\rho_\mathrm{G}(t)\Vert_1\;\leq\;|s|\,C_Vg\sqrt{n_{\max}+1}\;t ,
\end{align}
an inequality running in one direction only and containing no spectral quantity: contractivity of a quantum channel is available without a gap, and it is all that is used. The bound grows linearly in $t$, which is why the horizon is a hypothesis of the lemma rather than a remark upon it.

\emph{Passage to the kernels.} The output Eq.~\eqref{s:eq:output} is, for each fixed weight set, a bounded linear functional of the trajectory on the sector: writing $O$ for the readout observable and $\Vert O\Vert_\infty$ for its operator norm there, duality of the trace and operator norms gives $|\hat y_k(s)-\hat y_k(0)|\leq\Vert O\Vert_\infty\Vert\rho_\mathrm{full}-\rho_\mathrm{G}\Vert_1$, which Eq.~\eqref{s:eq:duhamel} bounds at $t=(M{+}1)T$. Both sides are polynomials in the input word, and the input monomials are linearly independent on $\mathcal{K}_{u_{\max}}$, a set with nonempty interior; the coefficient functionals are therefore bounded, with a constant $C_\mathrm{ext}(n,u_{\max})$ depending on the monomial degree and the input bound alone. Matching coefficients yields Eq.~\eqref{s:eq:satconst}. No spectral quantity enters at any step, and the constant depends on $K$ only through $n_{\max}$. This proves the lemma: the Volterra kernels move at most linearly in the saturation parameter, at a rate fixed by the input bound and the sector size alone. The Gaussian-limit kernels are therefore the leading term of a controlled expansion rather than a separate object.
\end{proof}
\paragraph{The invariant bounded-excitation sector.}
The proof above requires the residual superoperator $\mathcal{V}$ to be bounded, and it is bounded only on a sector of finitely many excitations. We supply that sector here.
\begin{lemma}[Invariant sector]
\label{s:lem:sector}
Let $\hat{N}=\sigma^+\sigma^-+\sum_k a_k^\dagger a_k$ and let $\mathcal{E}(n)$ denote the sector spanned by density operators supported on eigenspaces of $\hat{N}$ with eigenvalue $\leq n$. Work in the displaced frame $\varrho\mapsto D^\dagger(\beta_{\mathrm d}(t))\varrho D(\beta_{\mathrm d}(t))$ with $\beta_{\mathrm d}(t)=-\tfrac{i}{g}\alpha u(t)$, in which the input enters only through the displacement. The residual generator conserves $\hat{N}$ up to the dissipative channels, which are excitation-non-increasing; the frame displacement is uniformly bounded, $\sup_t\|\beta_{\mathrm d}(t)\|\leq u_{\max}/g$, while the coherent excitation of the state itself is carried by the driven trajectory, $\sup_t\|\zeta(t)\|\leq\zeta_{\max}=\eta u_{\max}\|\mathbf{\Gamma}_g^{-1}\alpha\|$; it is the latter that the sector must contain. Hence for inputs $|u_k|\leq u_{\max}$ the physical state remains, up to a tail bounded by the Chernoff estimate $\mathbb{P}(N\geq n)\leq e^{-\lambda}(e\lambda/n)^n$ at $\lambda=\zeta_{\max}^2$, in $\mathcal{E}(n_{\max})$ with $n_{\max}$ the least integer making that tail at most the target accuracy $\delta$ (equivalently $n_{\max}=\max\{\lceil 4\zeta_{\max}^2\rceil+1,\ \lceil\log_2\delta^{-1}\rceil\}$ suffices), and on $\mathcal{E}(n_{\max})$ the residual atom--field coupling satisfies $\|\mathcal{V}\|_{\mathcal{E}(n_{\max})}\leq C_V\, g\sqrt{n_{\max}+1}$. The earlier prescription $n_{\max}=\lceil4\zeta_{\max}^2\rceil+1$ is recovered whenever $\zeta_{\max}^2\gtrsim1$, where the exact Poisson tail at that $n_{\max}$ is already below $10^{-3}$; only in the weak-drive corner $\zeta_{\max}^2\lesssim1/2$ must $n_{\max}$ be raised, and then by one, costing a factor $\sqrt{4/3}$ in the bound on $\|\mathcal{V}\|$.  
\end{lemma}
\begin{proof}
Under the rotating-wave interaction of Eq.~\eqref{s:eq:LBDNfull}, $[\hat{H}_\mathrm{int},\hat{N}]=0$; the drive term is removed by the displacement; the Lindblad channels $v_\mathrm{geo}^*\cdot a$ and $\alpha^*\cdot a$ annihilate excitations. The displaced dynamics therefore maps $\mathcal{E}(n)$ into itself for every $n$. Undoing the displacement adds a coherent amplitude of at most $\beta_{\max}$ per mode direction, whose Poissonian number tails give the stated bound. On $\mathcal{E}(n_{\max})$, $\|a_k\|\leq\sqrt{n_{\max}}$ and $\|\sigma^\pm\|=1$, bounding $\mathcal{V}$.
\end{proof}

\begin{remark}[Status of the constant $\beta_{\max}$]
\label{s:rem:betamax}
The constant $\beta_{\max}=\eta u_{\max}\|\mathbf{\Gamma}_g^{-1}\alpha\|/g$ is an operating-point quantity with a physical name: it is the device's closed-loop gain, the steady displacement a bounded input can sustain against the loop's damping. Three properties fix its status. It is finite exactly when every dressed mode is damped ($\mathrm{Re}\,\lambda_k>0$ for all $k$), which the delay-stability and overlap conditions guarantee; at the dark-state configuration $\phi=\pi$ with unit feedback an undamped root appears, $\mathbf{\Gamma}_g^{-1}$ ceases to exist, and a bounded resonant input does grow the excitation without bound, which is the same excluded set as everywhere else in the paper and the configuration probed by the $\phi=\pi$ task-level floor of Sec.~\ref{s:sec:robustness}. A slowly damped, strongly driven mode makes $\beta_{\max}$ large rather than infinite: growth is transient over $\sim1/\mathrm{Re}\,\lambda_k$ and saturates at gain times input, which is what the sector bound then prices. Finally, $\beta_{\max}$ grows along the mode-number family as the slow modes' response grows (the trace identity closes the slowest rate as $1/K$, and a near-resonant component of the gain grows accordingly), so the sector bound is a per-operating-point statement, and no uniformity in $K$ is claimed or needed: the kernel-continuity lemma of this section is invoked at fixed operating points only.
\end{remark}
This is the sector on which Lemma~\ref{s:lem:saturation} applies, and the constant of Eq.~\eqref{s:eq:satconst} depends on $n_{\max}$, i.e.\ only on $u_{\max}$, the fixed reservoir parameters and the comparison horizon, as stated there.

\begin{remark}[Scope of the constant]
\label{s:rem:fixedK}
The constant of Eq.~\eqref{s:eq:satconst} is $K$-dependent: it enters through $n_{\max}$, which grows along the mode-number family with $\beta_{\max}$ (Remark~\ref{s:rem:betamax}). The estimate is therefore a per-operating-point statement and no uniformity in $K$ is claimed or needed; the uniform-in-$K$ content of the paper lives entirely in Lemma~\ref{s:lem:kerneldecay}, which is proven on the scalar kernel. The steady-trajectory version of the comparison is not established here, and the obstruction is structural rather than technical: the full generator Eq.~\eqref{s:eq:LBDNfull} carries a single dissipative channel, the $\gamma_g$ term of Eq.~\eqref{s:eq:LBDNMarkov} being produced by the adiabatic elimination rather than inherited from it, so the saturable device is less dissipative than its Gaussian limit and no fixed-$K$ contraction transfers to it. That extension belongs to the finite-$s$ problem stated in the Discussion.
\end{remark}

\begin{remark}
Lemma~\ref{s:lem:saturation} does not extend Theorem~1 to finite $s$ (that remains the open problem stated in the Discussion) but it changes the category of the theorem--device relation: the physical device's kernels lie within an explicitly controlled, continuously tunable distance of a family proven universal, rather than in a separate regime. For the simulations of the main text the operating values of $s$ are computable from the stated parameters. A direct numerical closure of this bound would extract the first- and second-order Volterra kernels of the simulated saturable device and overlay them on the closed-form Gaussian-limit kernels of Eq.~\eqref{s:eq:kernel}, checking agreement to $O(s)$ at the operating drive; that is the natural verification, and the mode-number convergence study of the main text (Fig.~2, saturable overlay) already exhibits the physical device converging along the Gaussian-limit axis, consistent with the continuity asserted here.
\end{remark}

\section{Genericity of the Universality Conditions}
\label{s:sec:genericity}

The proof requires that the spectrum of $\mathbf{\Gamma}_g$ satisfy the linear-independence (non-resonance) and overlap conditions. We show these hold for almost every mirror distance and verify them numerically for the simulated devices. We first record the equivalence used implicitly above.

The two non-resonance conditions are generic. The argument has two halves: the conditions are first recast as a $\mathbb{Q}$-independence statement, which is then shown to hold for almost every delay.

\begin{lemma}[Genericity of the non-resonance conditions]
\label{s:lem:Q}\label{s:lem:generic} 
\begin{enumerate}
\item[(a)]
For $K$ modes, the conditions $\sum_k\mathrm{Re}[\lambda_k]n_k\neq0$ and $\sum_k\mathrm{Im}[\lambda_k]n_k\neq0$ for all integer $\{n_k\}$ with $\sum_k n_k=0$, $\sum_k n_k^2>0$, are equivalent to linear independence over $\mathbb{Q}$ of $\{\mathrm{Re}[\lambda_k-\lambda_K]\}_{k<K}$ and of $\{\mathrm{Im}[\lambda_k-\lambda_K]\}_{k<K}$ respectively. Either one alone implies the complex non-degeneracy $\sum_k\lambda_k n_k\neq0$ used in Theorem~1 (if a real or imaginary part is nonzero the complex sum cannot vanish), so the genericity result below, proven for the real part and hence for the complex sums, delivers exactly what the theorem's hypothesis requires; the imaginary-part condition, which the theorem does not consume, is scoped separately in Remark~\ref{s:rem:impart}.
\item[(b)]
Regard the eigenvalues $\lambda_k(\tau)$ of $\mathbf{\Gamma}_g$ as functions of the delay $\tau$. Away from the isolated $\tau$ at which eigenvalues cross, each $\lambda_k(\tau)$ is real-analytic. For every integer tuple $\{n_k\}$ with $\sum_k n_k=0$, $\sum_k n_k^2>0$, the resonance function $R_{\{n\}}(\tau)=\sum_k n_k\lambda_k(\tau)$ is not identically zero, and neither is its real part $\sum_k n_k\,\mathrm{Re}\,\lambda_k(\tau)$. This holds for \emph{every} comb with pairwise-distinct frequencies---equidistant, chirped, or defected alike---with no disjointness condition on any frequency set. Consequently the set of $\tau$ violating the complex and real-part non-resonance conditions is a countable union of discrete sets and therefore has Lebesgue measure zero, and by Lemma~\ref{s:lem:Q} this alone discharges every non-resonance hypothesis of Theorem~1.
\end{enumerate}
\end{lemma}
\begin{proof}
\emph{(a)}
The constraint $\sum_k n_k=0$ lets one eliminate $n_K=-\sum_{k<K}n_k$, giving $\sum_{k<K}n_k(\lambda_k-\lambda_K)$; vanishing for some nonzero integer tuple is exactly $\mathbb{Q}$-linear dependence (integer and rational dependence coincide after clearing denominators).

\emph{(b)}
Consider the one-parameter family
\begin{align}
    \mathbf{\Gamma}_g(\tau,t)\;=\;i\mathbf{\Omega}\;+\;t\,\mathbf{E}(\tau),
    \quad
    \mathbf{E}(\tau)=\tfrac{\gamma}{2}\big(v_\mathrm{geo}\otimes v_\mathrm{geo}^*\big)+\tfrac{\gamma_g}{2}\big(\alpha(\tau)\otimes\alpha(\tau)^*\big),
    \quad t\in[0,1],
\end{align}
so that $t=1$ is the device and $t=0$ the bare comb; the ray carries the \emph{full} physical dressing, both channels at their physical rates. The entries are jointly analytic in $(\tau,t)$ wherever the normalization $D(\tau)=\sum_j\sin^2(\omega_j\tau/2)$ is positive, which excludes only a discrete set of delays. At $t=0$ the operator is diagonal with eigenvalues $i\omega_k$: pairwise distinct for any comb with distinct frequencies, independent of $\tau$, and simple, so each $\lambda_k(\tau,t)$ is jointly analytic near $t=0$ with no crossing analysis required at the expansion point, and first-order perturbation theory at a diagonal unperturbed operator \cite{Kato1995} returns exactly the diagonal entries of the perturbation:
\begin{align}
    \lambda_k(\tau,t)\;=\;i\omega_k\;+\;t\,E_{kk}(\tau)\;+\;O(t^2),
    \qquad
    E_{kk}(\tau)\;=\;\frac{\gamma}{2K}\;+\;\frac{\gamma_g}{2}\,|\alpha_k(\tau)|^2 ,
\end{align}
using the geometry-fixed flat coupler for the first term. For any admissible tuple the flat-coupler contribution cancels, $\sum_k n_k\,\gamma/2K=0$, leaving
\begin{align}
    \frac{\partial R_{\{n\}}}{\partial t}\Big|_{t=0}
    \;&=\;\frac{\gamma_g}{2}\sum_k n_k\,|\alpha_k(\tau)|^2
    \;=\;\frac{\gamma_g}{2}\,\frac{N(\tau)}{D(\tau)},\nonumber\\
    N(\tau)\;&=\;\sum_k n_k\sin^2\!\big(\tfrac{\omega_k\tau}{2}\big)\;=\;-\tfrac12\sum_k n_k\cos(\omega_k\tau),
\end{align}
where $\sum_k n_k=0$ removed the constant term. The frequencies $\omega_k$ being pairwise distinct, the functions $\cos(\omega_k\tau)$ are linearly independent as functions of $\tau$, so $N\not\equiv0$: the first-order coefficient is a not-identically-zero analytic function of $\tau$, for every comb with distinct frequencies and with no disjointness of any frequency set invoked anywhere. Joint analyticity and the identity theorem then give $R_{\{n\}}\not\equiv0$ on the connected domain, so $R_{\{n\}}(\cdot,t)\not\equiv0$ for every ray parameter $t$ outside a discrete exceptional set, and away from that set the zero set of $R_{\{n\}}(\cdot,t)$ in $\tau$ is discrete. The union over the countably many integer tuples, together with the countable crossing set at $t=1$, is measure zero, and the countably many finite $K$ contribute countably many such sets. Finally, the first-order coefficient is \emph{real}, so the identical argument applies verbatim to $\sum_k n_k\,\mathrm{Re}\,\lambda_k(\tau,t)$: real-part genericity is delivered outright, and with it, by Lemma~\ref{s:lem:Q}, the complex non-degeneracy Theorem~1 consumes. This proves the lemma: the non-resonance conditions fail only on a set of mirror distances of measure zero, so a device drawn at a generic geometry satisfies them, and Theorem~1 may consume them as a hypothesis without excluding any but an exceptional set of devices.
\end{proof}

\begin{remark}[Scope of the imaginary-part condition]
\label{s:rem:impart}
We first locate where the condition is consumed. Writing a conjugation-resolved sum point over a multiset $\mathbf m$ with pattern $p$ as $\Lambda(p)=\sum_k m_k\,\mathrm{Re}\,\lambda_k+i\sum_k(2p_k-m_k)\,\mathrm{Im}\,\lambda_k$, the real part depends on $\mathbf m$ alone; two conjugation patterns on a common multiset therefore have identically equal real parts and are separated only by $\sum_k b_k\,\mathrm{Im}\,\lambda_k$ with $b=p-p'$, $|b|_1\leq n$. This is exactly the separation Lemma~\ref{s:lem:distinct}(a) supplies at the designed point, through the graded defects and with no sum-zero restriction on $b$; it is what the defect hierarchy is for. Away from the design, at first order along the ray the imaginary part of $R_{\{n\}}$ is $\sum_k n_k\omega_k+O(t^2)$, which on the \emph{equidistant} comb vanishes identically for every tuple with $\sum_k n_k\,k=0$: the bare equidistant comb does not satisfy imaginary-part genericity at leading order, a further instance of the regularity-versus-selection theme of Sec.~S.2.5. The imaginary-part condition therefore requires either higher-order terms or a comb whose frequencies avoid the cross-frequency coincidences, and a generic chirp restores it at leading order: for $\omega_k=\omega_0+\Delta_0k+\Delta_2k^2$, a coincidence $\tfrac12(\omega_j+\omega_{j'})=\omega_m$ requires both $(j+j')/2=m$ and $(j^2+j'^2)/2=m^2$, impossible for $j\neq j'$ by strict convexity of the square, so the collisions occupy finitely many $\Delta_2$ values at each $K$ and almost every chirp avoids them all. None of this is consumed by Theorem~1, whose hypotheses need only the complex distinctness that the real part alone supplies (Lemma~\ref{s:lem:Q}); the numerical verification below checks the operating combs directly.
\end{remark}
\begin{remark}[Why functional rather than pointwise, independence]
A tempting shortcut would infer pointwise $\mathbb{Q}$-linear independence of the perturbative coefficients from their distinctness; this is false in general (distinctness does not imply $\mathbb{Q}$-independence, and for an equidistant comb the relevant consecutive differences are in fact rationally dependent). The proof above therefore requires no arithmetic condition at any point: non-vanishing \emph{as a function of $\tau$}, supplied by the linear independence of cosines at distinct frequencies, is all the identity theorem needs. The bare equidistant comb is itself the cautionary case twice over: its arithmetic regularity defeats not only the pointwise shortcut but the leading-order imaginary-part expansion (Remark~\ref{s:rem:impart}), which is why the argument is arranged, through the diagonal expansion point, to need neither.
\end{remark}

\begin{remark}
Lemma~\ref{s:lem:generic} is what a naive cardinality argument misses: the device spectrum is not a free point in $\mathbb{R}^K$ but an analytic curve $\tau\mapsto\{\lambda_k(\tau)\}$, and genericity along that curve is what must be, and is, established. The bare equidistant comb $\mathbf{\Omega}=\mathrm{diag}(\omega_0+k\Delta)$ is maximally resonant on the imaginary axis; the dressing by $\tfrac{\gamma}{2}(v_\mathrm{geo}\otimes v_\mathrm{geo}^*)$ and $\tfrac{\gamma_g}{2}(\alpha\otimes\alpha^*)$ lifts the degeneracy of the complex sums for almost every $\tau$ through the real parts, per the proof above, while the imaginary-axis coincidences themselves persist at leading order on the equidistant comb (Remark~\ref{s:rem:impart}).
\end{remark}

\paragraph{Numerical verification.}
This section concerns generically fabricated devices, which do not satisfy condition D3 and are the devices the simulations of this paper use. It carries no hypothesis of Theorem~1: every condition of the theorem is discharged at the designed operating point by Lemmas~\ref{s:lem:profile}, \ref{s:lem:radial}, \ref{s:lem:distinct} and \ref{s:lem:strictdiss}. We diagonalized the exact $\mathbf{\Gamma}_g$ over a dense sweep of $\tau$ for the mode numbers used in the main text. Across the swept range the eigenvalues are distinct with $\mathrm{Re}[\lambda_k]\geq\varepsilon_D>0$; the overlaps $|v_\mathrm{geo}^*\cdot v_{R,k}|$ and $|v_{L,k}^*\cdot\alpha|$ are bounded away from zero except at the isolated node distances where $\sin(\omega_k\tau/2)=0$; and the finite-order non-resonance surrogate $\min_{\{n\}:\sum n_k=0,\,\sum|n_k|\leq6}|\sum_k n_k\lambda_k(\tau)|$ stays strictly positive away from a discrete set of $\tau$. The parameter values used for every figure in the main text and this Supplement satisfy all conditions, so those simulated reservoirs satisfy the overlap and non-resonance conditions in the linear-transducer limit; they do not satisfy the weak-dressing entry of D1, and the explicit constants that entry supplies are correspondingly unavailable at these operating points. In addition, the delay-stability condition $D(c_0)$ of Lemma~\ref{s:lem:kerneldecay} is verified directly on the characteristic roots of Eq.~\eqref{s:eq:charroot}: at the simulated phase $\phi=\pi/3$ the margin is $c_0=0.62\gamma$ ($\gamma\tau=0.5$) and $c_0=0.36\gamma$ ($\gamma\tau=1.0$), and the worst case over all phases remains strictly stable for $\gamma\tau\leq3$ (Supplementary Fig.~\ref{s:fig:delayroots}).

\section{Shot-Noise-Corrected Moment Readout}
\label{s:sec:shotnoise}

\paragraph{From the intracavity quadrature to the detected field.}
The proof is written in terms of $q_\mathrm{geo}=\mathrm{Re}[v_\mathrm{geo}^*\cdot a]$, a quadrature of the \emph{standing-wave} mode combination selected by the measured channel, whereas the laboratory observable is the \emph{traveling} field at the detector. The two are related by the input--output boundary condition on the open end of the waveguide. The mirror terminates one side and supports no outgoing channel, so the detected continuum is the unidirectional field past the atom, and for the measured channel $v_\mathrm{geo}$ the standard relation reads
\begin{align}
    \label{s:eq:inout}
    \hat b_\mathrm{out}(t) \;=\; \hat b_\mathrm{in}(t) \;+\; \sqrt{\gamma}\;\big(v_\mathrm{geo}^*\cdot a(t)\big),
\end{align}
with $\hat b_\mathrm{in}$ the vacuum input to that channel, $[\hat b_\mathrm{in}(t),\hat b_\mathrm{in}^\dagger(t')]=\delta(t-t')$. Balanced homodyne detection of $\hat b_\mathrm{out}$ at local-oscillator phase $\theta$ measures $\hat b_\mathrm{out}e^{-i\theta}+\hat b_\mathrm{out}^\dagger e^{i\theta}$, i.e.\ $2\sqrt{\gamma}\,q_\mathrm{geo}$ plus the vacuum contribution of $\hat b_\mathrm{in}$. Equation~\eqref{s:eq:inout} is what licenses reading the binned outcomes $I_j$ below as estimates of the windowed $q_\mathrm{geo}$: the signal term is the intracavity quadrature the proof uses, scaled by $\sqrt{\gamma}$, and the $\hat b_\mathrm{in}$ term integrates to the noise contribution $\xi_j$. We stress that Eq.~\eqref{s:eq:inout} is an \emph{operator} identity and that its two terms are correlated: the same vacuum input $\hat b_\mathrm{in}$ that appears explicitly also drives $a(t)$ through the Langevin equation of the measured channel, so the fluctuation part of the windowed quadrature and the integrated shot noise of the same bin are not independent. The statistics of the binned outcomes are therefore not asserted here; they are \emph{derived} in Lemma~\ref{s:lem:shotnoise}, where the measurement back-action is accounted for exactly and the measured output field is shown to be a displaced vacuum in the linear-transducer limit. Because $\hat b_\mathrm{in}$ acts on an incoming continuum that never returns to the atom (the mirror side carries no outgoing channel) Eq.~\eqref{s:eq:inout} introduces no additional feedback and no non-Markovian structure beyond that already carried by $(\mathbf{\Omega},\alpha)$.  

\paragraph{Detection loss is not feedback loss.}
Two distinct efficiencies appear in this device and should not be conflated. The feedback transmissivity $\eta$ swept in Sec.~\ref{s:sec:robustness} attenuates the \emph{returning} field and therefore alters the dynamics---it changes $r=\gamma_R/\gamma$ in the characteristic equation Eq.~\eqref{s:eq:charroot} and with it the stability margin $c_0$, which is why severing it ($\eta=0$) collapses the device to the feedback-free plateau. Detector inefficiency $\eta_\mathrm{det}<1$, by contrast, acts only on the \emph{outgoing} channel Eq.~\eqref{s:eq:inout} after the physics has happened: it admixes vacuum, $\hat b_\mathrm{det}=\sqrt{\eta_\mathrm{det}}\,\hat b_\mathrm{out}+\sqrt{1-\eta_\mathrm{det}}\,\hat v$, leaving the generator Eq.~\eqref{s:eq:LBDNMarkov}, the kernel $h_K$, and the rate $c_0$ untouched. Because a beamsplitter with a vacuum ancilla is itself a passive map, it carries the displaced vacuum of Lemma~\ref{s:lem:shotnoise} to a displaced vacuum---the mean rescaled by $\sqrt{\eta_\mathrm{det}}$, the fluctuations still exactly vacuum---so the binned outcome becomes $I_j=\sqrt{\eta_\mathrm{det}\gamma}\,\bar q_j+\xi_j$ with $\mathrm{Var}[\xi_j]=v_0$ unchanged, and the triangular relation of Lemma~\ref{s:lem:shotnoise} holds verbatim with $\gamma\mapsto\eta_\mathrm{det}\gamma$: the ideal moments are still recovered exactly in expectation. The cost is entirely in the shot budget: the effective signal-to-noise scale $\bar v=v_0+q_{\max}^2$ becomes $v_0+\eta_\mathrm{det}q_{\max}^2$, so by Lemma~\ref{s:lem:shotbudget} the shot number for a target precision at order $n$ grows by at most $\eta_\mathrm{det}^{-n}$---a benign factor at the $n=1$ readout used for every real-world benchmark, and the reason imperfect detection degrades the \emph{statistics} of this device rather than its \emph{dynamics}.  

\paragraph{Shot-noise-corrupted moments.}
The proof uses ideal moments $\langle q_\mathrm{geo}^n\rangle$; experimentally one has the balanced-homodyne photocurrent, which in continuous measurement contains white shot noise. Pointwise powers of the raw current are therefore not well-defined objects (they would involve powers of white noise) so all moment estimation below is built on \emph{time-binned} outcomes: for each readout sample time $t_j$ the current is first integrated against a square-integrable window $w_j$ matched to the discretized weight grid,  
\begin{align}
    I_j \;=\; \int w_j(t)\, dI(t),
\end{align}
with the windows $w_j$ square-integrable and pairwise orthogonal (in practice, non-overlapping bins). Two quantities must be kept distinct in what follows. The \emph{windowed quadrature} is the operator $\int w_j(t)\,q_\mathrm{geo}(t)\,\dd t$, whose fluctuation part is correlated with the same-bin shot noise, as noted above. The \emph{windowed amplitude} $\bar q_j$ is the c-number window average of $\mathrm{Re}[v_\mathrm{geo}^*\cdot\zeta(t)]$ on the coherent driven trajectory of Sec.~\ref{s:sec:proof}---a deterministic functional of the input history, and the quantity in terms of which the statistics of $I_j$ are stated below. Powers and empirical moments are taken of the binned outcomes $I_j$, one per window per experimental repetition; with this definition in place, the statements below are well-posed (a loose phrasing in terms of ``powers of each shot'' would not be, precisely because of the white-noise issue), and $v_0\propto\|w_j\|^2$ denotes the vacuum variance of a binned homodyne outcome.
The readout is a moment estimator, and its two costs---what back-action does to the estimate, and how many shots the estimate needs---are one statement.

\begin{lemma}[Moment readout and shot budget]
\label{s:lem:shotnoise}\label{s:lem:shotbudget}  
\begin{enumerate}
\item[(a)]
In the linear-transducer (Gaussian) limit, for pairwise orthogonal windows, and provided readout begins after a washout interval long compared with $1/c_0$---so that the state has relaxed to the coherent driven trajectory of Sec.~\ref{s:sec:proof} and any correlations carried by the initial state have decayed---the binned outcomes $I_j$ are jointly Gaussian with mean $\sqrt{\gamma}\,\bar q_j$ and covariance exactly $v_0\,\delta_{jj'}$, independently of the drive. Consequently the empirical moments of $I$ are linear combinations of the ideal moments $\langle q_\mathrm{geo}^j\rangle$, $j\leq n$, with fixed, known coefficients; any output linear in $\{\langle q_\mathrm{geo}^j\rangle\}$ is realized by a linear reprocessing of $\{\langle I^j\rangle\}$, and the required weights are obtained by an invertible relabeling.  
\item[(b)]
Let $\hat\mu_n$ denote the empirical $n$-th moment of the homodyne outcome over $N_\mathrm{shots}$ repetitions, with vacuum-noise variance $v_0$ and signal bounded by $q_{\max}$. Then
\begin{align}
    \mathrm{Var}[\hat\mu_n] \;\leq\; \frac{(2n-1)!!\,(v_0+q_{\max}^2)^{n}}{N_\mathrm{shots}} ,
\end{align}
so precision $\sigma$ on the order-$n$ moment requires $N_\mathrm{shots}\geq(2n-1)!!\,(v_0+q_{\max}^2)^n/\sigma^2$. Writing $\bar v\equiv v_0+q_{\max}^2$ for this second-moment scale, the requirement reads $N_\mathrm{shots}\geq(2n-1)!!\,\bar v^{\,n}/\sigma^2$. Composing this with the approximation bound is one line, and we record it because the two are otherwise stated separately. Writing $\hat y^{(R)}_k$ for the $R$-shot estimate of the output Eq.~\eqref{s:eq:output} and $\Vert W_n\Vert_1=\int_0^{T_\mathrm{off}}|W_n(t)|\dd t$, linearity gives $|\hat y^{(R)}_k-\hat y_k|\leq\sum_n\Vert W_n\Vert_1\sigma_n$, so an end-to-end accuracy $\epsilon$ requires  
\begin{align}
    R\;\geq\;\max_{0\leq n\leq N}\frac{4(N{+}1)^2\,(2n-1)!!\;\bar v^{\,n}\,\Vert W_n\Vert_1^{2}}{\epsilon^{2}} .
\end{align}
The weight norm does not enter the approximation bound, which is the content of Proposition~\ref{s:prop:extrapolation}; it does enter the sample complexity, quadratically. The reading is that the theorem's guarantee is on expectation values, and that the cost of realising it in finite time is set by the size of the weights that realise it, which at the designed operating point is governed by separations that are exponentially small in $M$. Bounding $\Vert W_n\Vert_1$ at a generic operating point is not attempted here. For the Gaussian-limit readout at order $N$ the shot cost thus grows as $(2N-1)!!\,v_0^{N}$; for the saturable device with a linear readout ($N{=}1$) it is $\sim v_0/\sigma^2$. The nonlinearity transfer of the main text is therefore not merely an apparatus simplification but a super-exponential reduction of the measurement budget.
\end{enumerate}
\end{lemma}
\begin{proof}
\emph{(a)}
\emph{Statistics of the binned outcomes.} The Gaussian-limit generator Eq.~\eqref{s:eq:LBDNMarkov} is passive---quadratic with no anomalous ($a^\dagger a^\dagger$) terms---and stable ($\mathrm{Re}\,\lambda_k\geq\gamma/18K>0$, Lemma~\ref{s:lem:strictdiss}), and its only dissipation channels are the readout channel $\sqrt{\gamma}\,v_\mathrm{geo}^*\cdot a$ and the drive channel $\sqrt{\gamma_g}\,\alpha^*\cdot a$, both fed by vacuum; the input enters solely as the c-number displacement. The input--output map is therefore linear scattering: in frequency, $\hat b_\mathrm{out}(\omega)=S(\omega)\,\hat b_\mathrm{in}(\omega)+d(\omega)$ with the $2\times2$ scattering matrix $S(\omega)=I-C\,(\mathbf{\Gamma}_g-i\omega)^{-1}C^\dagger$, $C=(\sqrt{\gamma}\,v_\mathrm{geo}^*;\ \sqrt{\gamma_g}\,\alpha^*)$, and $d(\omega)$ the c-number displacement driven by $u$. For a passive stable system with all dissipation channels listed, the scattering matrix is $\Sigma(\omega)=I-CG(\omega)C^\dagger$ with $G(\omega)=(\mathbf{\Gamma}_g-i\omega)^{-1}$ and $C$ the $2\times K$ coupling matrix with rows $\sqrt{\gamma}v_\mathrm{geo}^*$ and $\sqrt{\gamma_g}\alpha^*$, so that $\tfrac12C^\dagger C=\mathbf{E}$ and $\mathbf{\Gamma}_g+\mathbf{\Gamma}_g^\dagger=C^\dagger C$. Then $G^\dagger C^\dagger CG=G^\dagger[(\mathbf{\Gamma}_g-i\omega)+(\mathbf{\Gamma}_g^\dagger+i\omega)]G=G^\dagger+G$, whence $\Sigma^\dagger\Sigma=I-CG^\dagger C^\dagger-CGC^\dagger+C(G^\dagger+G)C^\dagger=I$ at every real $\omega$, the resolvent existing there by Lemma~\ref{s:lem:strictdiss}. The identity is exact. Unitarity of the measured row means that the measured output channel, fed by vacuum on both inputs, is again an exactly $\delta$-correlated vacuum field, displaced by $d$: the output state of the measured channel is a displaced vacuum. One hypothesis of the lemma enters exactly here: the frequency-domain scattering relation is a stationary statement, valid when the field is initially in vacuum or, equivalently for the readout, once the transient of the stable dynamics has decayed and the state has reached the coherent driven trajectory. At transient times the correlations carried by the initial state survive, and binned outcomes in different windows would in general remain correlated; this is why readout is taken to begin after the washout interval of the lemma's statement---a period standard in reservoir-computing practice, and included in every simulation of this paper (Sec.~\ref{s:sec:robustness}). With the transient washed out, integrating against orthogonal windows gives the first statement: $I_j$ jointly Gaussian, mean $\sqrt{\gamma}\,\bar q_j$, covariance $v_0\,\delta_{jj'}$, with $v_0$ the vacuum value regardless of the drive.  

\emph{Where the back-action went.} The correlation flagged before Eq.~\eqref{s:eq:inout} is real: writing the windowed quadrature as $\bar q_j+\delta\bar q_j$ with $\delta\bar q_j$ the operator fluctuation, $\mathrm{Var}[I_j]=\gamma\,\mathrm{Var}[\delta\bar q_j]+v_0+2\sqrt{\gamma}\,\mathrm{Cov}[\delta\bar q_j,\xi_j]$, and the first step fixes the total at $v_0$: the back-action cross term is negative and cancels the intracavity-fluctuation excess exactly. This is the content of vacuum-in/vacuum-out for a passive network rather than an accident, and it is why no independence assumption between signal and noise appears anywhere in this proof.

\emph{Moment recovery.} The measured moments $\langle I^n\rangle$ are the moments of a Gaussian with mean $\sqrt{\gamma}\,\bar q$ and variance $v_0$: $\langle I^n\rangle=\sum_{j=0}^n\binom{n}{j}\gamma^{j/2}\bar q^{\,j}\mu_{n-j}$ with $\mu_r$ the central Gaussian moments of variance $v_0$ ($\mu_0=1$, $\mu_1=0$, $\mu_2=v_0$, higher $\mu_r$ by Isserlis). This is lower-triangular with unit diagonal in the pair $(\langle I^n\rangle,\gamma^{n/2}\bar q^{\,n})$, hence invertible, recovering $\bar q$ and its powers exactly in expectation. The ideal moments $\langle q_\mathrm{geo}^n\rangle$ are, in the coherent state of the driven trajectory, the moments of a Gaussian with mean $\bar q$ and the state-independent vacuum quadrature variance fixed by the commutator; they are therefore obtained from the powers of $\bar q$ by a second triangular, unit-diagonal relation with fixed coefficients. Composing the two, the ideal moments are linear combinations of the measured ones with fixed, known coefficients, and these absorb into the readout weights precisely as the normal-ordering coefficients did in Eq.~\eqref{s:eq:return}. No apparatus beyond the homodyne detector is required. This proves the lemma: every ideal moment the construction needs is recovered exactly in expectation from the homodyne record, and the price of doing so is paid entirely in shot count rather than in bias.  
\end{proof}

\begin{remark}[Scope of the moment inversion; readout order for the saturable device]
\label{s:rem:momentscope}
The displaced-vacuum argument above is a statement about the linear-transducer limit and is used only there. For the fully saturable device the output field is not Gaussian, and the signal--noise correlation of the second proof step does not cancel against a vacuum total; no moment-inversion claim is made for it at order $n\geq2$. None is needed: every real-world benchmark in this paper operates the saturable device with a purely linear readout, $n=1$ (Table~\ref{s:tab:shots}), where linearity of expectation gives $\langle I_j\rangle=\sqrt{\gamma}\,\langle\bar q_j+\delta\bar q_j\rangle$ exactly (an identity that holds under any signal--noise correlation whatsoever) and the sole $n=2$ usage, the Gaussian-limit convergence experiment, is run where the lemma applies. Every use of measured moments in this paper is therefore inside the lemma's stated scope.
\end{remark}

Unbiasedness alone does not establish that finite-shot estimates suffice; the following bound does, and quantifies the measurement cost of the moment readout.
Applied to the experiments of this paper, the lemma yields the measurement budgets of Table~\ref{s:tab:shots}: every real-world benchmark operates the saturable device with a purely linear readout ($n=1$), where the budget is the bare homodyne cost, while only the Gaussian-limit convergence experiment uses $n=2$; a hypothetical Gaussian-limit implementation of the benchmark nonlinearity at $n=9$ would pay a $(2\cdot9-1)!!\approx3.4\times10^7$-fold factorial penalty. We stress that this number is \emph{comparator-relative}: $n=9$ is the readout order the Markovian atom--cavity comparator of main-text Fig.~3 requires on these tasks rather than an intrinsic scale of the problem, and the penalty should be read as the cost of matching that comparator in the Gaussian limit rather than as a property of the tasks themselves. In laboratory terms, at relative precision $\varepsilon_\mathrm{rel}$ on the second moment (i.e.\ $\sigma=\varepsilon_\mathrm{rel}\bar v$) the requirement is $N_\mathrm{shots}\geq3/\varepsilon_\mathrm{rel}^{2}$, independent of $\bar v$: $3\times10^{4}$ repetitions at one-percent precision, or about $30$\,ms of wall clock per output sample at the $\sim\mu$s symbol clock of the main text.
\begin{table}[h]
\centering
\small
\setlength{\tabcolsep}{4pt}
\renewcommand{\arraystretch}{1.2}
\begin{tabular}{lccc}
\hline
Experiment & order $n$ & $N_\mathrm{shots}$ & penalty vs.\ $n{=}1$\\
\hline
Real-world benchmarks (saturable, linear) & 1 & $\bar v/\sigma^2$ & $1$\\
Convergence experiment (Gaussian limit) & 2 & $3\bar v^2/\sigma^2$ & $3\bar v$\\
Ninth-order comparator (Fig.~3), if Gaussian & 9 & $17!!\,\bar v^9/\sigma^2$ & $\sim3.4\times10^{7}\bar v^{8}$\\
\hline
\end{tabular}
\caption{Shot budgets by experiment (Lemma~\ref{s:lem:shotbudget}), from $N_\mathrm{shots}\geq(2n-1)!!\,\bar v^{\,n}/\sigma^2$ with $\bar v\equiv v_0+q_{\max}^2$ the second-moment scale and $\sigma$ the target precision. The saturable device's nonlinearity transfer keeps all real-world tasks at the $n=1$ cost.}
\label{s:tab:shots}
\end{table}
\begin{proof}
$\mathrm{Var}[\hat\mu_n]=\mathrm{Var}[I^n]/N_\mathrm{shots}\leq\mathbb{E}[I^{2n}]/N_\mathrm{shots}$; conditioning on the bounded signal and applying Isserlis' theorem to the centered Gaussian part, $\mathbb{E}[I^{2n}]\leq(2n-1)!!\,(v_0+q_{\max}^2)^n$ using $\mathbb{E}[\xi^{2m}]=(2m-1)!!\,v_0^m$ and the binomial expansion.
\end{proof}

\section{Prior Constructions Compared}
\label{s:sec:incumbents}
This table, referenced from the main text, locates the present result against prior reservoir-computing constructions.

\begin{table}[h]
\centering
\small
\setlength{\tabcolsep}{4pt}
\renewcommand{\arraystretch}{1.25}
\caption{Reservoir-computing constructions compared on the properties this work establishes. ``Physical reservoir'' asks whether the reservoir is realized as hardware dynamics (with its active component count) or computed digitally. ``Single-member universal?'' asks whether one fixed system carries a universal-approximation guarantee: ``No (class)'' means the guarantee attaches only to a class ranged over, never to an individual member. ``Strict scaling'' asks whether increasing a single physical resource of one device is proven never to reduce its approximation capability, with the span of exactly matchable kernels strictly enlarged; a ``No'' records that no such statement has been made for that construction rather than that it fails---for classical network classes containment across sizes is trivially available by zero-padding weights, and what is proven here is containment for a non-nested spectral family, which zero-padding does not supply. The universality entry for this work holds under the constructive conditions of the main text's universality theorem (Supplement Sec.~S.2.5), all proven satisfiable with explicit constants; delay stability of the feedback loop, verified with margin $c_0\geq0.36\gamma$ throughout the simulated regime, is separately the condition for the device's fading memory. The ``No'' entry for the single-node delay-line reservoir reflects a literature search across the delay-reservoir theory line: capacity analyses (including the master memory function) quantify memory rather than universal approximation \cite{grigoryeva2014tdr, GRIGORYEVA2015capacity, koster2021master}; the unified single-neuron framework of Ort\'in, Pesquera and co-workers shows the delay node \emph{implements} echo-state and extreme-learning schemes \cite{ortin2015unified}, whose universality is class-level \cite{GRIGORYEVA2018495, Gonon:2020}, so the equivalence transfers a class guarantee rather than a single-member one; and folded-in-time architectures inherit the guarantees of the network they emulate \cite{stelzer2021deep}. For the classical network classes, capability containment across sizes is trivially available by zero-padding weights; the ``No'' entries record that no such statement is proven in the cited works, and the nontrivial content of the present entry is containment for a family whose spectra are not nested, where zero-padding is unavailable (Sec.~\ref{s:sec:errorbound}). The search underlying the delay-line entries is current at the time of submission and includes the folded-in-time and deep delay-architecture line. The single-member and strict-scaling entries for this work hold on the operating class of Definition~\ref{s:def:opclass}, whose conditions a fabricated device can be checked against; the constructive certificate of Sec.~\ref{s:sec:design} supplies them by design and additionally specifies the resonator spectrum.}  
\label{s:tab:incumbents}
\begin{tabularx}{\linewidth}{@{}>{\raggedright\arraybackslash}X >{\raggedright\arraybackslash}X c c@{}}
\hline
Construction & Physical reservoir (components) & \shortstack{Single-member\\universal?} & \shortstack{Strict\\scaling?}\\
\hline
Echo-state network classes \cite{GRIGORYEVA2018495, GONON202110} & Realization ($O(\text{nodes})$) & No (class) & No\\
State-affine system classes \cite{JMLR:v19:18-020} & Abstract ($O(\text{dimension})$) & No (class) & No\\
Ising / spin-ensemble quantum reservoir classes \cite{chen2019learning, chen2020temporal} & Spin ensemble ($O(\text{spins})$) & No (class) & No\\
Gaussian quantum-optics classes \cite{nokkala2021gaussian} & Optical network ($O(\text{modes})$) & No (class) & No\\
Boyd--Chua / Volterra density \cite{Boyd:1985} & Abstract (functional class) & N/A (existence only) & No\\
Signature / next-generation RC \cite{Cuchiero, gauthier2021next} & No (digital feature map) & Yes & No\\
Single-node delay-line RC \cite{appeltant2011, larger2017high} & \textbf{Yes} (1 node + 1 delay loop) & No & No\\
Dissipative quantum systems~\cite{chen2019learning} & Yes & No (class of systems) & No\\
\textbf{This work} & \textbf{Yes} (1 atom, 1 mirror, 1 detector, 1 fixed band envelope) & \textbf{Yes} & \textbf{Yes}\\
\hline
\end{tabularx}
\end{table}

\section{Benchmark Tasks}
\label{s:sec:benchmarks}

We report no tuned, noise-matched comparison against an echo-state network; the echo-state figures that appear in the financial study below are a noiseless classical reference rather than a matched baseline, and are labelled as such. A matched comparison requires two things this manuscript does not supply: injection into the classical reservoir state of the shot noise implied by the device's finite measurement budget, and independent hyperparameter selection for both families at each noise level. Without both, a noiseless and separately tuned classical reservoir is not a matched baseline, and a comparison against one is uninformative in either direction. The comparators we do report are matched by construction: the atom--cavity comparator and the Markovian-limit ablation run through the same generator, the same readout and the same measurement statistics as the device, and they are what the nonlinearity-transfer claim rests on. The question a matched classical comparison would answer is also not the one this paper asks: the claim is minimality of the physical architecture rather than computational advantage, and the evidentiary standard for the real-world tasks is parity with standard classical methods.

\paragraph{Financial forecasting study.}
The financial study is reported here rather than in the main text: its one-step-ahead structure on normalized prices is closely tracked by a persistence predictor, which makes it a weak discriminator between methods; we therefore treat it as a consistency check. We used daily closing prices for Apple, the S\&P~500, and NASDAQ from Yahoo Finance over 1~January 2014 to 1~January 2024, normalized before injection and evaluated with a rolling two-year-train / one-year-test protocol. With ten input masks the reservoir achieved NRMSE values of $0.075\pm0.058$, $0.059\pm0.025$, and $0.079\pm0.070$ (mean $\pm$ SD over rolling windows) for the S\&P~500, Apple, and NASDAQ, against an echo-state network at $0.076\pm0.066$, $0.077\pm0.048$, and $0.072\pm0.043$; the differences lie well within one standard deviation, and we make no formal equivalence claim.
Under the range normalization used throughout, the persistence baseline $\hat{y}_t=y_{t-1}$ scores $\approx0.037$ on these identical windows (all three series; computed on the stated protocol by the deposited script \texttt{b6\_persistence\_check.py}), \emph{below} both models: one-step-ahead prediction of slowly varying normalized prices is dominated by persistence, which is precisely why this task is retained as a consistency check between methods rather than as a benchmark, and no claim of beating persistence is made. Varying the delay over $\tau=5,10,15,20$ gave $0.042\pm0.008$, $0.047\pm0.011$, $0.075\pm0.058$, and $0.101\pm0.102$, indicating that a moderate memory length is preferable and that excessively long feedback retains stale information.  

\paragraph{Trivial baselines.}
For the tumor task of the main text the $48$-sample FFPE set is near-evenly split between tumor and normal, so a majority-class classifier scores near chance---well below the reported $87.5\%$ ($42/48$). The feature subsets were selected on the full sample by the classical pipeline of the main text's Methods (a protocol optimized for, and shared with, the LDA baseline), so the selection step is common to both classifiers and the comparison between them is on equal footing; the Wilson 95\% interval on $42/48$ is $[75.3\%,94.1\%]$. Because that subset search is performed on the full sample, the absolute accuracies of both classifiers, and the interval, carry selection bias and should not be read as generalization estimates; the reservoir--LDA \emph{comparison}, whose selection step is shared, is the valid object. The paired comparison is moreover exact without any per-sample data: both classifiers scoring $42$ of $48$ forces the discordant pairs into balance ($42=n_\mathrm{both}+b$ and $42=n_\mathrm{both}+c$ give $b=c$), and the exact two-sided McNemar test at $b=c$ yields $p=1.0$ identically---no detectable difference at this sample size, whatever the discordant count. For the speech task, no paired significance is attainable at the reported design as a matter of arithmetic: with five paired folds, the exact two-sided Wilcoxon signed-rank test's smallest achievable $p$-value is $2/2^{5}=0.0625$, so the fold count itself precludes significance at the conventional level regardless of the per-fold values; the deposited standby script \texttt{b\_paired\_stats.py} is the protocol of record for any expanded-fold analysis.
\section{Robustness to Dephasing, Phase Jitter, and Feedback Loss}
\label{s:sec:robustness}
This section quantifies the sensitivity of task performance to the three imperfections most relevant to hardware: pure dephasing of the atom, jitter of the round-trip phase, and loss in the delayed-feedback path. Protocol common to all three sweeps: NARMA10 at the main-text parameters ($\gamma=0.1$, $\tau=10$, $\varepsilon_\mathrm{INPUT}=0.1$, $\phi=\pi/3$ unless stated); noise simulated by stochastic unraveling of pure-state trajectories; one ``node'' denotes one delayed quadrature feature of the single measured output (successive columns of the delay-embedded $(P,Q)$ record; the readout dimension, as defined for Fig.~3 of the main text); features are averaged over 40 independent noise realizations before the readout is trained, matching the ensemble averaging inherent to experimental moment estimation. Throughout this section NRMSE is normalized by the target range (max--min), the convention of the benchmark literature; uncertainty bands are 95\% bootstrap confidence intervals obtained by resampling the 40 noise realizations with replacement before feature averaging and retraining. The three sweeps report slightly different noiseless reference errors---$0.078$ (dephasing), $0.0860$ (jitter), and $0.084$ at the loss-sweep optimum---because they were run with different training and fading-memory washout points and independently configured node grids and noise-ensemble handling (the dephasing and jitter sweeps average over 40 stochastic trajectories, whereas the loss sweep uses single deterministic-input realizations, as noted below); the values are therefore internally consistent within each sweep but are not a single shared baseline, and comparisons are made within a sweep rather than across the three reference numbers.
Table~\ref{s:tab:robbaselines} collects the three reference points and their ensemble conventions in one place.
\begin{table}[h]
\centering
\footnotesize
\setlength{\tabcolsep}{4pt}
\renewcommand{\arraystretch}{1.2}
\begin{tabular}{lccc}
\hline
Sweep & Noiseless ref. & Ensemble & Uncertainty\\
\hline
Dephasing (Fig.~\ref{s:fig:rob-dephasing}) & $0.078$ & 40 noise realizations & 95\% bootstrap CI\\
Phase jitter (Fig.~\ref{s:fig:rob-jitter}) & $0.0860$ & 40 per-trajectory draws & 95\% bootstrap CI\\
Feedback loss (Fig.~\ref{s:fig:rob-loss}) & $0.084$ (optimum) & single deterministic run & none (no seed spread)\\
\hline
\end{tabular}
\caption{Reference baselines of the three robustness sweeps. The three noiseless references differ because the sweeps were run with different training and fading-memory washout points and independently configured node grids and ensemble handling; comparisons are made within a sweep only. The absence of a confidence interval on the loss sweep is a stated limitation of that sweep rather than of the other two.}
\label{s:tab:robbaselines}
\end{table}

\begin{figure}[h]
    \centering
    \includegraphics[width=.66\linewidth]{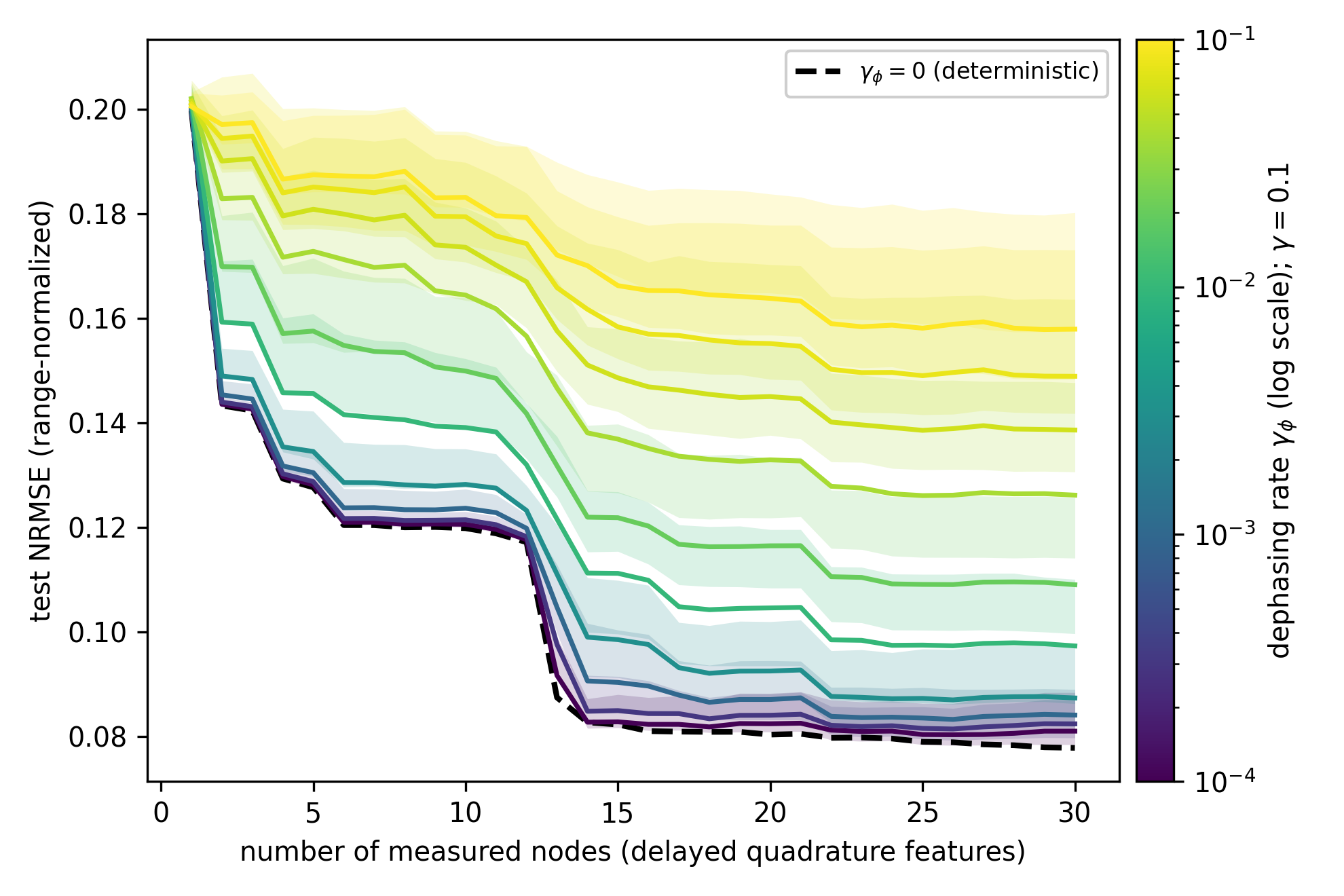}\\[2pt]
    \includegraphics[width=.55\linewidth]{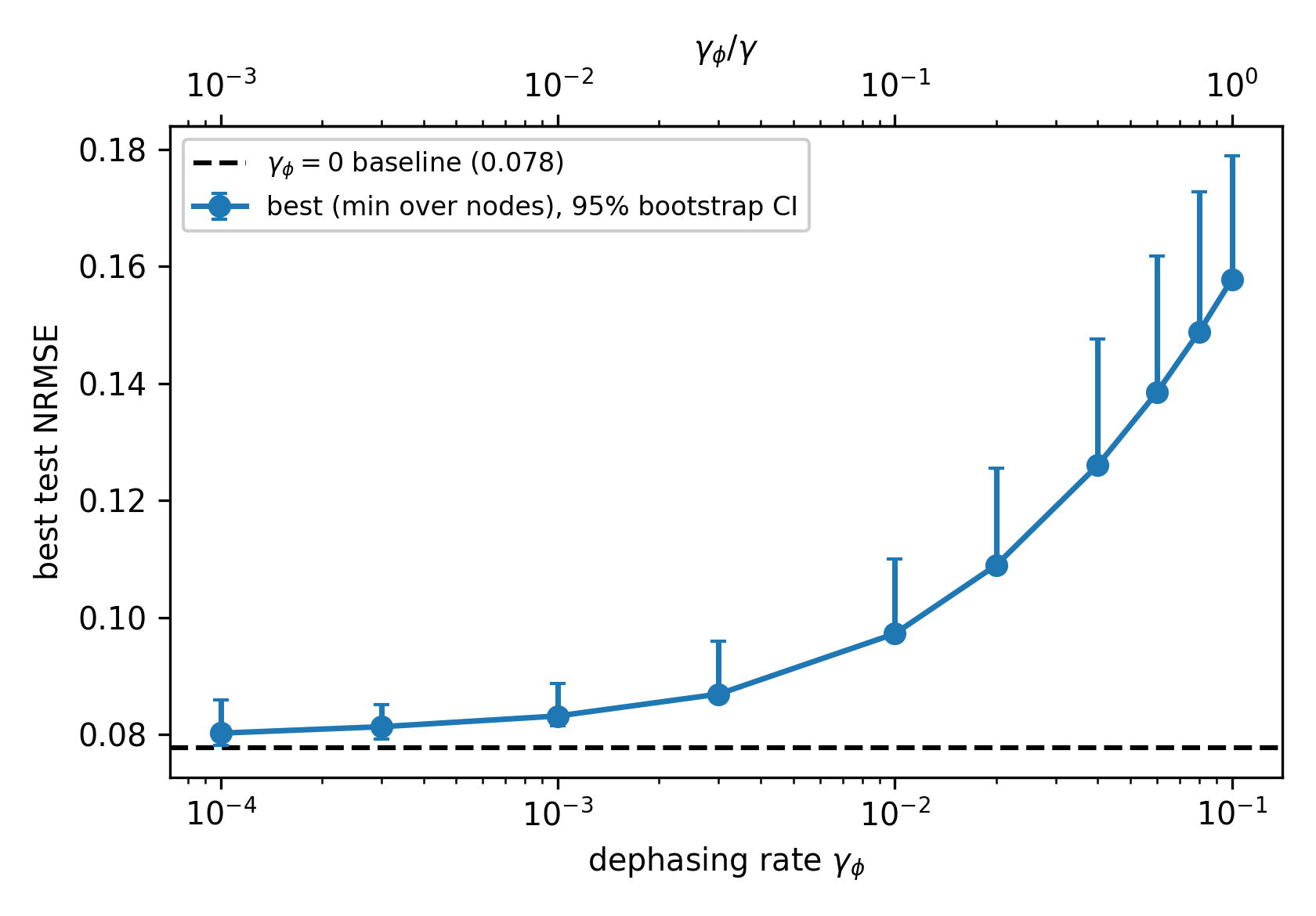}
    \caption{Pure dephasing. Top: NARMA10 test NRMSE versus number of measured nodes for logarithmically spaced dephasing rates (color bar); shaded bands, 95\% bootstrap CI over the 40 noise realizations. The $\gamma_\phi=0$ baseline (dashed) is deterministic and carries no band. Bottom: best (minimum-over-nodes) NRMSE versus $\gamma_\phi$ on a logarithmic axis, with bootstrap CIs; top axis, the ratio $\gamma_\phi/\gamma$. Degradation is graceful, and the useful node count shrinks as dephasing broadens the dressed modes.}
    \label{s:fig:rob-dephasing}
\end{figure}

\paragraph{Pure dephasing.}
Figure~\ref{s:fig:rob-dephasing} sweeps the dephasing rate over a logarithmically spaced grid $\gamma_\phi\in\{0,10^{-4},3\times10^{-4},10^{-3},3\times10^{-3},10^{-2},\ldots,10^{-1}\}$. Degradation is graceful, with no cliff: relative to the noiseless best error $0.078$, dephasing at $\gamma_\phi/\gamma=10^{-2}$ raises the error from $0.078$ to $0.083$, $\gamma_\phi/\gamma=10^{-1}$ costs $\sim$25\% ($0.097$), and at $\gamma_\phi=\gamma$ the error reaches $0.158$, roughly double this sweep's own noiseless reference: full-rate dephasing erases the non-Markovian benefit. (The severed-feedback plateau is a quantity of the loss sweep and is not compared against here, per the within-sweep rule stated above.) Two structural features are visible in the node-resolved curves. First, node-scaling saturates progressively earlier as $\gamma_\phi$ grows, consistent with the mode-resolution picture: dephasing broadens the dressed lines, and once $\gamma_\phi$ exceeds the relevant splittings, additional measured nodes carry no new information, giving a maximum useful node count that shrinks with $\gamma_\phi$. Second, the step structure of the noiseless curve (in particular the drop near 13 nodes) survives at small $\gamma_\phi$ and washes out at large $\gamma_\phi$, as expected for features that live in inter-mode coherences; whether these steps admit the same polynomial-degree-boundary interpretation as the staircase of main-text Fig.~3 is suggestive but not exact. For the platforms discussed in the main text, atomic dephasing ratios of $\gamma_\phi/\gamma\sim10^{-3}$--$10^{-2}$ are routine, placing them in the $\lesssim$10\% penalty regime.

\begin{figure}[h]
    \centering
    \includegraphics[width=.66\linewidth]{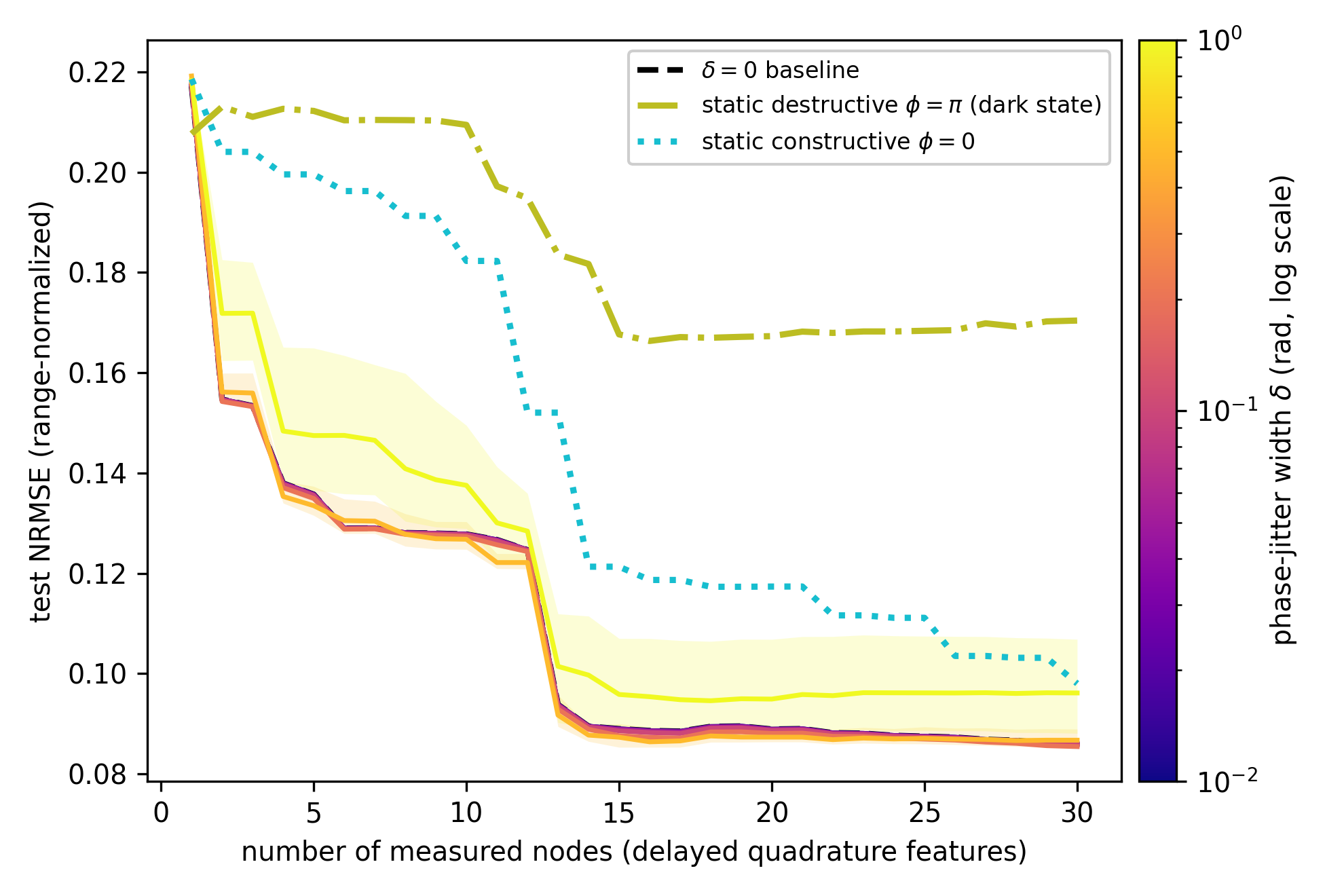}\\[2pt]
    \includegraphics[width=.55\linewidth]{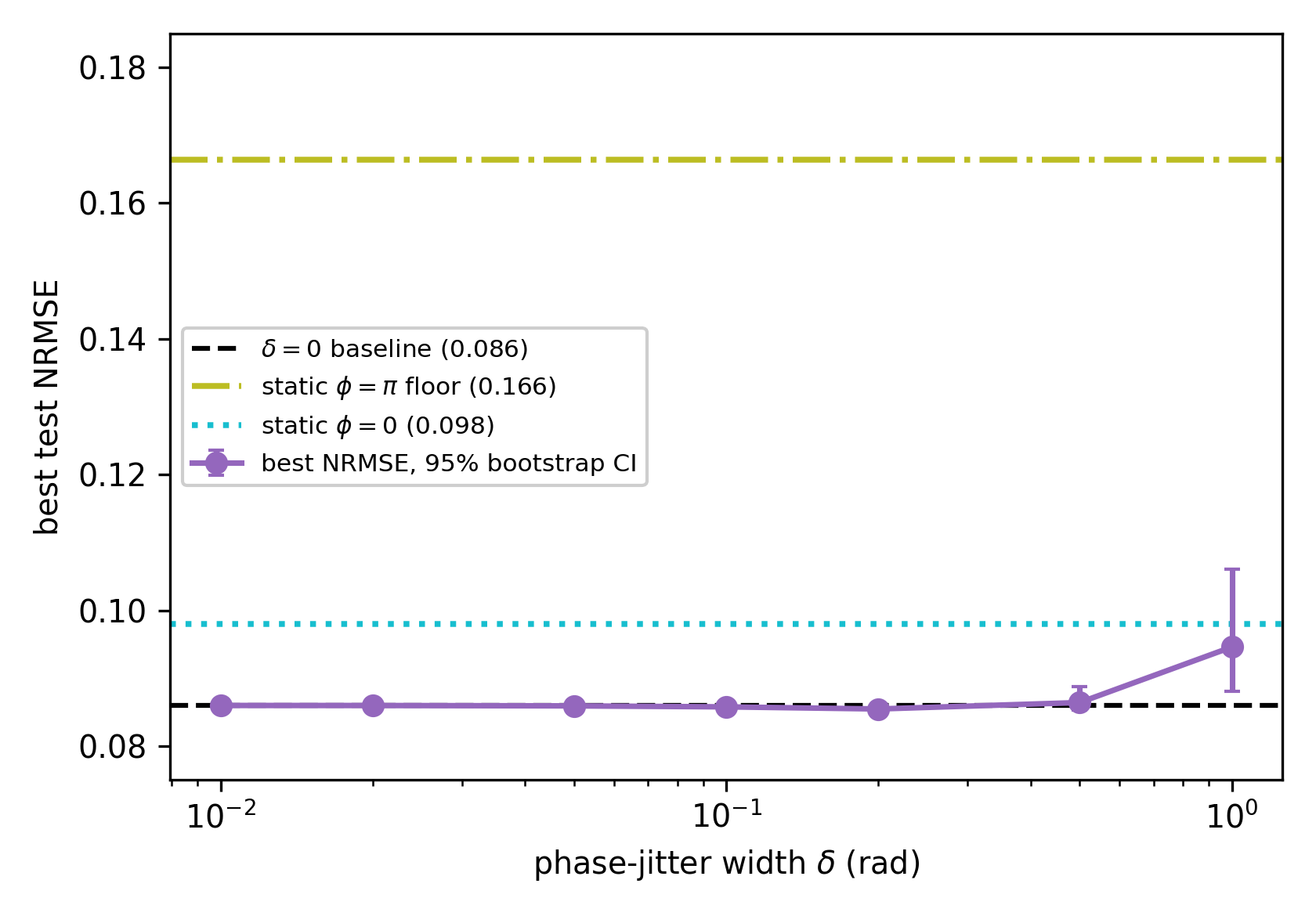}
    \caption{Round-trip phase jitter. Top: test NRMSE versus measured nodes for jitter widths $\delta$ (per-trajectory Gaussian draws; bands, 95\% bootstrap CI over 40 realizations), with the static constructive ($\phi=0$) and destructive ($\phi=\pi$) anchors. Bottom: best NRMSE versus $\delta$ (log axis) with bootstrap CIs against the baseline and static-anchor reference lines. The performance floor is the dark-state phase $\phi=\pi$ predicted by the delay-stability analysis; jitter is tolerated to at least $\delta\approx0.5$\,rad, the CIs overlapping the baseline throughout that range.}
    \label{s:fig:rob-jitter}
\end{figure}

\paragraph{Round-trip phase jitter.}
Figure~\ref{s:fig:rob-jitter} adds Gaussian jitter of standard deviation $\delta$ (radians) to the round-trip phase, drawn per trajectory (modeling slow interferometric drift of the atom--mirror distance, i.e.\ drift on timescales long compared with a round trip; fast intra-round-trip phase noise is outside the simulated regime), together with two static anchors: the fully constructive phase $\phi=0$ and the fully destructive phase $\phi=\pi$. Performance is statistically flat through $\delta=0.5$\,rad---the 95\% bootstrap CIs overlap the baseline across that entire range, and the shallow minimum at $\delta=0.2$ ($0.0855$ against baseline $0.0860$) is well within CI, so we do not interpret it---and loses $\sim$10\% at $\delta=1$ ($0.095$). The static anchors carry the theoretical content: the performance floor is the destructive phase $\phi=\pi$---precisely the dark-state configuration at which the delay-stability condition $D(c_0)$ of Theorem~1 becomes marginal (Sec.~\ref{s:sec:kerneldecay})---while the constructive phase $\phi=0$, which maximizes the collective decay and thus shortens the memory, is also inferior to the intermediate operating point $\phi=\pi/3$. The task-level simulation thus reproduces, dynamically, the stability landscape derived analytically from the loop characteristic equation. As an engineering statement: interferometric stability of the atom--mirror path to $\sim$0.5\,rad suffices ($\sim\lambda/13$ of the \emph{round-trip} path $2L$, equivalently $\sim\lambda/25$ of the one-way atom--mirror distance $L$), a comfortable requirement on the platforms considered. The per-trajectory (slow-drift) model is the appropriate regime for those platforms: the round trip is $\tau=2L/v$---nanoseconds or less for the superconducting transmission lines and trapped-atom geometries cited in the main text---whereas the dominant phase noise in both, thermal and mechanical drift of the atom--mirror path length, occurs on millisecond-and-slower timescales. The round-trip phase is therefore effectively static within any single trajectory and varies between them, which is exactly the sampling this sweep implements; phase noise with appreciable spectral weight at the round-trip frequency would require a separate treatment and is not claimed here.

\begin{figure}[h]
    \centering
    \includegraphics[width=.94\linewidth]{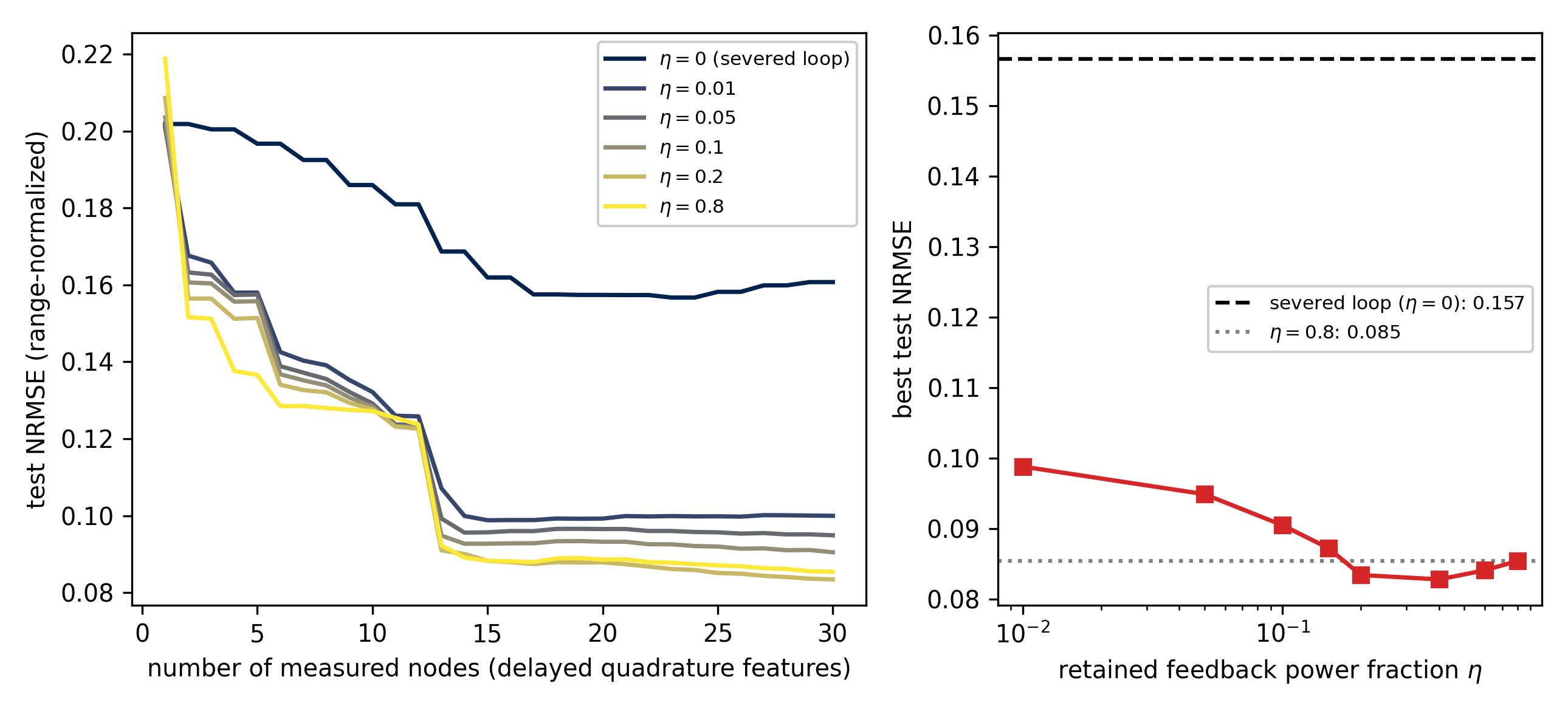}
    \caption{Feedback loss. Left: test NRMSE versus measured nodes as the retained feedback power fraction $\eta$ is varied (deterministic-input runs, one realization per $\eta$; no noise ensemble, hence no bands). Right: best NRMSE versus $\eta$ over the full nine-point grid. $\eta=0$ (severed loop) collapses to the feedback-free plateau; $\eta=\NumAPAB$ (a 20\,dB round-trip loss) already recovers most of the severed-to-optimum gap, returns saturate beyond $\eta\approx0.1$, and the shallow minimum near $\eta\approx0.2$--$\NumAPE$ is left uninterpreted (single realizations).}
    \label{s:fig:rob-loss}
\end{figure}

\paragraph{Feedback loss.}
Figure~\ref{s:fig:rob-loss} attenuates the delayed-feedback path, with $\eta$ the retained power fraction of the returning field ($\eta=0$ severs the loop entirely). Severing the feedback collapses performance to a high plateau ($0.157$, the level of the feedback-free device), directly executing the falsifier stated in the main text: if performance were insensitive to the feedback, the trained readout rather than the physics would be doing the computation. It is not---and the recovery is strikingly nonlinear in $\eta$: restoring just $\eta=0.01$ (a 20\,dB round-trip loss) already recovers most of the severed-to-optimum gap (best error $0.099$, against $0.084$ at the optimum and $0.157$ severed; recovering $0.157-0.099$ of the $0.157-0.084$ severed-to-optimum gap, a comparison we quote only as an order-of-magnitude indication because this sweep, unlike the dephasing and jitter sweeps, uses single deterministic-input realizations and therefore carries no seed spread or confidence interval), returns saturate beyond $\eta\approx0.1$, and the shallow minimum near $\eta\approx0.2$--$0.4$ visible in the nine-point grid is left uninterpreted, these being single deterministic-input realizations. The non-Markovian resource is therefore not only necessary for the performance but remarkably loss-tolerant: even one percent of the returning power carries most of the computational benefit, which substantially relaxes the insertion-loss budget of any experimental implementation.

\section{Numerical Verification of the Universality Properties}
\label{s:sec:univexamples}

\begin{figure}
    \centering
    \includegraphics[width=1\linewidth]{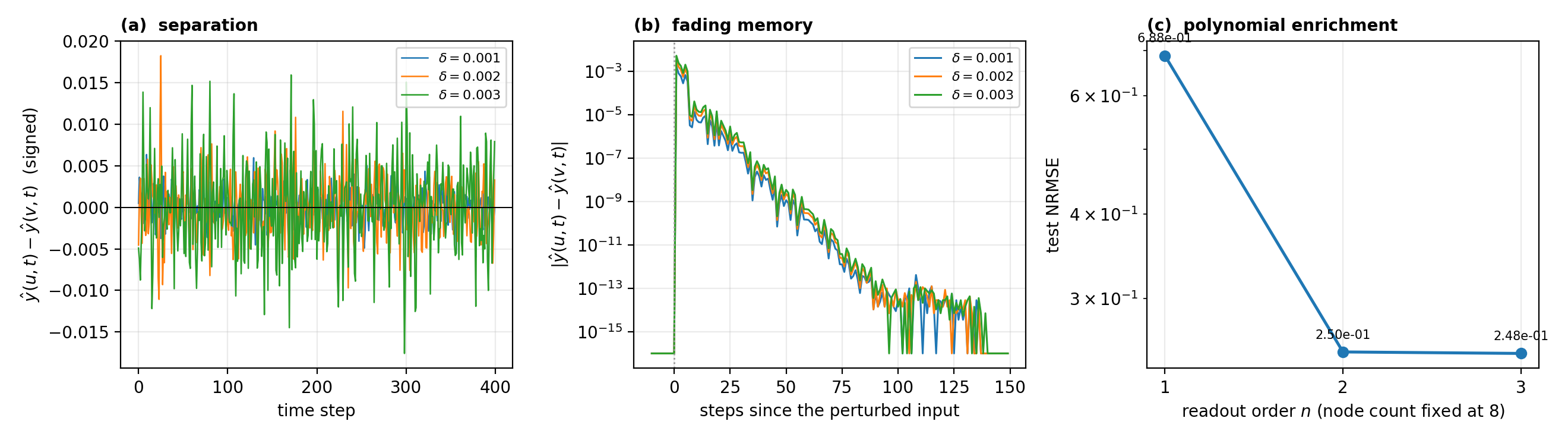}
    \caption{Numerical demonstration of separation, fading memory, and polynomial enrichment. Parameters: $\Delta t=1$, $\gamma=\NumAPB$, $\epsilon=\NumAPBF$, $\theta=\phi=\pi/3$, $\tau=10$.}
    \label{s:fig:universality}
\end{figure}

Figure~\ref{s:fig:universality} reports the three classical sufficient conditions for reservoir universality (separation, fading memory, and polynomial enrichment) for the non-Markovian reservoir. They are an independent sanity check rather than ingredients of the proof of Theorem~1, which rests instead on the Volterra expansion, the resolution of the eigenvalue sums, the Vandermonde inversion, the extrapolation identity and the tail bound; fading memory of the device is established separately in Sec.~\ref{s:sec:fadingmemory}. Separation is shown by applying small perturbations to the entire input: even minute perturbations yield distinguishable output trajectories, so nearby input histories map to distinct reservoir states. Fading memory is shown by perturbing a single input point near $t=100$: the response rises then decays, so recent inputs dominate over distant ones. Polynomial enrichment is shown by forming nonlinear combinations of collected states, which enlarge the readout space and improve approximation with saturating gains as the node count grows. These are numerical evidence for the ingredients that the proof establishes analytically in the linear-transducer limit.

\section{Supplementary Discussion: the Stone--Weierstrass route, and what the atom contributes}
\label{s:sec:suppdisc}
This section carries, at full length, two discussions summarized in the main text: why the theorem is proven by the constructive Volterra-kernel route although the abstract density route is also available, and the strongest form of the classical-linear-optics objection together with our answer.

We emphasize that universality \emph{itself} is also within reach of the standard, shorter route: the reservoir outputs contain the constants, form an algebra---sums are realized by adding weights, and products of outputs are outputs of higher readout order, under the same spectral non-degeneracy the theorem's third condition supplies---and, ranged over the nested family of operating points, they separate distinct input histories, so the Stone--Weierstrass theorem delivers density in the continuous fading-memory functionals on $\mathcal{K}_{u_{\max}}$, exactly as in prior universality results for reservoir classes~\cite{GRIGORYEVA2018495, chen2019learning}. We prove Theorem~1 of the main text constructively not because the abstract route fails here, but because of what it cannot say. The Volterra-kernel route provides this where a Stone--Weierstrass argument cannot: it certifies that scaling the one device up is never wasted, rather than merely asserting that some adequate member of a class exists---and it does so with explicit rates and constants, which no density argument produces. The containment is what is proven outright; the strictness is generic in the same sense as the theorem's conditions. Separation, fading memory, and polynomial enrichment are additionally verified numerically for the simulated device in Sec.~S.11.  

We are careful about what does the work in the proof, since it is easy to overstate. The essential ingredient is a \emph{linear multimode} structure with generically non-resonant frequencies and a tunable readout---and linear multimode structure is not the exclusive property of a quantized field: classical wave optics supplies it too, in the independent spatial or spectral modes of a linear optical network. The relevant dichotomy is therefore not quantum-versus-classical but \emph{nonlinear single node with virtual (time-multiplexed) nodes} versus \emph{linear field with genuinely independent modes}. The single-node delay-line reservoir \cite{appeltant2011, larger2017high}, the atom--mirror system's closest \emph{architectural} relative, sits on the first side: its virtual nodes are time-multiplexed samples of one nonlinear trajectory rather than independent degrees of freedom, which is precisely why it has resisted a universality theorem. The atom--mirror device sits on the second: the theorem becomes available for this instance because its delay loop decomposes into independent linear modes---the structure the classical device's nonlinear node destroys (the mode-space section of the main text). What quantization contributes specifically, in one atom, is the packaging of three ingredients into a single passive component: input encoding on the atomic drive, the mirror-selected independent mode structure that carries the recurrence, and (beyond the Gaussian limit) the atom's saturable nonlinearity, which moves the nonlinearity into the hardware and collapses the Gaussian readout's $n=9$ shot budget to $n=1$ (Sec.~S.7). We prove the theorem for the quantized instance and identify why the classical delay-line original cannot inherit it; we do not claim quantization is the only route to a linear multimode reservoir.

That concession invites the sharpest form of the objection, and we state it in full rather than leave it implicit. If linear multimode structure is what the proof needs, and classical linear optics supplies it, then in the regime where our theorem holds (the Gaussian limit, where the atom is a linear transducer) a classical linear optical network with a comparable mode structure would appear to be an equally good instance; while in the regime where the atom's own nonlinearity matters, we have no theorem. Our answer is not that the atom is doing something a classical field cannot, because in the Gaussian limit it demonstrably is not. The answer is that the two regimes are the same device at two settings of one dial, and that the claim being made is about \emph{packaging} rather than about quantum supremacy of any kind. A classical linear multimode network can host the recurrence, but it must import its nonlinearity and its input encoding from separate components---a modulator, a nonlinear element, or a polynomial readout paid for in the factorial shot budget of Sec.~S.7. The atom--mirror device carries all three in one passive object: the drive encodes, the mirror-selected modes recur, and the same atom that acts as a linear transducer at small drive becomes the saturable nonlinearity at larger drive, continuously and with a bounded, computable gap between the two (Sec.~S.5). What the theorem certifies is that this packaging is not paid for in expressivity: at the setting where the device is simplest to analyze, it is already universal. The minimality claim is therefore about component count for a fixed capability, and it survives the concession intact---but it is a claim about one atom replacing an assembly rather than about quantum mechanics enabling a computation classical optics could not perform.

\bibliography{SRC}